\documentclass[sn-mathphys-num]{sn-jnl}

\usepackage{graphicx}%
\usepackage{multirow}%
\usepackage{amsmath,amssymb,amsfonts}%
\usepackage{amsthm}%
\usepackage{mathrsfs}%
\usepackage[title]{appendix}%
\usepackage{xcolor}%
\usepackage{textcomp}%
\usepackage{manyfoot}%
\usepackage{booktabs}%
\usepackage{algorithm}%
\usepackage{algorithmicx}%
\usepackage{algpseudocode}%
\usepackage{listings}%
\usepackage{wrapfig}

\usepackage{enumitem}
\usepackage{times,xcolor,stackengine}
\usepackage{url} 
\usepackage{comment}
\usepackage{titlesec} 
\usepackage{bm,dsfont}
\usepackage[scr=boondoxo,scrscaled=1.05]{mathalfa}
\usepackage{subcaption}
\usepackage{placeins}
\usepackage{epstopdf}
\usepackage{siunitx}
\usepackage{comment}
\colorlet{darkred}{red!85!black}
\colorlet{darkgreen}{green!50!black}
\colorlet{darkblue}{blue!60!black}

\theoremstyle{thmstyleone}%
\theoremstyle{thmstyletwo}%

\theoremstyle{thmstylethree}%

\DeclareRobustCommand{\vect}[1]{
  \ifcat#1\relax
    \boldsymbol{#1}
  \else
    \mathbf{#1}
  \fi}

 \newcommand{\ti}{t_{\iota}}
 \newcommand{\tf}{t_{\styleB{f}}}
 
\newcommand{\styleA}[1]{\mathrm{#1}} 
\newcommand{\styleB}[1]{\mathfrak{#1}} 
\newcommand{\styleC}[1]{{#1}}  
\newcommand{\styleD}[1]{\mathscr{#1}}
 
 \newcommand{\tti}{\styleA{t}_{\iota}}
\newcommand{\ttf}{\styleA{t}_{\styleB{f}}}
\newcommand{\nt}{\styleA{t}}
\newcommand{\ns}{\styleA{s}}
\newcommand{\nq}{\styleA{q}}
\newcommand{\np}{\styleA{p}}
\newcommand{\nx}{\styleA{x}}
\newcommand{\nv}{\styleA{V}}
\newcommand{\nU}{\styleA{U}}
\newcommand{\nf}{\styleA{f}}
\newcommand{\nl}{\styleA{L}}

 \newcommand{\st}[1]{\styleA{t}_{#1}}
 \newcommand{\slt}[1]{\bm{\styleA{t}}_{#1}}

 \newcommand{\av}[1]{\left\langle#1\right\rangle}
 \newcommand{\cbr}[1]{\left(#1\right)}
 \newcommand{\bcbr}[1]{\big(#1\big)}

\newcommand{\vv}{;}

\newcommand{\syma}{\mbox{\textbf{[KL]}}}
\newcommand{\symb}{\mbox{\textbf{[EP]}}}
  
\titleformat{\subsubsection}[runin]
  {\normalfont\normalsize\bfseries}{\thesubsubsection}{1em}{}

\begin{document}

\title[Optimal Control of Underdamped Systems: An Analytic Approach]{Optimal Control of Underdamped Systems: An Analytic Approach}


\author[1]{\fnm{Julia} \sur{Sanders}}
\equalcont{These authors contributed equally to this work.}

\author*[2]{\fnm{Marco} \sur{Baldovin}}\email{marco.baldovin@cnr.it}
\equalcont{These authors contributed equally to this work.}

\author[1]{\fnm{Paolo} \sur{Muratore-Ginanneschi}}
\equalcont{These authors contributed equally to this work.}
\affil[1]{\orgdiv{Department of Mathematics and Statistics}, \orgname{University of Helsinki}, \orgaddress{\city{Helsinki}, \postcode{00014}, \country{Finland}}}

\affil*[2]{\orgdiv{Institute for Complex Systems}, \orgname{CNR}, \orgaddress{ \city{Rome}, \postcode{00185}, \country{Italy}}}


\abstract{
Optimal control theory deals with finding 
protocols to steer a system between assigned initial and final states, such that a trajectory-dependent cost function is minimized. The application of optimal control to stochastic systems is an open and challenging research frontier, with a spectrum of applications ranging from stochastic thermodynamics to biophysics and data science. Among these, the design of nanoscale electronic components motivates the study of underdamped dynamics, leading to practical and conceptual difficulties. 

In this work, we develop analytic techniques to determine protocols steering finite time transitions at a minimum thermodynamic cost for stochastic underdamped dynamics. As cost functions, we consider two paradigmatic thermodynamic indicators. The first is the Kullback-Leibler divergence between the probability measure of the controlled process and that of a reference process. The corresponding optimization problem is the underdamped version of the Schr\"odinger diffusion problem that has been widely studied in the overdamped regime. The second is the mean entropy production during the transition, corresponding to the second law of modern stochastic thermodynamics.

For transitions between Gaussian states, we show that optimal protocols satisfy a Lyapunov equation, a central tool in stability analysis of dynamical systems. For transitions between states described by general Maxwell-Boltzmann distributions, we introduce an infinite-dimensional version of the Poincar\'e-Linstedt multiscale perturbation theory around the overdamped limit.  
This technique fundamentally improves the standard multiscale expansion. Indeed, it enables the explicit computation of momentum cumulants, whose variation in time is a distinctive trait of underdamped dynamics and is directly accessible to experimental observation. 
Our results allow us to numerically study cost asymmetries in expansion and compression processes and make predictions for inertial corrections to optimal protocols in the Landauer erasure problem at the nanoscale.  
  }




\maketitle

\section{Introduction}
\label{sec:introduction}

In his remarkable paper \cite{SchE1931} (English translation in \cite{ChMGSc2021}), Schr\"odinger addresses the problem of statistical reversibility
of a physical system in contact with an environment. In doing so, he puts forward the idea of using entropic indicators 
to quantify deviations from thermodynamic equilibrium and, therefore, dissipation.  
	Schr\"odinger identifies 
what
is now commonly known as the \emph{Kullback-Leibler divergence} or \emph{relative entropy}~\cite{CoTh2006} as a quantifier between the joint probability distribution of the system's end states and those of a free diffusion.

In the last decades of the 20th century, Schr\"odinger's trailblazing idea was reformulated into the language of stochastic optimal control~\cite{FoeH1988,DaiP1991,FoGa1997}, refining  Schr\"odinger's original ``static bridge problem''~\cite{SchE1931} into a 
``dynamic Schr\"odinger bridge", where the relative entropy is computed between probability measures over the systems' pathspace \cite{LeoC2014,ChGePa2021}. 
Schr\"odinger bridges have active research interest because they allow computational optimal transport methods to be applied to dynamical models. This enables efficient computation in fields such as neuroscience 
\cite{TodE2004,TodE2009b};  data science and machine learning \cite{PeCu2019}; and generative modeling, sampling, and dataset imputation \cite{DeBoThHeAr2021,VaOvFeGiLa2021}. 
Technological advances in the last two decades have paved the way for the observation and manufacturing of nanomachines. At nanoscale, random fluctuations of thermal and topological origin may swamp out any mechanical behavior \cite{KaLeZe2007}. A fundamental question is, therefore, how natural or artificial nanosystems can efficiently harness randomness in order to generate controlled motion or perform thermodynamic work on larger scales. 
Schr\"odinger bridges find an optimal control protocol to rectify a system obeying stochastic dynamics, thus making it possible to devise systematic methods characterizing the efficiency of nanomachines~\cite{FiHo2005}. 

In addition, the discovery of fluctuation relations (see Chapter~4 of \cite{PePi2020} for a thorough conceptual and historical account) introduces a substantial development with respect to \cite{SchE1931}. For Markov stochastic processes, fluctuation relations stem from considering the relative entropy of probability measures connected by a time reversal \cite{MaeC1999,MaReMo2000,ChGa2008}: the corresponding generalized Schr\"odinger bridges minimize the thermodynamic cost associated with the transition. This means that we can consider thermodynamic cost functionals other than the Kullback-Leibler divergence~\cite{schmiedl2007,GoScSe2008,EsVadeBr2011,AuMeMG2011,SiCr2012,RoCrVE2017,baldovin2023,ChRo2023}.
Because the Kulllback-Leibler divergence is non-negative, 
the minimum mean entropy production in finite time transitions between assigned probability distributions is strictly larger than zero. Remarkably, the overdamped dynamics minimizer \cite{AuGaMeMoMG2012,GawK2013}  turns out to be the solution of a system of Monge-Amp\`ere-Kantorovich optimal mass transport equations \cite{VilC2009}.
Two results of \cite{AuGaMeMoMG2012,GawK2013} stand out. First, minimizers can be determined by efficient numerical algorithms even in the multi-dimensional case~\cite{BrFrHeLoMaMoSo2003}. 
Second, the minimum entropy production is proportional to the squared Wasserstein distance between the probability distributions of the end states divided by the duration of the control horizon. 
This relation between mean entropy production and squared Wasserstein distance continues to hold as scaling limits for Markov jump processes \cite{MGMePe2012} and underdamped dynamics \cite{PMG2014,MGSc2014}, see also \cite{ShFuSa2018,ReSe2021}. 

A detailed description of optimal protocols in the underdamped regime is urgent for several reasons. Optically levitated nanoparticles have become a common tool to study transitions in stochastic thermodynamics. Stable confinement and manipulation of nanoparticles within optical traps requires an account of the momentum dynamics. For instance, particle-environment energy exchanges during isoentropic (isochoric) transition within Brownian Carnot (Stirling) engines occurs through the momentum degrees of freedom \cite{MaRoDiPePaRi2016,DiMaRoPaRi2016}. Understanding how to simultaneously control particles' position and momentum is required to devise robust shortcuts to equilibration protocols \cite{guery2019,RaGuGuOdTr2023,plata2021,baldovin2022,guery2023driving}. 

A further motivation comes from the design of nanoscale electronic components~\cite{lopez2016sub, Deshpande2017, ciampini2021experimental}. 
Increasing the efficiency of such operations toward the bound prescribed by Landauer is a non-trivial task, with potentially relevant consequences for the design of information and computation technology~\cite{LeOrPoSn2019}.
 The presence of inertia has been shown to lower the energetic cost needed to perform logic operations on bits~\cite{lopez2016sub, ray2021, dago21}. This has sparked interest in the control of underdamped stochastic systems, with particular emphasis on the non-linear case, needed for the description of information bits \cite{PrEhBe2020,PrEhBe2020b,ZhEgMoDa2021,DeBoThHeAr2021}. Ad-hoc experimental solutions have been found to realize controlled protocols for stochastic dynamics with inertia, confirming that inertial effects allow for fast and precise bit operations~\cite{dago21, GonzalezBallestero2021, dago22, dago2023virtual}.  

With these motivations in mind, we introduce a systematic analytical derivation of optimal protocols for the 
underdamped dynamics. We consider two paradigmatic cases of running costs:
\begin{itemize}
	\item The underdamped version of the Schr\"odinger dynamic bridge problem, referred to as \hyperref[itm:a]{KL}. 
 The cost functional in this case is the Kullback-Leibler divergence between the probability measure of the controlled process and that of an assigned reference process.  In \cite{SchE1931}, Schr\"odinger motivates this cost functional as a quantifier of the likelihood of non-equilibrium fluctuations, i.e., 
 a large deviation functional in the current literature. Since then, Schr\"odinger dynamic bridge problems have emerged as relevant efficiency measure for diffusion-mediated transport processes with applications ranging from cybernetics~\cite{TodE2009b} to molecular-scale engines~\cite{FiHo2005}. More recently, it has been realized that when the reference process is a diffusion subject to inertia in the absence of a confining potential, Schr\"odinger dynamic bridge problems provide a viscous regularization of optimal mass transport \cite{MikT2004,PMG2013,ChGePa2016}. This regularization has applications in machine learning~\cite{PeCu2019}. Finally, when the reference process describes motion in a confining potential equal to that in the Maxwell-Boltzmann distribution of the final state, we obtain a model of an optimally controlled shortcut to adiabaticity.  
	\item The minimization of the mean entropy production, referred to as \hyperref[itm:b]{EP}. This is the cost functional characterizing the Second Law of thermodynamics and Landauer's principle \cite{LeOrPoSn2019}. We study this problem in the most general formulation compatible with detailed balance, and which can be self-consistently derived from Hamiltonian mechanics of a system coupled to an infinite bath described by harmonic oscillators. In this generalized formulation, the underdamped entropy production explicitly depends upon the control via a non-dimensional parameter $g$. Physically, $g$ describes the intensity of momentum coupling between the system and bath. Interactions of these type have been recently observed in Josephson junctions \cite{CuFuToVa2001,AnPo2007}.
\end{itemize}

Besides the cost functional, the specification of an optimal control problem requires a definition of the functional space over which to carry out the optimization. This functional space is called the class of admissible controls. Our focus is on the class of admissible controls described by functions that are sufficiently regular to be differentiable in space and continuous time. Physically, this corresponds to the requirement that the control be slow with respect to the fastest time scale in the problem set by the Wiener process modeling the interaction to the bath. 
Under these hypotheses, we derive the stationary equations for the cost functionals
by taking variations over the class of admissible controls specified by confining mechanical smooth potentials. In such a case \cite{MGSc2014}, extremals of the cost solve a set of integro-differential equations, with features reminiscent of the Vlasov-Poisson-Fokker-Planck problem \cite{BoCaSo1997}.

We obtain the following main results:
\begin{enumerate}
	\item[I] We show that the cumulants of the probability measure describing transitions between Gaussian states are amenable to the solution of a Lyapunov system of equations~\cite{BhEl2002} in any number of dimensions. This immediately yields a body of rigorous results concerning existence, uniqueness and, when applicable, positivity of solutions (Section~\ref{sec:G}).
	\item[II] For transitions between states described by Maxwell-Boltzmann distributions in phase space, we introduce an infinite dimensional extension of Poincar\'e-Lindstedt multiscale perturbation theory \cite{VerF2005} around the overdamped limit. This method allows us to treat all cumulants of the system probability measure on the same footing in the renormalization group fashion \cite{AmiD2005}. We hence obtain explicit predictions for the behavior in time of all phase space cumulants within second order accuracy. The method builds on ideas introduced in \cite{WyBa1987,WyBa1987c} for dissipative and \cite{GenG2010} for conservative dynamics. Although we restrict our analysis to a two-dimensional phase space, the analysis of the Gaussian case shows that extension to higher dimensional phase spaces is possible, albeit cumbersome (Section~\ref{sec:ms}).
	\item[III] In the case of mean entropy production by an underdamped dynamics with purely mechanical coupling, our results support tightness of the lower bound provided by the overdamped dynamics \cite{ChCoGrRe2021,MGPe2023}. For more general couplings, both the mean entropy production and the cost of the dynamic Schr\"odinger bridge receive strictly positive corrections in the presence of inertia (Section~\ref{sec:mcd}). 
    \item[IV] The cost of expansion is higher than that of compression when the initial states are thermodynamically equidistant (section~\ref{sec:exco}). This result is a manifestation of intrinsic asymmetries in thermal kinematics, recently pointed out in~\cite{LaGo2020,IbDiLaGoRi2024}.
\end{enumerate}
The structure of the paper is as follows. In section~\ref{sec:model}, we introduce the model of underdamped dynamics of a nanosystem weakly coupled to an environment by both mechanical and momentum dissipation interaction. When the intensity of the momentum coupling $g$ vanishes, the model recovers the most widely applied underdamped dynamics. Next, we consider two thermodynamic cost functionals, \hyperref[itm:a]{KL} and \hyperref[itm:b]{EP}, and motivate their broad interest for applications to physics and other applied sciences.
Our goal is to minimize these functionals towards the mechanical potential $U_{t}$ governing the underdamped dynamics conditioned on the system's initial and final probability distributions. For this reason, we present a brief overview of the mathematical results leading to known bounds for the cost functionals \hyperref[itm:a]{KL} and \hyperref[itm:b]{EP} in the second half of the section. 
 
In Section~\ref{sec:oc}, we introduce the Pontryagin-Bismut functional and derive its stationary equations. The Pontryagin-Bismut functional provides a description of optimal control dual to Bellman's principle. 

Section~\ref{sec:G} focuses on the Gaussian case in a phase space of arbitrary dimension, and we derive our first main result here. 

In Section~\ref{sec:problem}, we set the stage for multiscale perturbation theory presented in Section~\ref{sec:ms}. As usual, the idea is to use slow scales to cancel secular terms. Our main goal is to obtain a detailed analytical description of experimentally measurable indicators. We therefore summarize the logic of the derivation and the results before proofs. Readers
only interested in our results may thus skip the second part of Section~\ref{sec:ms}.

In Section~\ref{sec:gauss_results}, we briefly return to the Gaussian case and provide the analytic expression of the solution of the cell problem of the multiscale expansion \cite{PaSt2008}. The solution of the cell problem allows us to determine all cumulants within second order accuracy in the overdamped expansion. 

Section~\ref{sec:num} applies the results with some numerical computations. We have emphasized the Gaussian case for two reasons. Firstly, methods for accurate numeric integration of the exact optimal control equations are immediately available, meaning we can compare the perturbative approach with exact numeric predictions in the case of Gaussian boundary conditions. Secondly, transitions between Gaussian states are well adapted to model Brownian engines \cite{ScSe2008,MGSC2015,DeKiLu2017,LaGo2020,IbDiLaGoRi2024}. We therefore also study the cost of optimal protocols driving isothermal expansions and compressions of a system to an equilibrium state, which are modelled by a dynamic Schr\"odinger bridge. 
Additionally, we solve the cell problem in the case of Landauer's erasure problem numerically and thus find inertial corrections to the erasure protocol, as well as predictions for the system's probability measure cumulants.

The final section is devoted to conclusions and outlook. We defer further supplementary material to the Appendices.

\section{Underdamped control model}
\label{sec:model}

We consider the dynamics of a nanosystem with mass $m$, whose position $\bm{\mathscr{q}}_{t}$ and momentum $\bm{\mathscr{p}}_{t}$ obey the Langevin--Kramers stochastic differential
equations in $\mathbb{R}^{2 d}$
\begin{subequations}
	\label{model:sde}
	\begin{align}
		&\label{model:sde1}\mathrm{d}\bm{\mathscr{q}}_{t}=\left(\dfrac{\bm{\mathscr{p}}_{t}}{m}
		-\dfrac{g\,\tau}{m}\,(\bm{\partial}U_{t})(\bm{\mathscr{q}}_{t})\right) \mathrm{d}t
		+\sqrt{ \dfrac{2\,g\,\tau}{m\,\beta}}\,\mathrm{d}\bm{\mathscr{w}}^{(1)}_{t}
		\\
		&\label{model:sde2}
		\mathrm{d}\bm{\mathscr{p}}_{t}=-\left(\dfrac{\bm{\mathscr{p}}_{t}}{\tau}
		+(\bm{\partial}U_{t})(\bm{\mathscr{q}}_{t})\right)\mathrm{d}t
		+\sqrt{ \dfrac{2\,m}{\tau\,\beta}}\,\mathrm{d}\bm{\mathscr{w}}^{(2)}_{t}\,.	
	\end{align}
\end{subequations}
In Eqs.~\eqref{model:sde}, $\bm{\mathscr{w}}^{(1)}_{t}$ and $\bm{\mathscr{w}}^{(2)}_{t}$ denote two $d$-dimensional independent Wiener processes. The Stokes time  $\tau$ is a constant parameter specifying the characteristic time scale of dissipation. 

In~\eqref{model:sde1}, a non-dimensional constant $g$ couples the mechanical force 
$\bm{\partial}U$ and the fluctuating environment modeled by the Wiener process $\bm{\mathscr{w}}^{(1)}_{t}$ to the nanosystem position dynamics. For any $g\geq 0$, Eq. \eqref{model:sde} guarantees convergence towards a Maxwell-Boltzmann equilibrium whenever the potential $U$ is time independent, confining and sufficiently regular. Setting $g$ to zero recovers the standard Langevin-Klein-Kramers model~\cite{PavG2014}. 

We emphasize that the dynamics described by~\eqref{model:sde} are consistent with the general analysis \cite{CaLe1983} of the conditions guaranteeing the self-consistency of the harmonic environment hypothesis. In fact,~\eqref{model:sde} can be obtained from a microscopic Hamiltonian dynamics, in which the system interacts with a {\textquotedblleft}bipartite  harmonic{\textquotedblright} environment \cite{MaAnRa2020}. By  bipartite harmonic environment, we mean an environment modeled by two kinds of oscillators: one type interacts with the system via the commonly assumed position-coupling \cite{ZwaR1973}, and the other via a linear momentum coupling \cite{CuFuToVa2001,AnPo2007}. Linear momentum coupling models \emph{momentum dissipation} observed e.g. in a single Josephson junction interacting with the blackbody electromagnetic field.

As for the force in~\eqref{model:sde}, we only assume that it is the negative gradient of a confining and sufficiently regular mechanical potential, i.e. a potential depending only on the system position. We suppose that potentials of this type give rise to an open set of controls. Within this set, the controls ensure that at every instant of time $t$ in a given time horizon $[\ti,\tf]$ the probability density of the system 
\begin{align}
	\operatorname{Pr}\left (\bm{x}\,\leq\,\begin{bmatrix}
		\bm{\mathscr{q}}_{t}	\\  \bm{\mathscr{p}}_{t}
	\end{bmatrix}< \bm{x}+\mathrm{d}^{2\,d}\bm{x}\right )=\styleC{f}_{t}(\bm{x})\,\mathrm{d}^{2 \,d}\bm{x}
	\nonumber
\end{align}
is well defined and satisfies the Fokker-Planck equation.

At an initial time $t=\ti $, we posit that the state of the nanosystem is statistically described by an assigned Maxwell-Boltzmann distribution at inverse temperature $\beta$:
\begin{align}
	\label{model:ini}
	\styleC{f}_{\ti}(\bm{q},\bm{p})=Z_{\iota}^{-1}\exp\left(
	-\dfrac{\beta\,\|\bm{p}\|^{2}}{2\,m}-\beta\,U_{\iota}(\bm{q})\right)
\end{align}
Furthermore, we require that at the end of the control horizon $t=\tf$ the probability density of the system 
satisfies the boundary condition
\begin{align}
	\label{model:fin}
	\styleC{f}_{\tf}(\bm{q},\bm{p})
	=Z_{\styleB{f}}^{-1}\exp\left(-\dfrac{\beta\,\|\bm{p}\|^{2}}{2\,m}-\beta\,U_{\styleB{f}}(\bm{q})\right)\,.
\end{align}
These assumptions on the probability distributions of the system end states are not necessary for the considerations that follow. They have the merit, however, to be both physically admissible and to lead to simplifications in the multiscale analysis of section~\ref{sec:ms}.

\begin{figure}
\centering
\includegraphics[width=0.7\linewidth]
{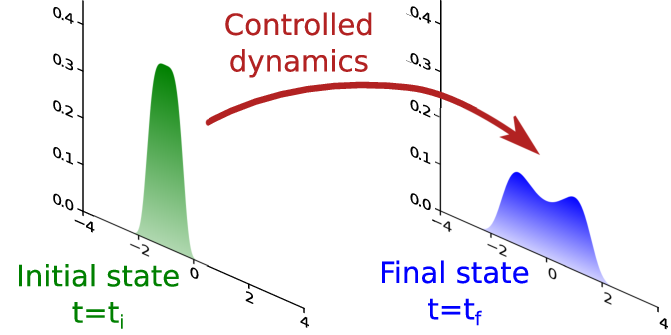}
\centering
\caption{Stylized representation of a Schr\"odinger bridge modeling Landauer's erasure of one bit of memory at minimum dissipation.}
\label{fig:bvp_landauer}
\end{figure}


The set of confining potentials $U_{t}$ that  give rise to phase space diffusions with probability marginals~\eqref{model:ini},~\eqref{model:fin} define the class of \emph{admissible controls} of~\eqref{model:sde}. 

Our aim is to determine the optimal mechanical potentials $U_{t}$ among the admissible ones that minimizes the thermodynamic cost functionals defined below conditioned on the initial and final probability distributions (\ref{model:ini}) and (\ref{model:fin}).

\subsection{Thermodynamic cost functionals}

We focus our attention on two physically relevant cases, hereafter referred to as~\textbf{KL}~and~\textbf{EP}. 
\medskip
\subsubsection*{KL:}\label{itm:a}
Underdamped dynamic Schr\"odinger bridge~ \cite{SchE1931}. The thermodynamic cost functional to minimize is the Kullback--Leibler divergence of the measure $\mathcal{P}=\mathcal{P}_{\iota}^{\styleB{f}}$ generated by~\eqref{model:sde} subject to~\eqref{model:ini} and~\eqref{model:fin}, 
    from the measure $\mathcal{Q}=\mathcal{Q}_{\iota}$ 
    generated by~\eqref{model:sde} when the mechanical force is $ \bm{\partial} U_{\star}$ and only the initial density~\eqref{model:ini} is assigned. The cost functional reads (see Appendix~\ref{app:KL})
	\begin{align}
 \begin{split}
 \operatorname{K}(\mathcal{P}\mathrel{\Vert}\mathcal{Q})&:=\operatorname{E}_{\mathcal{P}}\ln \dfrac{\mathrm{d}\mathcal{P}}{\mathrm{d}\mathcal{Q}} \\
 &=\dfrac{\beta\,\tau\,(1+g)}{4\,m}\operatorname{E}_{\mathcal{P}}\int_{\ti}^{\tf}\hspace{-0.2cm}\mathrm{d}t\, \|(\bm{\partial}U_{t})(\bm{\mathscr{q}}_{t})-(\bm{\partial}U_{\star})(\bm{\mathscr{q}}_{t})\|^{2}\,.
	\label{model:KL}
  \end{split}
\end{align}
	The notation $\operatorname{E}_{\mathcal{P}} $ emphasizes that the expectation value over the diffusion path is with respect to the measure $\mathcal{P}$, and  $\mathrm{d}\mathcal{P}/\mathrm{d}\mathcal{Q}$ denotes the Radon-Nikodym derivative between $\mathcal{P}$ and $\mathcal{Q}$. 
 
	In the mathematics literature, the minimization of~\eqref{model:KL} at $$U_{\star} =0$$
 is referred to as \emph{entropic interpolation} \cite{LeoC2014} or \emph{entropic transportation cost} \cite{CoRi2020}. This terminology is due to the discovery \cite{MikT2004} that  the minimization of the overdamped counterpart of~\eqref{model:KL} yields a viscous regularization of the Monge-Amp\`ere-Kantorovich optimal transport problem (see \cite{VilC2009}). Finally, \cite{FiHo2005} supports the use of the cost of a Schr\"odinger bridge as a natural efficiency measure for nano-engines in highly fluctuating environments; see \cite{TodE2009b,ThTo2012} for a wider class of applications.  In section~\ref{sec:exco} we show how the optimization of (\ref{model:KL}) provides a plausible model of shortcut to equilibration.
	\medskip
 \subsubsection*{EP:}\label{itm:b}
 Mean entropy production. In stochastic thermodynamics, the average entropy production
	is identified with the Kullback-Leibler divergence of the forward measure $\mathcal{P}$ from a measure $\mathcal{P}_{\mathcal{R}} $ obtained by a 
	combined time-reversal and path-reversal operation (see e.g. \cite{MaReMo2000,ChGa2008,GawK2013} and Appendix~\ref{app:KL} for further details):
	\begin{align}
\mathcal{E}&=\operatorname{E}_{\mathcal{P}}\ln \dfrac{\mathrm{d}\mathcal{P}}{\mathrm{d}\mathcal{P}_{\mathcal{R}}}\nonumber\\
&=\operatorname{E}_{\mathcal{P}}\ln\dfrac{\styleC{p}_{\ti}(\bm{\mathscr{q}}_{\ti},\bm{\mathscr{p}}_{\ti})}{\styleC{p}_{\tf}(\bm{\mathscr{q}}_{\tf},\bm{\mathscr{p}}_{\tf})}+\operatorname{E}_{\mathcal{P}}\int_{\ti}^{\tf}\mathrm{d}t\,
\left(\dfrac{\beta\,\|\bm{\mathscr{p}}_{t}\|^{2}}{m\,\tau}-\dfrac{d}{\tau}\right)
\nonumber\\
&\qquad+\dfrac{\beta\,g\,\tau}{m}\operatorname{E}_{\mathcal{P}}\int_{\ti}^{\tf}\mathrm{d}t\,\left(\left\|(\bm{\partial} U)(\bm{\mathscr{q}}_{t})\right\|^{2}-\dfrac{(\bm{\partial}^{2} U)(\bm{\mathscr{q}}_{t})}{\beta}\right)\,.
\label{model:ep}
	\end{align}
	Some observations are in order. To start with, the identification of (\ref{model:ep}) as mean entropy production of during a thermodynamic transition is legitimate in consequence of its relation with the heat release by the system evolving according to  (\ref{model:sde}). This is a consequence of the general theory expounded in \cite{ChGa2008}.	In appendix~\ref{app:st} for reader convenience, we reproduce the calculation that justifies the identification.
	
	For any $g$, the entropy production vanishes for a system in a Maxwell-Boltzmann equilibrium. For a bridge process, equilibrium means that the boundary conditions~\eqref{model:ini},~\eqref{model:fin} are specified by the same Maxwell-Boltzmann distribution. The corresponding optimal control problem becomes trivial. In any non-trivial case, the Gibbs-Shannon entropy difference appearing in the first row of~\eqref{model:ep} does not play a role in the optimization as it is fully specified by the boundary conditions. 
 
 Finally, the entropy production is \emph{non-coercive}, i.e it is not a convex functional of the control at $g$ equal zero. As a practical consequence, none of the infinitely many time-dependent protocols that connect the selected end states can be said to minimize the entropy production. Precise treatments of the optimal control problem in such a case are possible either by regularizing the problem \cite{MGSc2014}, or in special cases \cite{DeKiLu2017}, by considering non-purely mechanical controls \cite{PMG2014}. Studying, as we propose here, the mean entropy production at finite $g$ has the advantage of making the cost functional coercive with respect to the mechanical force.

At this point, it is worth commenting on our working hypotheses. The cost functionals in both case~\hyperref[itm:a]{KL} and \hyperref[itm:b]{EP} are readily convex in the mechanical potential. We surmise the existence of an open set of admissible potentials that allows us to look for a minimum in the form of a regular extremal of a variational problem \cite{BoSiSu2021}. 
To justify this assumption we recall that
H\"ormander's theorem (see e.g \cite{ReyL2006}) ensures that any potential $U_{t}$~\eqref{model:sde} that is sufficiently regular, bounded from below, and growing sufficiently fast at infinity results in a smooth density. 

\subsection{Bounds of the thermodynamic cost functionals}
  \label{sec:FG}
  In practice, the cost functionals~\eqref{model:KL} and~\eqref{model:ep} are the limit of Riemann sums on ratios of transition probability densities evaluated over increasingly small time increments.
  This construction is recalled in appendix~\ref{app:KL}. The construction immediately implies that~\eqref{model:KL} is bounded from below by the Kullback--Leibler divergence of the joint probability distribution of the system state at the end-times of the control horizon. 
  
  The measure theoretic analysis in Section~3 of \cite{FoGa1997} permits drawing more precise qualitative conclusions without making direct reference to the details of the dynamics. 
 To summarize them, let us denote by $\mathbb{S}$ the state space of dimension $d_{S}$, where the stochastic process  $\big{\{}\bm{\mathscr{x}}_{t},t\in [\ti,\tf]\big{\}}$ with probability measure $\mathcal{P}$ takes values. We also denote by $\mathcal{P}_{\bm{y}}^{\bm{x}}$ ($\mathcal{Q}_{\bm{y}}^{\bm{x}}$) the probability measures subject to the bridge conditions 
  \begin{align}
  	&\bm{\mathscr{x}}_{\ti}=\bm{y}\quad\&\quad\bm{\mathscr{x}}_{\tf}=\bm{x} \,.
  	\nonumber
  \end{align}
   Under technical hypotheses guaranteeing that the optimization problem is well-posed, the main takeaways of \cite{FoGa1997} are the following. 
 First, the Kullback-Leibler divergence is always amenable to the decomposition \cite{FoeH1988}
  \begin{align}
  \operatorname{K}(\mathcal{P}\mathrel{\Vert}\mathcal{Q})=\operatorname{K}(\wp\mathrel{\Vert}\wp_{\star})
 + \int_{\mathbb{S}^{2}} \mathrm{d}^{d_{S}}\bm{x}\,\mathrm{d}^{d_{S}}\bm{y}\,\wp(\bm{x},\bm{y})\, \operatorname{K}(\mathcal{P}_{\bm{y}}^{\bm{x}}\mathrel{\Vert}\mathcal{Q}_{\bm{y}}^{\bm{x}})\,.
  	\label{FG:KLd}
  \end{align}
  The first addend is the quantity originally considered by Schr\"odinger in \cite{SchE1931}, namely the \emph{static Kullback-Leibler divergence} 
  \begin{equation}
	\operatorname{K}(\wp\mathrel{\Vert}\wp_{\star})=
	\int_{\mathbb{S}^{2}} \mathrm{d}^{d_{S}}\bm{x}\,\mathrm{d}^{d_{S}}\bm{y}\,\wp(\bm{x},\bm{y})
	\ln\dfrac{\wp(\bm{x},\bm{y})}{\wp_{\star}(\bm{x},\bm{y})}
  	\label{FG:static}
  \end{equation}
  of the joint probability density $\wp$ of $\bm{\mathscr{x}_{\ti}}$ and $\bm{\mathscr{x}}_{\tf}$ from the two point probability 
  \begin{align}
  	\wp_{\star}(\bm{x},\bm{y})=\styleC{T}_{\tf,\ti}^{(\mathcal{Q})}(\bm{x}\mid\bm{y})\styleC{f}_{\ti}(\bm{y})\,,
  	\nonumber
  \end{align}
  which is uniquely defined by the transition probability density  $\styleC{T}_{\tf,\ti}^{(\mathcal{Q})}(\cdot\mid\cdot)$ of the reference process and the probability distribution of $\bm{\mathscr{x}_{\ti}}$.

  Both addends in~\eqref{FG:KLd} are positive. Furthermore, the static divergence~\eqref{FG:static} vanishes if and only if
  \begin{align}
  	\wp(\bm{x},\bm{y})=\wp_{\star}(\bm{x},\bm{y}).
  	\nonumber
  \end{align}
Many possible $\mathcal{P}$ are compatible with the same $\wp$. Once $\wp$ is fixed, $\operatorname{K}(\mathcal{P}\mathrel{\Vert}\mathcal{Q})$ attains an infimum, in fact a minimum, for the $\mathcal{P}$ that makes the second term of~\eqref{FG:KLd} vanish. A necessary condition \cite{FoeH1988,FoGa1997} enforcing this requirement is that for any $\ti\,\leq\,s\,\leq\,t\,\leq\,\tf$
 \begin{align}
	\mathcal{P}(\bm{\mathscr{x}}_{t}=\bm{x}\mid\bm{\mathscr{x}}_{s}=\bm{y},\bm{\mathscr{x}}_{\ti}=\bm{z})
=\dfrac{h_{t,\ti}(\bm{x},\bm{z})\,\styleC{T}_{t,s}^{(\mathcal{Q})}(\bm{x}\mid\bm{y})}{h_{s,\ti}(\bm{y},\bm{z})}
 \label{FG:tpdNM}
 \end{align}
 where the function $h$ is defined by 
 \begin{align}
 	h_{t,\ti}(\bm{x},\bm{y})=
 		\int_{\mathbb{S}}\mathrm{d}^{d_{S}}\bm{z}\,	h_{\tf,\ti}(\bm{z},\bm{y})\,\styleC{T}_{\tf,t}^{(\mathcal{Q})}(\bm{x}\mid\bm{z})\,.
 	\nonumber
 \end{align} 
Once~\eqref{FG:tpdNM} holds true, the control of the abstract optimization problem is $ h_{\tf,\ti}$. Correspondingly,~\eqref{FG:KLd} reduces to~\eqref{FG:static}, which in turn we can couch into the form
 \begin{align}
 	\operatorname{K}_{\text{opt}}(\mathcal{P}\mathrel{\Vert}\mathcal{Q})=&\operatorname{K}_{\text{opt}}(\wp\mathrel{\Vert}\wp_{\star})
= 	\int_{\mathbb{S}^{2}} \mathrm{d}^{d_{S}}\bm{x}\,\mathrm{d}^{d_{S}}\bm{y}\, h_{t,\ti}(\bm{x},\bm{y}) \,\styleC{T}_{\tf,\ti}^{(\mathcal{Q})}(\bm{x}\mid\bm{y})\,\styleC{f}_{\ti}(\bm{y})\,\ln h_{\tf,\ti}(\bm{x},\bm{y})\,.
\nonumber
 \end{align}
    The general form (\ref{FG:tpdNM}) of the necessary condition for the reduction to a static problem does not require the optimal process to enjoy the Markov property; the transition probability (\ref{FG:tpdNM}) may carry memory of the value taken by $\bm{\mathscr{x}}_{\ti}$. 
    The results of \cite{FoGa1997} ensure that (\ref{FG:tpdNM}), under further regularity assumptions, reduces for any $s\,\leq\,t \in \left[\ti,\tf\right]$ to a Markov transition probability density
    \begin{align}
    	\styleC{T}_{t,s}^{(\mathcal{P}_{\iota}^{\styleB{f})}}(\bm{x}\mid\bm{y})
    	=\dfrac{h_{t}(\bm{x})\,\styleC{T}_{t,s}^{(\mathcal{Q})}(\bm{x}\mid\bm{y})}{h_{s}(\bm{y})}\,.
    	\nonumber
    \end{align}
    From the physics point of view, the assumptions leading to Markov transition probability densities immediately include an overdamped dynamics \cite{DaiP1991} or an underdamped dynamics driven by force field depending on both the position and momentum of the system \cite{ChCoGrRe2021,CaHa2022}, and thus distinct from (\ref{model:sde}).
    
    Our discussion so far refers to case~\hyperref[itm:a]{KL}.  The connection to case~\hyperref[itm:b]{EP} stems from the Talagrand-Otto-Villani inequalities \cite{TalM1996,OtVi2000}. These inequalities show that the static Kullback-Leibler divergence between probability densities is bounded from below by the squared Wasserstein distance between the densities multiplied by a proportionality factor. 
    For the overdamped dynamics considered in \cite{DaiP1991}, Mikami \cite{MikT2004} (see also \cite{LeoC2012,PMG2013,CoRi2020}) later proved that the bound becomes tight in a suitable scaling limit and the proportionality factor reduces to the inverse of the duration of the control horizon. More explicitly, the entropic transport cost ($U_{\star}=0$) multiplied by the viscosity becomes equal to the  cost of a Monge-Amp\`ere-Kantorovich optimal mass transport problem \cite{VilC2009} in the limit of vanishing viscosity. 
    
    The connection to problem~\hyperref[itm:b]{EP} consists in the proof \cite{AuGaMeMoMG2012,GawK2013} and \cite{ChGePa2016} that the minimization of the mean entropy production by bridge processes obeying the overdamped dynamics can be exactly mapped into a Monge-Amp\`ere-Kantorovich optimal mass transport. The reason is that the optimal control problem admits an equivalent reformulation, in which the current velocity of the admissible processes \cite{NelE2001} play the role of control instead of the drift.
    
 In the underdamped case, the presence of inertial effects complicates the picture. The mean entropy production cannot be written as the square of the current velocity. This prevents a direct application of the Benamou-Brenier inequality \cite{BeBr2000} (see also appendix~\ref{app:ep}). The Benamou-Brenier inequality allows one to couch the minimum mean overdamped entropy production into the squared Wasserstein distance between the densities at the end of the control horizon. It is, however, possible to show \cite{PMG2014,MGSc2014} that the underdamped mean entropy production admits its overdamped counterpart as a lower bound. In particular, for (\ref{model:ep}), the following inequality 
 \begin{align}
 	\mathcal{E}_{\ti}^{\tf}\,\geq\,\dfrac{m\,\beta}{(1+g)\,\tau} \dfrac{\operatorname{E}_{\mathcal{P}}\left\|\bm{\mathscr{q}}_{\tf}-\bm{\mathscr{q}}_{\ti}\right\|^{2}}{\tf-\ti}
 	\label{FG:ep}
 \end{align}
holds true. Bounds of the type (\ref{FG:ep}) for the mean entropy production appeared in \cite{PMG2014,MGSc2014} and later in \cite{DeSa2018}. The proof of \eqref{FG:ep} presented in \cite{MGPe2023} is motivated by \cite{GawK2021}. For reader convenience, we reproduce the proof in appendix~\ref{app:ep}. 
 

The above considerations suggest that for both the over- and underdamped dynamics (\ref{model:sde}), the inequality
\begin{align}
	\operatorname{K}(\mathcal{P}\mathrel{\Vert}\mathcal{Q})
\,\geq\,\dfrac{\mathcal{C}_{\text{TOV}}\,m\,\beta}{(1+g)\,\tau} \dfrac{\operatorname{E}_{\mathcal{P}}\left\|\bm{\mathscr{q}}_{\tf}-\bm{\mathscr{q}}_{\ti}\right\|^{2}}{\tf-\ti}
	\label{FG:TOV}
\end{align}
should also hold true with $\mathcal{C}_{\text{TOV}}$ a positive constant in agreement with the Talagrand-Otto-Villani theory \cite{TalM1996,OtVi2000}. We refer to  \cite{ChCoGrRe2021} for a mathematical proof of the bound for the underdamped dynamics, and for the overdamped dynamics \cite{MiTh2006} (see also \cite{ChGePa2016,GeLeRi2017,CoRi2020}), including an explicit prediction of the constant $\mathcal{C}_{\text{TOV}}$.

\section{Optimal control formulation}
\label{sec:oc}

Optimal control problems can be turned into variational problems by coupling the dynamics to the cost functional through
Lagrange multipliers. In hydrodynamics, such an approach is referred to as the adjoint equation method, and has a long history going back to \cite{SerJ1959,SeWh1968}. By themselves alone, solutions of the variational equations only provide a necessary condition for the existence of regular extremals of an optimal control problem: extremals continuously satisfying (partial) differential equations. 
Optimality follows from convexity of the cost, as in our case, or from the study of the second variation. 
A mathematical formulation of the adjoint method in the context of stochastic optimal control is due to Bismut see e.g. \cite{BisJ1979}. Bismut theory may be regarded as extending the Pontryagin principle of deterministic optimal control \cite{BecJ2021} see also \cite{LibD2012,BoSiSu2021}. Based on these considerations, we construct from \hyperref[itm:a]{KL} and \hyperref[itm:b]{EP} variational functionals by imposing that the mean forward derivative \cite{NelE2001} of a Lagrange multiplier is an exact differential when the density $\styleC{f}_{t}(\bm{x})$ with respect to the Lebesgue measure satisfies a Fokker-Planck equation associated with probability preserving boundary conditions. As in \cite{PMG2014,MGSc2014} we refer to these functionals as Pontryagin-Bismut.
In such a setup, we look for variational extremals of
\begin{align}
	\hspace{-0.3cm}	\mathcal{A}[\styleC{f},U,V]&=\int_{\mathbb{R}^{2d}}\mathrm{d}^{2d}\bm{x} \Big{(}V_{\ti}(\bm{x})\styleC{f}_{\iota}(\bm{x})
	-V_{\tf}(\bm{x})\styleC{f}_{\styleB{f}}(\bm{x})\Big{)}+
	\nonumber\\
	&\qquad\int_{\ti}^{\tf}\mathrm{d}t\,\int_{\mathbb{R}^{2d}}\mathrm{d}^{2\,d}\bm{x} \,\styleC{f}_{t}(\bm{x})\Big{(}C_{t}^{(U_{t})}(\bm{x})+(\partial_{t}+\styleB{L}_{\bm{x}})V_{t}(\bm{x})\Big{)}\,.
	\label{oc:BP}
\end{align} 
Here we collectively denote phase space coordinates as $$\bm{x}=(\bm{q},\bm{p})$$ and define the running cost functional as
\begin{align}
 C_{t}^{(U_{t})}(\bm{x})=\begin{cases}
	\dfrac{\beta\,\tau\,(1+g)}{4\,m} \|(\bm{\partial}U_{t})(\bm{q})-(\bm{\partial}U_{\star})(\bm{q})\|^{2} & \syma
	\\[0.3cm]
	\dfrac{\beta\,\|\bm{p}\|^{2}}{m\,\tau}-\dfrac{\,d\,}{\tau}+\dfrac{\beta\,g\,\tau}{m}\left(\left\|(\bm{\partial} U_{t})(\bm{q})\right\|^{2}-\dfrac{(\bm{\partial}^{2} U_{t})(\bm{q})}{\beta}\right)\,. &\symb
		\end{cases}
	\nonumber
\end{align}
In writing (\ref{oc:BP}) we conceptualize the fields $\styleC{f}$, $V$, and $U$ as unknown variational fields. The existence of the functional requires integrability with respect to $\styleC{f}_{t}$ which we assume to be a probability density taking the values $\styleC{f}_{\iota}$ and $\styleC{f}_{\styleB{f}} $ at the start and end of the control horizon respectively, fixed by~\eqref{model:ini},~\eqref{model:fin}.  
The field $V$ becomes the value function of Bellman's formulation of optimal control theory \cite{BecJ2021}. In~\eqref{oc:BP}, it plays the role of a Lagrange multiplier enforcing the dynamics. 

Accordingly, we denote by $\styleB{L}_{\bm{x}}$ the differential generator of the dynamics determined by~\eqref{model:sde}:
\begin{equation}
\label{eq:FPoperator}
\begin{aligned}
\styleB{L}_{\bm{x}}=	\dfrac{\bm{p}-\tau\,g\,(\bm{\partial}U_{t})(\bm{q})}{m}\cdot\partial_{\bm{q}}
-\left(\dfrac{\,\bm{p}\,}{\tau}+(\bm{\partial}U_{t})(\bm{q})\right)\cdot\partial_{\bm{p}}+\dfrac{g\,\tau}{m\,\beta}\,\partial_{\bm{q}}^{2}
+\dfrac{m}{\tau\,\beta}\,\partial_{\bm{p}}^{2}\,.
\end{aligned}    
\end{equation}
Thus, if $\styleC{f}_{t}$ is the instantaneous density of (\ref{model:sde}), then 
\begin{align}
	(\operatorname{D}V)_{t}(\bm{x})=(\partial_{t}+\styleB{L}_{\bm{x}})V_{t}(\bm{x})
	\nonumber
\end{align}
is the mean forward derivative of $V$ along the paths of (\ref{model:sde}) and by definition
 \begin{align}
 	\operatorname{E}_{\mathcal{P}}\cbr{V_{\tf}(\bm{\mathscr{x}}_{\tf})-V_{0}(\bm{\mathscr{x}}_{\ti})}=\int_{\ti}^{\tf}\mathrm{d}t \operatorname{E}_{\mathcal{P}}(\operatorname{D}V)_{t}(\bm{\mathscr{x}}_{t})\,.
 	\nonumber
 \end{align}
 This observation justifies the introduction of the value function as a Lagrange multiplier.

Our definition of the value function in \hyperref[itm:b]{EP} omits the contribution from the variation of the Gibbs-Shannon entropy to the mean entropy production from the Pontryagin-Bismut functional. This is because the Gibbs-Shannon entropy in (\ref{model:ep}) is fully specified by the assigned boundary conditions and therefore does not enter the determination of the optimal control.

\subsection{Variational equations}

We determine the optimal control equations by a stationary variation of  (\ref{oc:BP}). As expected, the variation with respect to the value function yields the Fokker-Planck equation for the probability density 
\begin{align}
	(\partial_{t}-\styleB{L}_{\bm{x}}^{\dagger})\,\styleC{f}_{t}(\bm{x})=0\,.
	\label{oc:FP}
\end{align}
The variation with respect to the probability density yields the \emph{dynamic programming equation} \cite{BecJ2021}
\begin{align}
	(\partial_{t}+\styleB{L}_{\bm{x}})V_{t}(\bm{x})+C_{t}^{(U_{t})}(\bm{x})=0\,.
		\label{oc:dp}
\end{align}
In the overdamped case \cite{DaiP1991,AuGaMeMoMG2012,ChGePa2016}, and in the case when the control is a function of both position and momentum \cite{ChGePa2015,ChCoGrRe2021,CaHa2022},  the variation with respect to the potential yields a local, exactly integrable condition for the optimal control. In stark contrast, we find that the optimal control potential in the underdamped case must solve an integral equation coupled to the Fokker-Planck and dynamic programming equations \cite{MGSc2014}:
\begin{align}
&\bm{\partial}_{\bm{q}}\cdot\int_{\mathbb{R}^{d}} \mathrm{d}^{d}\bm{p}\,\styleC{f}_{t}(\bm{q},\bm{p})\left(\dfrac{g\,\tau}{m}\partial_{\bm{q}} V_{t}(\bm{q},\bm{p})+\partial_{\bm{p}} V_{t}(\bm{q},\bm{p})\right)=\bm{\partial}_{\bm{q}}\cdot \Big{(}\styleC{\tilde{f}}_{t}(\bm{q})\,\bm{b}_{t}(\bm{q})
   \Big{)}
	\label{oc:pot}
\end{align}
with $\tilde{\styleC{f}}_{t}(\bm{q})$ as the position marginal of $\styleC{f}_{t}(\bm{q},\bm{p}) $ (see Eq.~\eqref{app:ep:mp}~) and
\begin{align}
	\bm{b}_{t}(\bm{q})= \begin{cases}
		\dfrac{\beta\,\tau\,(1+g)}{2\,m} \big{(}(\bm{\partial}U_{t})(\bm{q})-(\bm{\partial}U_{\star})(\bm{q})\big{)} & \syma
		\\[0.3cm]
		\dfrac{\beta\,g\,\tau}{m}\left(2\,(\bm{\partial} U_{t})(\bm{q})+\dfrac{(\bm{\partial} \ln \styleC{\tilde{f}}_{t} )(\bm{q})}{\beta}\right)\,. &\symb
	\end{cases}
	\nonumber
\end{align}
Finding regular extremals  amounts to finding the simultaneous solutions of Eqs.~\eqref{oc:FP}, \eqref{oc:dp} and~\eqref{oc:pot}.
The integro-differential stationary condition (\ref{oc:pot}) is hard to approach due to its non-local nature (in momentum space). These issues are to some extent reminiscent of the Vlasov-Poisson-Fokker-Planck (see e.g.~\cite{BoCaSo1997}) and the McKean–Vlasov (see e.g.~\cite{ CaHa2020}) equations. The condition somewhat simplifies when the configuration space is one dimensional. We can write
\begin{align}
\int_{\mathbb{R}} \mathrm{d} p\,\dfrac{\styleC{f}_{t}(q,p)}{\styleC{\tilde{f}}_{t}(q)}\left(\dfrac{g\,\tau}{m}\partial_{q} V_{t}(q,p)+\partial_{p} V_{t}(q,p)\right)=\begin{cases}
	\dfrac{\beta\,\tau\,(1+g)}{2\,m} \Big{(}(\partial U_{t})(q)-(\partial U_{\star})(q)\Big{)} & \syma
	\\[0.3cm]
	\dfrac{\beta\,g\,\tau}{m}\left(2\,(\partial U_{t})(q)+\dfrac{(\partial \ln \styleC{\tilde{f}}_{t} )(q)}{\beta}\right)\,. &\symb
\end{cases}
	\label{oc:pot1d}
\end{align}

\subsection{Dual expression of the optimal cost}
\label{sec:dual}
When the dynamic programming equation (\ref{oc:dp}) holds, the Pontryagin-Bismut functional (\ref{oc:BP}) reduces to
\begin{equation}
	\mathcal{A}[\styleC{f},U,V]\Big{|}_{\mathscr{d.p.}}=\int_{\mathbb{R}^{2d}}\mathrm{d}^{2d}\bm{x} \Big{(}V_{\ti}(\bm{x})\styleC{f}_{\iota}(\bm{x})
		-V_{\tf}(\bm{x})\styleC{f}_{\styleB{f}}(\bm{x})\Big{)}\,.
	\label{oc:dual}
\end{equation}
The optimum value of the cost hence coincides with the minimum, or at least infimum of (\ref{oc:dual}), taken over all value functions satisfying the dynamic programming equation.
This observation is the basis for the aforementioned duality relation used in \cite{MiTh2006}, and later in \cite{ChGePa2016,GeLeRi2017,CoRi2020,ChCoGrRe2021}. In what follows, we use (\ref{oc:dual}) to compute the expression of minimum costs predicted by multiscale perturbation theory.  

\section{Gaussian Case}
\label{sec:G}

In view of the complexity of the optimal control condition (\ref{oc:pot}), it is instructive to analyze the case of Gaussian boundary conditions.
A similar analysis was performed in \cite{MGSc2014} for a case closely related to~\hyperref[itm:b]{EP}, but only in one dimensional configuration space.

Gaussian boundary conditions lead to major simplifications. The structure of the Fokker-Planck and dynamic programming equations preserve the space of Gaussian probability densities and second order polynomials in phase space for any at most quadratic control 
\begin{align}
	U_{t}(\bm{q})=\mathscr{u}_{t}+\bm{u}_{t}\cdot\bm{q}
	+\dfrac{1}{2}\bm{q}^{\top}\styleD{U}_{t}\bm{q}
	\nonumber
\end{align}  
and reference 
\begin{align}
		U_{\star}(\bm{q})=\mathscr{u}_{\star}+\bm{u}_{\star}\cdot\bm{q}
	+\dfrac{1}{2}\bm{q}^{\top}\styleD{U}_{\star}\bm{q}
	\nonumber
\end{align}
potentials. In the above expressions $\bm{u}_{t} $, $ \bm{u}_{\star}$ are vectors in $\mathbb{R}^{d}$ and $ \styleD{U}_{t}$, $ \styleD{U}_{\star}$  are $d\,\times\,d$ real symmetric matrices. Thus, the probability density is fully specified by the set of 
first and second order cumulants
\begin{align}
	&	\styleD{Q}_{t}=\operatorname{E}_{\mathcal{P}}(\bm{\mathscr{q}}_{t}\,\otimes\,\bm{\mathscr{q}}_{t})-\operatorname{E}_{\mathcal{P}}\bm{\mathscr{q}}_{t}\,\otimes\,\operatorname{E}_{\mathcal{P}}\bm{\mathscr{q}}_{t}
	\nonumber\\
	&	\styleD{C}_{t}=\operatorname{E}_{\mathcal{P}}(\bm{\mathscr{q}}_{t}\,\otimes\,\bm{\mathscr{p}}_{t})-\operatorname{E}_{\mathcal{P}}\bm{\mathscr{q}}_{t}\,\otimes\,\operatorname{E}_{\mathcal{P}}\bm{\mathscr{p}}_{t}
	\nonumber\\
	&	\mathscr{P}_{t}=\operatorname{E}_{\mathcal{P}}(\bm{\mathscr{p}}_{t}\,\otimes\,\bm{\mathscr{p}}_{t})-\operatorname{E}_{\mathcal{P}}\bm{\mathscr{p}}_{t}\,\otimes\,\operatorname{E}_{\mathcal{P}}\bm{\mathscr{p}}_{t}\,.
	\nonumber
\end{align}
Here and below, we use $\otimes$ to denote the outer product of vectors in $\mathbb{R}^{d}$.
Correspondingly, a value function of the form
\begin{align}
&	V_{t}(\bm{q},\bm{p})=\mathscr{v}_{t}+\bm{v}_{t}^{(q)}\cdot\bm{q}+\bm{v}_{t}^{(p)}\cdot\bm{p}+\dfrac{\bm{q}^{\top}\styleD{V}_{t}^{(q,q)}\bm{q}+\bm{p}^{\top}\styleD{V}_{t}^{(p,p)}\bm{p}
		+\bm{p}^{\top}\styleD{V}_{t}^{(p,q)}\bm{q}+\bm{q}^{\top}\styleD{V}_{t}^{(q,p)}\bm{p}
	}{2}
	\label{G:value}
\end{align}
satisfies the dynamic programming equation (\ref{oc:dp}). In (\ref{G:value}), $\styleD{V}_{t}^{(q,q)}$ and $ \styleD{V}_{t}^{(p,p)}$ are $d\,\times\,d$ symmetric matrices, and
\begin{align}
	\styleD{V}_{t}^{(q,p)}={\styleD{V}_{t}^{(p,q)}}^{\top}\,.
	\nonumber
\end{align}
The Fokker-Planck equation (\ref{oc:FP}) reduces to a closed system of differential equations of first order for the second order cumulants of the Gaussian statistics
\begin{equation}
	\begin{split}
	&	\dfrac{\mathrm{d}}{\mathrm{d}t}\styleD{Q}_{t}=\dfrac{\styleD{C}_{t}+\styleD{C}_{t}^{\top}}{m}-\dfrac{g\,\tau}{m}\left(\styleD{U}_{t}\styleD{Q}_{t}+\styleD{Q}_{t}\styleD{U}_{t}\right)+\dfrac{2\,g\,\tau}{m\,\beta}\mathds{1}
	\\
	&\dfrac{\mathrm{d}}{\mathrm{d}t}\styleD{C}_{t}=-\dfrac{1}{\tau}\styleD{C}_{t}-\dfrac{g\,\tau}{m}\styleD{U}_{t}\styleD{C}_{t}-\styleD{U}_{t}\styleD{Q}_{t}+\dfrac{1}{m}\mathscr{P}_{t}
	\\
	&\dfrac{\mathrm{d}}{\mathrm{d}t}\mathscr{P}_{t}=-\dfrac{2}{\tau}\mathscr{P}_{t}-\styleD{U}_{t}\styleD{C}_{t}-\styleD{C}_{t}^{\top}\styleD{U}_{t}+\dfrac{2\,m}{\beta\,\tau}\mathds{1}
	\end{split}
	\label{G:c2}
\end{equation}
and to a system of differential equations of first order for the first order cumulants sustained by the solution of second order ones:
\begin{equation}
	\begin{split}
	&	\dfrac{\mathrm{d}}{\mathrm{d}t }\operatorname{E}_{\mathcal{P}}\bm{\mathscr{q}}_{t}=\dfrac{\operatorname{E}_{\mathcal{P}}\bm{\mathscr{p}}_{t}}{m}
		-\dfrac{g\,\tau}{m}\left(\bm{u}_{t}+\styleD{U}_{t}\operatorname{E}_{\mathcal{P}}\bm{\mathscr{q}}_{t}\right )
		\\
	&	\dfrac{\mathrm{d}}{\mathrm{d t} }\operatorname{E}_{\mathcal{P}}\bm{\mathscr{p}}_{t}=-\left(\dfrac{\operatorname{E}_{\mathcal{P}}\bm{\mathscr{p}}_{t}}{\tau}
	+ \bm{u}_{t}+\styleD{U}_{t}\operatorname{E}_{\mathcal{P}}\bm{\mathscr{q}}_{t}\right )\,.
	\end{split}
		\label{G:c1}
\end{equation}
The full system of cumulant equations \eqref{G:c2}-\eqref{G:c1} is complemented by boundary conditions at both ends of the control horizon: 
\begin{align}
	&	\styleD{Q}_{0}=\styleD{Q}_{\iota}\quad\&\quad\styleD{Q}_{\tf}=\styleD{Q}_{\styleB{f}}
	\nonumber\\[3pt]
	& \styleD{C}_{0}=\styleD{C}_{\tf}=0
	\nonumber\\[3pt]
	& \mathscr{P}_{0}=\mathscr{P}_{\tf}=\dfrac{m}{\beta}\,\mathds{1}_{d}
	\nonumber
\end{align}
and
\begin{align}
&	\operatorname{E}_{\mathcal{P}}\bm{\mathscr{q}}_{\ti} = \boldsymbol{q}_{\iota} \quad\&\quad\operatorname{E}_{\mathcal{P}}\bm{\mathscr{q}}_{\tf} = \boldsymbol{q}_{\styleB{f}}
	\nonumber\\[3pt]
&	\operatorname{E}_{\mathcal{P}}\bm{\mathscr{p}}_{\ti} =\operatorname{E}_{\mathcal{P}}\bm{\mathscr{p}}_{\tf} =0\,.
	\nonumber
\end{align}
The boundary conditions can be satisfied because the potential couples the cumulant equations to a first-order differential system of equal size for the coefficients  of the value function in (\ref{G:value}). 

\subsection{Analysis of case~\hyperref[itm:a]{KL}}
\label{sec:G:KL}

For case~\hyperref[itm:a]{KL}, we get  
\begin{subequations}
	\label{G:v2}
\begin{align}
	\dfrac{\mathrm{d}}{\mathrm{d} t}\styleD{V}_{t}^{(q,q)}&=\big{(}\styleD{U}_{t}\styleD{V}_{t}^{(p,q)}+\styleD{V}_{t}^{(q,p)}\styleD{U}_{t}\big{)}+\dfrac{g\,\tau}{m}\left(\styleD{U}_{t}\styleD{V}_{t}^{(q,q)}+\styleD{V}_{t}^{(q,q)}\styleD{U}_{t}\right)
  \nonumber\\& 
  \qquad-\dfrac{\beta\,\tau\,(1+g)}{2\,m}\left(\styleD{U}_{t}-\styleD{U}_{\star}\right)\left(\styleD{U}_{t}-\styleD{U}_{\star}\right)
		\label{G:v2qq}\\[5pt]
	\dfrac{\mathrm{d}}{\mathrm{d} t}\styleD{V}_{t}^{(q,p)}&=\dfrac{\styleD{V}_{t}^{(q,p)}}{\tau}+\styleD{U}_{t}\styleD{V}_{t}^{(p,p)}+\dfrac{g\,\tau}{m}\styleD{U}_{t}\styleD{V}_{t}^{(q,p)}-\dfrac{\styleD{V}_{t}^{(q,q)}}{m}
	\label{G:v2pq}\\[5pt]
	\dfrac{\mathrm{d}}{\mathrm{d} t}\styleD{V}_{t}^{(p,p)}&=\dfrac{2}{\tau}\styleD{V}_{t}^{(p,p)}-\dfrac{\styleD{V}_{t}^{(p,q)}+\styleD{V}_{t}^{(p,q)\top}}{m}
\label{G:v2pp}
\end{align}
\end{subequations}
and
\begin{subequations}
	\label{G:v1}
	\begin{align}
\dfrac{\mathrm{d}}{\mathrm{d} t}\bm{v}_{t}^{(q)}&=\styleD{V}_{t}^{(p,q)\top}\bm{u}_{t}+\styleD{U}_{t}\bm{v}_{t}^{(p)} +\dfrac{g\,\tau}{m}\left(\styleD{U}_{t}\bm{v}_{t}^{(q)}+\styleD{V}_{t}^{(q,q)}\bm{u}\right)
\nonumber\\&  
		\qquad-\dfrac{\beta\,\tau\,(1+g)}{2\,m}\left(\styleD{U}_{t}-\styleD{U}_{\star}\right)\left(\bm{u}_{t}-\bm{u}_{\star}\right)
	\label{G:v1q}	\\[5pt]
	\dfrac{\mathrm{d}}{\mathrm{d} t}\bm{v}_{t}^{(p)}&=\dfrac{\bm{v}_{t}^{(p)}}{\,\tau\,}-\dfrac{\bm{v}_{t}^{(q)}}{m}+\styleD{V}_{t}^{(pp)}\bm{u}_{t}+\dfrac{g\,\tau}{m}\styleD{V}_{t}^{(p q)}\bm{u}\,.
		\label{G:v1p}
\end{align}
\end{subequations}
Finally, we find
\begin{align}
\dfrac{\mathrm{d}}{\mathrm{d} t}\mathscr{v}_{t}&=\bm{v}_{t}^{(p)}\cdot\bm{u}_{t}+\dfrac{g\,\tau}{m}\bm{v}_{t}^{(q)}\cdot \bm{u}_{t}-\operatorname{Tr}\left(\dfrac{m}{\beta\,\tau}\styleD{V}^{(p,p)}+\dfrac{g\,\tau}{m\,\beta}\styleD{V}^{(q,q)}\right)
\nonumber\\
&\qquad-\dfrac{(1+g)\,\beta\,\tau}{4\,m}\left\|\bm{u}_{t}-\bm{u}_{\star}\right\|^{2}\,.
\label{G:v0}
\end{align}
The structure of (\ref{G:v2})-(\ref{G:v1}) is analogous to that of the cumulant equations. The coefficients of second degree monomials in (\ref{G:value}) 
satisfy a closed system whose solution sustains the equation for the coefficients of first order monomial. 

We now turn to the solution of (\ref{oc:pot}).
A straightforward exercize in Gaussian integration yields the explicit expression of the \textquotedblleft osmotic force\textquotedblright\  \cite{NelE2001} or \textquotedblleft score function\textquotedblright\ \cite{PeCu2019} of the position marginal
\begin{align}
(\bm{\partial}\ln \widetilde{\styleC{f}}_{t})(\bm{q})=- \styleD{Q}_{t}^{-1}(\bm{q}-\operatorname{E}_{\mathcal{P}}\bm{\mathscr{q}}_{t})
\label{G:score}
\end{align}
as well as an explicit expression for the conditional expectation
\begin{align}
\operatorname{E}_{\mathcal{P}}(\bm{\mathscr{p}}_{t}\mid\bm{\mathscr{q}}_{t}=\bm{q})=\operatorname{E}_{\mathcal{P}}\bm{\mathscr{p}}_{t}+\styleD{C}_{t}\styleD{Q}_{t}^{-1}(\bm{q}-\operatorname{E}_{\mathcal{P}}\bm{\mathscr{q}}_{t})\,.
\label{G:ce}
\end{align}
Upon inserting (\ref{G:score}), (\ref{G:ce}) into  (\ref{oc:pot}) and matching the coefficients of monomials of same degree in $\bm{q}$, we arrive at the equations
\begin{subequations}
	\label{G:oc}
	\begin{align}		
	\styleD{Q}_{t}^{-1}\styleD{U}_{t}+\styleD{U}_{t}\styleD{Q}_{t}^{-1}&=\styleD{Q}_{t}^{-1}\styleD{M}_{t}+\styleD{M}_{t}^{\top}\styleD{Q}_{t}^{-1}
	\label{G:Lyap}
		\\[5pt]
		\operatorname{Tr}\left(\styleD{M}_{t}-\styleD{U}_{t}\right)&=0
		\label{G:trace}
	\end{align}
\end{subequations}
with
\begin{align}
	\styleD{M}_{t}&=\styleD{U}_{\star}+\dfrac{2\,m}{(1+g)\,\tau\,\beta}\left(\styleD{V}_{t}^{(p,p)}\styleD{C}_{t}\styleD{Q}_{t}^{-1}+\styleD{V}_{t}^{(p,q)}\right)
	+\dfrac{2\,g\,\tau}{(1+g)\,\beta}\left(\styleD{V}_{t}^{(q,p)}\styleD{C}_{t}\styleD{Q}_{t}^{-1}+\styleD{V}_{t}^{(q,q)}\right)
	\nonumber
\end{align}
and the dependent conditions
\begin{align}
		\bm{u}_{t}&=\bm{u}_{\star}+(\styleD{U}_{\star}-\styleD{U}_{t})\operatorname{E}_{\mathcal{P}}\bm{\mathscr{q}}_{t}	+\dfrac{2\,m}{(1+g)\,\beta\,\tau}
	\left(\bm{v}_{t}^{(p)}+\styleD{V}_{t}^{(p,q)}\operatorname{E}_{\mathcal{P}}\bm{\mathscr{q}}_{t}+\styleD{V}_{t}^{(p,p)}\operatorname{E}_{\mathcal{P}}\bm{\mathscr{p}}_{t}\right)
	\nonumber\\
	&\qquad+\dfrac{2\,g}{(1+g)\,\beta}
	\left(\bm{v}_{t}^{(q)}+\styleD{V}_{t}^{(q,q)}\operatorname{E}_{\mathcal{P}}\bm{\mathscr{q}}_{t}+\styleD{V}_{t}^{(q,p)}\operatorname{E}_{\mathcal{P}}\bm{\mathscr{p}}_{t}\right)\,.
\label{G:vec}
\end{align}
Clearly, the conditions (\ref{G:vec}) are always satisfied if (\ref{G:oc}) is solvable. In fact we recognize that equation (\ref{G:Lyap}) is in fact a Lyapunov equation. Uniqueness, symmetry and positivity of the solution are very well understood \cite{BhEl2002}. In particular, for every $t\in [\ti,\tf]$ we can write the solution of (\ref{G:Lyap}) as
\begin{align}
	\styleD{U}_{t}=\int_{0}^{\infty}\mathrm{d}s\, e^{-\styleD{Q}_{t}^{-1}\,s}\left(\styleD{Q}_{t}^{-1}\styleD{M}_{t}+\styleD{M}_{t}^{\top}\styleD{Q}_{t}^{-1}\right)e^{-\styleD{Q}_{t}^{-1}\,s}\,.
	\label{G:Lsol}
\end{align}
The solution is well defined because by definition $ \styleD{Q}_{t}$ is a positive matrix. Finally,  taking the trace of both sides of (\ref{G:Lsol}) readily recovers (\ref{G:trace}) thus completing the proof that the Gaussian case is solvable.

\subsection{Analysis of case~\hyperref[itm:b]{EP}}
\label{section:epgauss}

The equations that change are (\ref{G:v2qq}) 
\begin{align}
	\dfrac{\mathrm{d}}{\mathrm{d} t}\styleD{V}_{t}^{(q,q)}&=\Big{(}\styleD{U}_{t}\styleD{V}_{t}^{(p,q)}+\styleD{V}_{t}^{(q,p)}\styleD{U}_{t}\Big{)}+\dfrac{g\,\tau}{m}\left(\styleD{U}_{t}\styleD{V}_{t}^{(q,q)}+\styleD{V}_{t}^{(q,q)}\styleD{U}_{t}\right)
	-\dfrac{2\,\beta\,\tau\,g}{m}\styleD{U}_{t}\styleD{U}_{t}\,,
	\nonumber
\end{align}
Eq. \eqref{G:v1q} which is replaced by
\begin{align}
\dfrac{\mathrm{d}}{\mathrm{d} t}\bm{v}_{t}^{(q)}&=\styleD{V}_{t}^{(p,q)\top}\bm{u}_{t}+\styleD{U}_{t}\bm{v}_{t}^{(p)}+\dfrac{g\,\tau}{m}\left(\styleD{U}_{t}\bm{v}_{t}^{(q)}+\styleD{V}_{t}^{(q,q)}\bm{u}\right)
-\dfrac{2\,g\,\tau\,\beta}{m}\styleD{U}_{t}\bm{u}_{t}
	\nonumber
\end{align}
and, finally, Eq. \eqref{G:v0} which for the mean entropy production reads
\begin{align}
\dfrac{\mathrm{d}}{\mathrm{d} t}\mathscr{v}_{t}&=\dfrac{d}{\tau}+\bm{v}_{t}^{(p)}\cdot\bm{u}_{t}+\dfrac{g\,\tau}{m}\bm{v}_{t}^{(q)}\cdot \bm{u}_{t}
	\nonumber\\
	&\qquad-\operatorname{Tr}\left(\dfrac{m}{\beta\,\tau}\styleD{V}^{(p,p)}+\dfrac{g\,\tau}{m\,\beta}\styleD{V}^{(q,q)}\right)
-\dfrac{g\,\beta\,\tau}{m}\left(\left\|\bm{u}_{t}\right\|^{2}-\dfrac{1}{\beta}\operatorname{Tr}\styleD{U}_{t}\right)\,.
	\nonumber
\end{align}
A qualitative difference with case~\hyperref[itm:a]{KL} occurs for vanishing $g$ when the mean entropy production does not  explicitly depend upon the control potential.
This is most evident when inspecting (\ref{oc:pot}). We get
\begin{subequations}
	\label{G:epoc}
	\begin{align}	g\,\styleD{Q}_{t}^{-1}\styleD{U}_{t}+ g\,\styleD{U}_{t}\styleD{Q}_{t}^{-1}&=\styleD{Q}_{t}^{-1}\tilde{\styleD{M}}_{t}+\tilde{\styleD{M}}_{t}^{\top}\styleD{Q}_{t}^{-1}
		\label{G:epLyap}
		\\[5pt] 
  \operatorname{Tr}\left(\tilde{\styleD{M}}_{t}-g\styleD{U}_{t}\right)&=0
		\label{G:eptrace}
	\end{align}
\end{subequations}
where now
\begin{align}
	\tilde{\styleD{M}}_{t}&=\dfrac{g}{2}\styleD{Q}_{t}^{-1}+\dfrac{m}{2\,\tau\,\beta}\left(\styleD{V}_{t}^{(p,p)}\styleD{C}_{t}\styleD{Q}_{t}^{-1}+\styleD{V}_{t}^{(p,q)}\right)+\dfrac{g\,\tau}{2\,\beta}\left(\styleD{V}_{t}^{(q.p)}\styleD{C}_{t}\styleD{Q}_{t}^{-1}+\styleD{V}_{t}^{(q,q)}\right)
	\nonumber
\end{align}
and
\begin{align}
		g\,\bm{u}_{t}&=-g\,\styleD{U}_{t}\styleD{Q}_{t}^{-1}\operatorname{E}_{\mathcal{P}}\bm{\mathscr{q}}_{t}+\dfrac{m}{2\,\beta\,\tau}
	\left(\bm{v}_{t}^{(p)}+\styleD{V}_{t}^{(p,q)}\operatorname{E}_{\mathcal{P}}\bm{\mathscr{q}}_{t}+\styleD{V}_{t}^{(p,p)}\operatorname{E}_{\mathcal{P}}\bm{\mathscr{p}}_{t}\right)
	\nonumber\\
	&\qquad+\dfrac{g}{2\,\beta}
	\left(\bm{v}_{t}^{(q)}+\styleD{V}_{t}^{(q,q)}\operatorname{E}_{\mathcal{P}}\bm{\mathscr{q}}_{t}+\styleD{V}_{t}^{(q,p)}\operatorname{E}_{\mathcal{P}}\bm{\mathscr{p}}_{t}\right)\,.
	\label{G:epvec}
\end{align}
Whereas for $g\,>\,0$ the optimal potential is uniquely determined by the solution of the Lyapunov equation (\ref{G:epLyap}), the limit $g\downarrow 0$ is singular.
The Lyapunov equation becomes a constraint imposed on the coefficients of the value function. The upshot is that for vanishing $g$ it is not possible to satisfy boundary conditions imposed on \emph{all} phase space cumulants. In other words, the problem is not generically solvable for a generic assignment of Gaussian probability densities (\ref{model:ini}), (\ref{model:fin}). The problem admits, however, a solution if boundary data are just the position marginals. A detailed slow manifold analysis performed in the one-dimensional case in the supplementary material of \cite{MGSc2014} shows that the equation for $g$ equal zero coincides with the slow manifold equations (see e.g. \cite{HaKaKoOn1998}) of the limit $g \downarrow 0$ optimal control equations. This gives a precise mathematical meaning to the idea of $\delta$-Dirac optimal control upheld in \cite{GoScSe2008}. It also shows that even if the optimal control does not exist for $g$ equal zero, the strictly positive lower bound on the mean entropy production is always in agreement with (\ref{FG:ep}).

\section{General case in one dimension}
\label{sec:problem}

A distinctive trait of the underdamped extremal equations~\eqref{oc:FP},~\eqref{oc:dp} and~\eqref{oc:pot} is the integral term in Eq.~\eqref{oc:pot}, which introduces a non-local condition in the momentum variable. 
This is in stark contrast with the overdamped counterpart of \eqref{oc:pot}. Indeed the latter is exactly integrable and thus reduces the extremal conditions to a pair of local hydrodynamics equations \cite{DaiP1991,AuGaMeMoMG2012}. In this section we construct a systematic multiscale expansion of  \eqref{oc:FP}~-~\eqref{oc:pot} around the overdamped limit.
By proceeding in this way we manage to reabsorb the non-locality in phase space into effective parameters of local equations --the cell problem-- in configuration space.
We perform our analysis in two-dimensional phase space. Extension to higher dimensional phase space is possible at the price of dealing with far more cumbersome algebra.

The approach we follow is inspired by~\cite{WyBa1987,WyBa1987c}. The first step is to project the momentum dependence in Eqs.~\eqref{oc:FP}~-~\eqref{oc:pot}  onto the basis of Hermite polynomials orthonormal with respect to the Maxwell thermal equilibrium distribution. We obtain an kinetic-theory-type hierarchy of coupled equations that do not depend on the momentum.  Despite the additional complication of dealing with an infinite number of equations, this description turns out to be the ideal starting point for a multiscale expansion approach (in time).

\subsection{Non-dimensional variables}

In order to neaten our notation, it is expedient to preliminary introduce non-dimensional variables:
\begin{equation} 
	\nonumber
	\begin{aligned}
		\nt=\dfrac{\,t\,}{\tau},\quad \nq=\dfrac{\,q\,}{\ell},\quad \np=\sqrt{\dfrac{ \beta}{  m}}\,p
	\end{aligned} 
\end{equation} 
where $\ell$ is the typical length-scale set by the mechanical potentials in the boundary conditions.

Next, we introduce the non-dimensional counterparts of the phase space density, value function and mechanical control potential: 
	\begin{align}
		\nf_{\nt}(\nq,\np)&=\ell\,\sqrt{\dfrac{m}{\beta} }\, \styleC{f}_{t}(q,p)
		\nonumber\\
		 \nv_{\nt}(\nq,\np)&=V_{t}(q,p)  
		\nonumber\\[0.2cm]
		\nU_{\nt}(\nq)&=\beta \,U_{t}(q)\,.  
	\nonumber
	\end{align} 
In non-dimensional variables, the generator of the phase space process ~\eqref{eq:FPoperator} becomes
\begin{equation} 
	\begin{aligned}
		\label{eq:fpo}
		\nl_{\bm{\nx}}&=-(\np-\partial_{\np})\,\partial_{\np}+\varepsilon\, \np \,\partial_{\nq}-\varepsilon\,(\partial_{\nq}\nU_{\nt})\,\partial_{\np}
		\\
  &\qquad-\varepsilon^2 g \cbr{\bcbr{\partial_{\nq}\nU_{\nt}}-\partial_{\nq}}\partial_{\nq} \\
  &=\tau \,\styleB{L}_{\bm{x}} 
	\end{aligned} 
\end{equation} 
where now the order parameter of the overdamped expansion
\begin{equation}
\varepsilon=\sqrt{\dfrac{\tau^2}{\beta \, \ell^2 \, m}}\, 
\label{eq:eps}
\end{equation}
explicitly appears. Equipped with these definitions, we rewrite the Fokker-Planck 
\begin{subequations}
\label{eq:control_dl}
\begin{equation} 
		\label{eq:fp_dl}
		\cbr{\partial_{\nt} - \nl_{\bm{\nx}}^{\dagger}}\nf_{\nt}=0\,,\\
  \end{equation}
  the dynamic programming 
  \begin{equation}
  	\begin{split}
		\label{eq:control1_dl}
		&\cbr{\partial_{\nt} + \nl_{\bm{\nx}}}\nv_{\nt}=\begin{cases}
		    -\varepsilon^2\,(1+g)\dfrac{\left (\partial_{\nq}\nU_{\nt}-\partial_{\nq}\nU_{\star}\right)^{2}}{4} \quad &\syma\\[0.4cm]
		    1-\np^{2}\,\varepsilon^2\,g \,\bcbr{(\partial_{\nq}  \nU_{\nt})^2 -\partial_{\nq} ^{2} \nU_{\nt}}\quad &\symb\,,\\
		\end{cases}
		\end{split}
  \end{equation}
  and the stationary condition equations
  \begin{equation}
  	\begin{split}
		\label{eq:control2_dl}
&	\int_{\mathbb{R}} \mathrm{d}\np\, \dfrac{\nf_{\nt}(\nq,\np)}{\nf_{\nt}^{(0)}(\nq)}\cbr{\partial_{\np}+\varepsilon\, g \,\partial_{\nq}}\nv_{\nt}(\nq,\np) =\begin{cases}
		    \dfrac{\varepsilon\,(1+g) }{2}\,\bcbr{\partial_{\nq} \nU_{\nt}(\nq)-\partial_{\nq}\nU_{\star}(\nq)} \quad &\syma\\[0.3cm]
      \varepsilon\, g \,\partial_{\nq} \big{(}2\,\nU_{\nt}(\nq)+\ln \nf_{\nt}^{(0)}(\nq)\big{)}\quad &\symb\,,
		\end{cases}
  \end{split}
\end{equation} 
\end{subequations}
where 
$$
\nf_{\nt}^{(0)}(\nq)=\int_{\mathbb{R}}\mathrm{d}\np \,\nf_{\nt}(\nq,\np)\,
$$
denotes the position marginal of the probability density function.

\subsection{Expansion in Hermite polynomials}

Calling $H_n$ the $n$-th Hermite polynomial (see Appendix~\ref{app:hermite} for details), we expand the probability density and the value function as
\begin{equation} 
\label{eq:pdfhermite}
	\nf_{\nt}(\nq,\np)
	=\dfrac{e^{-\frac{\np^{2}}{2}}}{\sqrt{ 2\,\pi}}\sum_{n=0}^{\infty}\nf_{\nt}^{(n)}(\nq)
	\,H_{n}(\np)
\end{equation} 
and
\begin{equation} 
\label{eq:valuehermite}
	\styleA{V}_{\nt}(\nq,\np)=\sum_{n=0}^{\infty} v_{\nt}^{(n)}(\nq)\, H_n(\np)\, .
\end{equation} 
The expansion coefficients $\nf_{\nt}^{(n)}$ and $\nv_{\nt}^{(n)}$ are scalar functions of the position and time. At equilibrium, the expansion for the probability density consists of the term $n=0$ only. The remaining contributions are non-zero only in out-of-equilibrium conditions. In particular, all $\nf_{\nt}^{(n)}$ and  $\nv_{\nt}^{(n)}$ for $n\,>\,0$ vanish at the beginning and at the end of the control horizon because of the boundary conditions~\eqref{model:ini} and~\eqref{model:fin}. The expansion in Hermite polynomials turns the extremal equations~\eqref{eq:control_dl} into an infinite hierarchy of equations whose $n$-th elements are:

\begin{subequations}
\label{eq:eqhermite}
\begin{align}
\cbr{\partial_{t}+n}&\,\nf_{\nt}^{(n)}+\varepsilon\,\big(\partial_{\nq}-\cbr{\partial_{\nq}  \nU_{\nt}}\big)\,\nf_{\nt}^{(n-1)}	+\varepsilon\,(n+1)\,\partial_{\nq}\nf_{\nt}^{(n+1)}\nonumber =\\ &\varepsilon^2\, g \,\cbr{\partial_{\nq} \bcbr{\cbr{\partial_{\nq} \nU_{\nt}}\nf_{\nt}^{(n)}}+\partial_{\nq}^{2} \nf_{\nt}^{(n)}}
\label{eq:eq1hermite}\\[10pt]
	\cbr{\partial_{t} - n }&v_{\nt}^{(n)} +\varepsilon(n+1)\bcbr{ \partial_{\nq} - (\partial_{\nq} \nU_{\nt})}v_{\nt}^{(n+1)}+\varepsilon \partial_{\nq} v_{\nt}^{(n-1)}\nonumber\\
 &=\begin{cases}
     -\delta_{n,0}\, \dfrac{\varepsilon^{2}\,(1+g)}{4}\,{\left (\partial_{\nq}\nU_{\nt}-\partial_{\nq}\nU_{\star}\right )^{2}}\quad &\syma\\[0.3cm]
-\delta_{n,0}\, g \,\varepsilon^2 \,\cbr{(\partial_{\nq} \nU_{\nt})^2-\partial_{\nq}^2\nU_{\nt}}-\delta_{n,2}\quad &\symb    
 \end{cases}
 \label{eq:eq2hermite}
\\[10pt]
\sum_{n=0}^{\infty}n!&\cbr{ (n+1)\,\nf_{\nt}^{(n)}\, v_{\nt}^{(n+1)}+\varepsilon\, g\, \partial_{\nq} v_{\nt}^{(n)}}=\begin{cases}
        \dfrac{\varepsilon\,(1+g)}{2}\, \nf_{\nt}^{(0)} \cbr{{\partial_{\nq}\nU_{\nt}-\partial_{\nq}\nU_{\star}}}\quad &\syma\\[0.3cm]
        \varepsilon \,g \, \nf_{\nt}^{(0)} \,\partial_{\nq} \big{(}2\,\nU_{\nt}+\ln \nf_{\nt}^{(0)}\big{)}\,. \quad &\symb
    \end{cases}
     \label{eq:eq3hermite}
\end{align}
\end{subequations}

More detail on the derivation of the above equations is given in Appendix~\ref{app:hermite}. The hierarchy is complemented by equilibrium boundary conditions on the probability density, that, in the non-dimensional variables, read:
\begin{subequations}
\label{eq:bchermite}
\begin{eqnarray}
     \nf_{\tti}^{(0)}(\nq)&=&\dfrac{\exp(-\nU_{\iota}(\nq))}{\int_{\mathbb{R}}\mathrm{d}y \,\exp(-\nU_{\iota}(y)) }\\[3pt]
      \nf_{\ttf}^{(0)}(\nq)&=&\dfrac{\exp(-\nU_\styleB{f}(\nq))}{\int_{\mathbb{R}}\mathrm{d}y \,\exp(-\nU_\styleB{f}(y)) }\\[3pt]
      \nf_{\tti}^{(n)}(\nq)&=&\nf_{\ttf}^{(n)}(\nq)=0\quad\quad n\ge 1\,.
\end{eqnarray}
\end{subequations}

\section{Multiscale perturbation theory}
\label{sec:ms}

The hierachy~\eqref{eq:eqhermite} is equivalent to the original extremal equations~\eqref{eq:control_dl}, and holds for any value of $\varepsilon$. We are interested in cases where $\varepsilon \ll 1$ in order to solve~\eqref{eq:eqhermite} with a perturbative strategy. The limit of vanishing $\varepsilon$  is, however, singular and cannot be handled by regular perturbation theory. We therefore resort to multiscale perturbation theory. In doing so, we need to take into account an essential difference with respect to the multiscale treatment of the overdamped limit of the underdamped dynamics \cite{WyBa1987, WyBa1987c}. The difference is that the mechanical potential is not assigned but must be determined by solving the stationary conditions (\ref{eq:eq3hermite}). In addition, we are dealing with a time-boundary value problem rather than with an initial data problem.
To overcome these difficulties, we formulate the multiscale expansion drawing from the Poincaré-Lindstedt technique~\cite{VerF2005} and renormalization group ideas that in recent years have been successfully applied to the resummation of perturbative series arising from Hamiltonian and dissipative dynamical systems \cite{GenG2010}. Our strategy is based on the following considerations.
\begin{itemize}
	\item We suppose that the time variation of all functions in the hierarchy (\ref{eq:eqhermite}) occurs through effective time variables 
\begin{align}
	\label{ms:multi}
&	\st{j}=\varepsilon^{j} \,\nt, && j \,\geq\, 0\,.
\end{align}
occasioned by the overdamped order parameter $\varepsilon$. As a consequence, the partial derivative with respect to $\nt$ breaks down into a differential operator
\begin{align}
	\partial_{\nt}=\partial_{\st{0}}+\varepsilon\,\partial_{\st{1}}+\varepsilon^{2}\,\partial_{\st{2}}+\dots
	\label{ms:td}
\end{align}
thus introducing a new dynamical variable at each order of the overdamped expansion.
\item  We assume that the mechanical potential has a finite limit when $\varepsilon$ tends to zero. This assumption \cite{PMG2014,MGSc2014} is central in order to recover the overdamped dynamics \cite{DaiP1991,AuGaMeMoMG2012}. The a priori justification of the assumption is that momentum marginals of the boundary conditions already describe a Maxwell thermal equilibrium. 
The need for a controlled dynamics only arises in consequence of the boundary conditions imposed on the position process. 
In the generator (\ref{eq:fpo}), the mechanical potential is coupled to the dynamics by the overdamped expansion order parameter $\varepsilon$. 
This fact leads to the inference that the control potential should admit a regular expansion in $\varepsilon$ as a function of the position variable, varying in time on scales set by $\varepsilon$.
\item The Poincaré-Lindstedt method is usually formulated for initial value problems. In such a context, the dynamics of the slow times $\st{j}$ with $j\,>\,0$ is fixed by canceling secular terms (equivalently: resonances), i.e. polynomial terms in the time variable which as times increases would lead to a breakdown of perturbation theory. Such a secular term subtraction scheme is equivalent to a renormalization group type partial resummation of the perturbative expansion \cite{GenG2010}. We need to adapt the subtraction scheme to a boundary problem. At $\varepsilon$ equal zero the Fokker-Planck hierarchy (\ref{eq:eq1hermite}) is decoupled from the dynamic programming one (\ref{eq:eq2hermite}). As a consequence, the boundary conditions (\ref{eq:bchermite}) cannot be satisfied at zero order of the regular perturbative expansion. We therefore use the boundary conditions to determine the slow time dependence of the $\nf^{(n)}$.
\item The value function expansion coefficients $\nf^{(n)} $ are not subject to anything other than satisfying the dynamics (\ref{eq:eq1hermite}). The logical basis for the resonance subtraction scheme is the duality relation (\ref{oc:dual}). We reason that a cost can only be generated on the same time scales over which the mechanical control potential varies. We thus require that the dependence of $\nf^{(0)} $ must be constant with respect to the fastest time $ \st{0}$. We also observe that, although physically motivated, a non-uniqueness is intrinsic in any secular term cancellation or finite renormalization scheme exactly because these techniques involve a partial and not a complete resummation of the perturbative expansion \cite{AmiD2005}.
Consistent alternative schemes may differ by higher order terms in the regular perturbative expansion.
\item The introduction of the slow time variables (\ref{ms:multi}) is justified under a sufficiently wide scale separation. In principle, the perturbative expansion only holds for $\varepsilon\ll 1$ and $\ttf \gg 1$ (i.e. $\tf \gg \tau$).  Yet, we hope that extrapolating the results for finite control horizons will give sufficiently accurate results if the resummation scheme correctly captures the \textquotedblleft turnpike behavior\textquotedblright\ of the exact solution of the optimal control. Turnpike behavior means the tendency of optimal controls to approximate the solution of the adiabatic limit, corresponding to a vanishing cost, as much as possible. 

We refer to \cite{ChCoGrRe2021} for further discussion and references on this point.
\end{itemize} 
In summary, our aim is to look for a solution of (\ref{eq:eqhermite}) in the form of multiscale power series
\begin{subequations}
\begin{align}   
\label{eq:expeps1}
        \nf_{\nt}^{(n)}(\nq)&=\sum_{i=0}^{\infty}\varepsilon^{i}\,\nf_{\slt{0}}^{(n:i)}(\nq)\\ 
\label{eq:expeps2}
        v_{\nt}^{(n)}(\nq)&=\sum_{i=0}^{\infty}\varepsilon^{i}\,v_{\slt{0}}^{(n:i)}(\nq)\,\\ 
\label{eq:expeps3}
    \nU_{\nt}(\nq)&=\sum_{i=0}^{\infty}\varepsilon^{i}\,\nU_{\slt{0}}^{(i)}(\nq)\,,
\end{align}
\end{subequations}
where each addend of the above series depends, \textit{a priori}, on all time scales
\begin{equation}
	\slt{{j}}=(\st{{j}},\,\st{{j}+1},\,\st{j+2},\,...)\,.
\end{equation}

\subsection{Results}
\label{sec:results}

We report the main results of the overdamped multiscale expansion, while deferring their derivation to Section~\ref{sec:maobo}. Without loss of generality, we set 
$$\ti=0$$ to neaten the notation.
Within second order in $\varepsilon$ in the multiscale expansion, the solution of the Fokker-Planck equation takes the form
\begin{align}\begin{split}
	&\nf_{\nt}(\nq,\np)=\cbr{\nf_{\st{0},\st{2}}^{(0:0)}(\nq)+\varepsilon\, \np\, \nf_{\st{0},\st{2}}^{(1:1)}(\nq)}\dfrac{e^{-\frac{\np^{2}}{2}}}{\sqrt{2\,\pi}}
	\\
	&+\varepsilon^2\cbr{\nf_{\st{0},\st{2}}^{(0:2)}(\nq)+\np\,\nf_{\st{0},\st{2}}^{(1:2)}(\nq)+(\np^{2}-1)\,\nf_{\st{0},\st{2}}^{(2:2)}(\nq)}\dfrac{e^{-\frac{\np^{2}}{2}}}{\sqrt{2\,\pi}}+\dfrac{e^{-\frac{\np^{2}}{2}}}{\sqrt{2\,\pi}}\,O(\varepsilon^{3})\,.
	\label{eq:fsol}
\end{split}\end{align}
We emphasize that $\st{0}=\nt$ and $\st{2}=\varepsilon^2 \,\nt$ and that Eq.~\eqref{eq:fsol} is independent of $\st{1}=\varepsilon\,\nt$. We also neglect slower time scales $\st{j}$, ${j}\,>\,2$, as they only provide higher order corrections. Hence, for all the results presented in this Subsection, we drop the explicit dependence on $\st{1}$ and $\slt{3}$. 

\subsubsection{Cell problem equations}

As customary \cite{PaSt2008}, we refer to the secular term subtraction conditions emerging at order $O(\varepsilon^{2})$ in regular perturbation theory as the \emph{cell problem}. 
Secular term subtraction fixes the functional dependence upon the slow time $\st{2}$. As a consequence, we find it expedient to denote the unknown quantities of the cell problem 
as 
\begin{equation}
	\label{eq:rho}
	\rho_{\st{2}}(\nq)=\nf_{\st{0},\st{2}}^{(0:0)}(\nq)
\end{equation}  
and as an auxiliary field $\sigma_{\st{2}}(\nq)$ related to  $ v_{\ttf,\st{2}}^{(0:0)}(\nq)$ and $\nf_{\st{0},\st{2}}^{(0:0)}(\nq)$ by equation (\ref{eq:sigma}) in section~\ref{sec:maobo} below. The formulation of 
the cell problem in terms of the pair $\rho_{\st{2}}$, $\sigma_{\st{2}}$ exactly recovers the optimal control equations governing the overdamped limit in \hyperref[itm:a]{KL} and \hyperref[itm:b]{EP}:
	\begin{subequations}
		\label{eq:sol_o2} 
		\begin{align} 
			\label{eq:sol_o2_f}
			\partial_{\st{2}}\rho_{\st{2}}&=\partial_{\nq}\bcbr{\rho_{\st{2}}\partial_{\nq} \sigma_{\st{2}}+\alpha\,  \partial_{\nq}\rho_{\st{2}}}\\
			\label{eq:sol_o2_v}
			\partial_{\st{2}}\sigma_{\st{2}}
			&=\frac{1}{2}(\partial_{\nq} \sigma_{\st{2}})^2-\alpha \,\partial_{\nq}^2 \sigma_{\st{2}}
			+\alpha^2\,\cbr{\partial_{\nq}^{2} \nU_{\star} -\dfrac{\bcbr{\partial_{\nq} \nU_{\star}}^2}{2}}-\alpha^{2} \,\chi_{\st{2}} \nq\\
			\chi_{\st{2}}&=\frac{B}{A}\,\int_{\mathbb{R}}\mathrm{d}\nq\,\rho_{\st{2}}(\nq)\,\partial_{\nq} \cbr{\partial_{\nq}^2 \nU_\star(\nq) - \dfrac{\bcbr{\partial_{\nq}\nU_{\star}(\nq)}^2}{2}}\,.
		\end{align}
	\end{subequations}	
We fully specify the cell problem by complementing (\ref{eq:sol_o2})  with the exact boundary conditions imposed by the position marginals of (\ref{model:ini}), (\ref{model:fin}) and written in non-dimensional variables as in (\ref{eq:bchermite})
	\begin{equation}
		\begin{split}
			\rho_{\st{2}}(\nq)\Big|_{\st{2}=0}&=\frac{e^{-\nU_{\iota}(\nq)}}{\int_{\mathbb{R}}\mathrm{d}y\, e^{-\nU_{\iota}(y)}} 
			\\
			\rho_{\st{2}}(\nq)\Big|_{\st{2}=\varepsilon^2 \ttf}&=\frac{e^{-\nU_{\styleB{f}}(\nq)}}{\int_{\mathbb{R}}\mathrm{d}y\, e^{-\nU_{\styleB{f}}(y)}}\,. 
		\end{split}
		\label{eq:bc}
	\end{equation}
In (\ref{eq:sol_o2}) we imply 
\begin{align}
	\nU_{\star}(\nq)=0
	\nonumber
\end{align}	
for \hyperref[itm:b]{EP}. Depending upon the problem under consideration, the constant $\alpha$ takes the values  
\begin{equation}
	\label{eq:alpha}
	\alpha=\begin{cases}
		\sqrt{(1+g)\,A}\quad&\syma\,\\[0.3cm]
		0\quad&\symb\,,
	\end{cases}
\end{equation}
while the constants $A$ and $B$ are given by (see Eq.~\eqref{eq:AB_constants} below)  
\begin{equation}
	\begin{split}
	&	\frac{A}{1+g}=1-\frac{(\omega^{2}-4)\tanh\frac{\omega\,\ttf}{2} \tanh \ttf }{\omega\,\ttf \left(\omega\,\tanh\frac{\omega\,\ttf}{2}-2 \tanh \ttf \right)}
	\\
	&	\frac{B}{1+g}=-\,\frac{\tanh\frac{\omega\,\ttf}{2}}{\ttf}
	\frac{\omega\,\tanh \ttf -2\tanh\frac{\omega\,\ttf}{2}}{ \omega\,\tanh\frac{\omega\,\ttf}{2}-2 \tanh \ttf }
	\end{split}
	\label{eq:A&B}
\end{equation}
with 
\begin{equation}
	\omega=\begin{cases}
		1 \quad &\syma\\[0.3cm]
		\sqrt{\dfrac{1+g}{g}}\,.\quad &\symb
	\end{cases}
	\label{eq:omega}
\end{equation}
$A$ and $B$ always admit a finite limit as $g$ tends to zero: by (\ref{eq:omega}) the limit of vanishing $g$ entails $\omega$ tending to infinity in case \hyperref[itm:b]{EP}. Furthermore, they depend upon the size of the control horizon $\ttf$  so that
\begin{align}
	\lim_{\ttf\nearrow\infty}\left(A-1-g\right)=\lim_{\ttf\nearrow\infty}B=0\,.
	\label{eq:infhor}
\end{align}
When $U_{\star}=0$, the cell problem reduces to a coupled system of a Fokker-Planck and Burgers' equation
\begin{subequations}
	\label{eq:cell} 
	\begin{align} 
		\label{eq:cellFP}
		\partial_{\st{2}}\rho_{\st{2}}&=\partial_{\nq}\bcbr{\rho_{\st{2}}\partial_{\nq} \sigma_{\st{2}}+\alpha\,  \partial_{\nq}\rho_{\st{2}}}\\
		\label{eq:cellB}
		\partial_{\st{2}}\sigma_{\st{2}}
		&=\frac{1}{2}(\partial_{\nq} \sigma_{\st{2}})^2-\alpha \,\partial_{\nq}^2 \sigma_{\st{2}}\,.
	\end{align}
\end{subequations}
By (\ref{eq:infhor}) and (\ref{eq:alpha}) in the limit of infinite control horizon ($\ttf \nearrow \infty$), we recover the result of \cite{PMG2013} that in the overdamped limit optimal entropic transport is a viscous regularization of the minimization of the mean entropy production. As the optimal control of the latter problem \cite{AuGaMeMoMG2012} is equivalent to optimal mass transport, we also recover Mikami's result \cite{MikT2004}. In fact, for $\ttf$ finite but sufficiently large to justify the scale separation required by the multiscale approach, the cell problem  (\ref{eq:cell}) allows us to extract information about corrections to the overdamped 
limit.

\subsubsection{Cumulants and marginal distribution}
\label{section:cumulants}

Solving the cell problem allows us to evaluate the leading order corrections to the overdamped limit of all phase space cumulants within order $O(\varepsilon^{2})$.  Namely, all cumulants turn out to be linear combinations of functionals of the pair $\rho_{\st{2}}$ and $\sigma_{\st{2}}$,  weighed by pure functions of the fast time $\st{0}$.

We denote the non-dimensional counterparts of the position and momentum processes with a tilde
\begin{align}
&	\tilde{\mathscr{q}}_{\nt}=\frac{\mathscr{q}_{\tau\,\nt}}{\ell},\quad\tilde{\mathscr{p}}_{\nt}=\sqrt{\frac{\beta}{m}}\,\mathscr{p}_{\tau\,\nt}\,.
	\nonumber
\end{align}
Unlike the cell problem, the cumulants are functions of both the fast $\st{0}$ and slow $\st{2}$ time variables. To neaten the expressions,
we denote the moments of the position process with respect to the probability density specified by the cell problem as
\begin{align}
\label{eq:mus}
&	\mu_{\st{2}}^{({n})}=\int_{\mathbb{R}}\mathrm{d}\nq\, \rho_{\st{2}}(\nq)\,\nq^{{n}},\quad {n}=1,2\,.
\end{align}
Correspondingly, we also write
\begin{align}
	\dot{\mu}_{\st{2}}^{({n})}=\partial_{\st{2}}\int_{\mathbb{R}}\mathrm{d}\nq \,\rho_{\st{2}}(\nq)\,\nq^{{n}}\,.
	\nonumber
\end{align}
In particular, 
\begin{align}
	\dot{\mu}_{\st{2}}^{({1})} =-\int_{\mathbb{R}}\mathrm{d}\nq \,\rho_{\st{2}}(\nq)\,(\partial \sigma_{\st{2}})(\nq)\,
	\nonumber
\end{align}
is a constant, i.e. 
\begin{align}
	\ddot{\mu}_{\st{2}}^{({1})}=0\,,
	\nonumber
\end{align}
for both the optimal entropic transport (\hyperref[itm:a]{KL} with zero reference potential) and minimum mean entropy production \hyperref[itm:b]{EP} problems. We justify this claim in appendix~\ref{app:oboe}. 

\begin{description}[style=unboxed,leftmargin=0cm]
	\item[Momentum mean] recalling~\eqref{eq:fsol}, the expectation value of the momentum process conditioned on the position process is
	\begin{align}
		\operatorname{E}_{\mathcal{P}}\left(\tilde{\mathscr{p}}_{\nt}\,\big{|}\,\tilde{\mathscr{q}}_{\nt}=\nq\right)\rho_{\nt}(\nq)=\int_{\mathbb{R}}\mathrm{d}\np\, \np \,\nf_{\mathrm{t}}(\nq,\np) = \varepsilon \,\nf_{\st{0},\st{2}}^{(1:1)}(\nq)+O(\varepsilon^2)\,.
		\label{eq:cmmv}
	\end{align}
	In section~\ref{sec:maobo}, we show how to compute $\nf_{\st{0},\st{2}}^{(1:1)}$ from the solution of the cell problem. We obtain
	\begin{equation}
		\label{eq:sol_o1_f}
		\begin{split}
			\nf_{\st{0},\st{2}}^{(1:1)}&=-a_{\st{0}}\,\rho_{\st{2}}\,\frac{\partial (
					\sigma_{\st{2}}+\alpha \,\ln \rho_{\st{2}})}{A}	+\rho_{\st{2}}\frac{B\,a_{\st{0}}-A \,b_{\st{0}}}{A\,(A-B)}\,\dot{\mu}_{\st{2}}^{({1})} \,.
	\end{split}
	\end{equation}		
   Here $ a_{\st{0}}$ and $ b_{\st{0}}$ are pure functions of the fast time $\st{0}$:
	\begin{subequations}
		\begin{align}
&a_{\st{0}}=1+\sinh (\omega \st{0}) \tanh \frac{ \omega \ttf}{2}-\cosh ( \omega \st{0})+b_{\st{0}}
\\
&b_{\st{0}}=\frac{\omega\,e^{-\ttf}\,
	\left(\cosh (\omega\st{0})-e^{2 \st{0}}\right)}{\omega  \cosh \ttf-2 \sinh  \ttf \coth \frac{ \omega \ttf}{2}}	+		\frac{\omega\,e^{-\ttf} \left(e^{2 \ttf}-\cosh (\omega \ttf)\right)\, \sinh ( \omega\st{0} )
	}{\left(\omega  \cosh \ttf-2 \sinh  \ttf \coth \frac{ \omega \ttf}{2}\right)\sinh( \omega\ttf)}		\,.
	\end{align}
	\label{eq:a&b}
	\end{subequations}
	It is straightforward to verify that
	\begin{align}
		a_{0}=a_{\ttf}=b_{0}=b_{\ttf}=0
		\nonumber
	\end{align}
	hence enforcing the boundary conditions imposed on (\ref{eq:sol_o1_f}). The derivation of an explicit expression of this quantity requires 
	the solution of the $O(\varepsilon^{4})$ cell problem in the same way as (\ref{eq:sol_o2_f}) specifies $\nf^{(1:1)}$.

    Equipped with the above definitions, by integrating~\eqref{eq:cmmv} in $\nq$ we arrive at
	\begin{align}
    \operatorname{E}_{\mathcal{P}}\tilde{\mathscr{p}}_{\mathrm{t}}=\varepsilon\, \frac{a_{\st{0}}-b_{\st{0}}}{A-B}\,\dot{\mu}_{\st{2}}^{({1})}+O(\varepsilon^{2})\,.
	\end{align}
	In order to interpret this result, we recall a standard result of multiscale analysis (see e.g. \S ~2.5.1 of  \cite{PaSt2008}) ensuring that
	\begin{align}
		\int_{0}^{\ttf}\mathrm{d}\nt\,\operatorname{E}_{\mathcal{P}}\tilde{\mathscr{p}}_{\mathrm{t}}\approx\varepsilon
		\int_{0}^{\ttf}\mathrm{d}\st{0}\,\frac{a_{\st{0}}-b_{\st{0}}}{A-B}
		\int_{0}^{\varepsilon^{2}\,\ttf}\mathrm{d}\st{2}\,\dot{\mu}_{\st{2}}^{({1})}
		\nonumber
	\end{align}
	when the separation of scales is sufficiently large: $\ttf\gg O(1)$ with $ \varepsilon^{2}\,\ttf =O(1)$. 
	 The relation (\ref{eq:AB_constants}) between the integral over the functions (\ref{eq:a&b}) and the constants $A$ and $B$ (\ref{eq:A&B})
	 implies that as the duration of the control horizon grows, the momentum expectation tends to
	\begin{align}
		\int_{0}^{\ttf}\mathrm{d}\nt\,\operatorname{E}_{\mathcal{P}}\tilde{\mathscr{p}}_{\mathrm{t}}
		\overset{\ttf  \gg 1}{\approx}\varepsilon\,\ttf\,\frac{\mu_{\varepsilon^{2}\ttf}^{({1})}-\mu_{0}^{({1})}}{1+g}\,.
		\nonumber
	\end{align}
	Once recast into dimensional quantities, the identity reads
\begin{align}
	\int_{0}^{\tf}\mathrm{d}t\,\operatorname{E}_{\mathcal{P}}\mathscr{p}_{t}
	\overset{\tf  \gg \tau}{\approx}\frac{\tau\,\tf}{\beta}\,\frac{\mu_{\varepsilon^{2}\ttf}^{({1})}-\mu_{0}^{({1})}}{\ell\,(1+g)}\,.
\end{align}
	
	\item[Correction to the position marginal distribution] upon integrating out the momentum variable in  (\ref{eq:fsol}), we get
	\begin{align}
		\int_{\mathbb{R}}\mathrm{d}\np\,\nf_{\nt}(\nq,\np)=\rho_{\st{2}}(\nq)+\varepsilon^{2}\,\nf^{(0:2)}_{\st{0},\st{2}}(\nq)+O(\varepsilon^{3})\,.
		\label{eq:qmar}
	\end{align}
In section~\ref{sec:maobo} we show that
	\begin{equation}
		\label{eq:f02}
		\begin{aligned}
			\nf^{(0:2)}_{\st{0},\st{2}}=-g \,\partial_{\nq} \nf_{\st{0},\st{2}}^{(1:1)}   +(1+g) \left(\dfrac{\st{0}}{\ttf}\int_{0}^{\ttf}\mathrm{d}\ns  -\int_{0}^{\st{0}}\mathrm{d}\ns \right)\partial_{\nq}\nf_{\ns,\st{2}}^{(1:1)}\,.
		\end{aligned}
	\end{equation}
	Inspection of (\ref{eq:f02}) reveals that the marginal (\ref{eq:qmar}) exactly satisfies the non-perturbative boundary conditions (\ref{eq:bc}) and  preserves normalization within accuracy. We avail ourselves of (\ref{eq:qmar}) to evaluate the remaining linear and second order cumulants.
	\item[Position mean] we readily obtain
 \begin{align}
	\lefteqn{\operatorname{E}_{\mathcal{P}}\tilde{\mathscr{q}}_{\mathrm{t}}=\mu_{\st{2}}^{({1})}	+	\varepsilon^{2}\,g\, \frac{a_{\st{0}}-b_{\st{0}}}{A-B}\,\dot{\mu}_{\st{2}}^{(1)}}
		\nonumber\\
		&\qquad-\frac{\varepsilon^{2}\,(1+g)\,\dot{\mu}_{\st{2}}^{(1)}}{A-B} \left( \frac{\st{0}}{\ttf}\int_{0}^{\ttf}\mathrm{d}\ns\,  -\int_{0}^{\st{0}}\mathrm{d}\ns\,   \right)(a_{\ns}-b_{\ns})	+O(\varepsilon^{3})
	\end{align}
	As expected, at the boundaries of the control horizon the cumulant are fully specified by the boundary conditions and so are independent of $\varepsilon$.
	\item[Position-momentum cross correlation] after straightforward algebra we find
	\begin{align}
    \operatorname{E}_{\mathcal{P}}\left(\tilde{\mathscr{q}}_{\mathrm{t}}\tilde{\mathscr{p}}_{\mathrm{t}}\right)-\operatorname{E}_{\mathcal{P}}\left(\tilde{\mathscr{q}}_{\mathrm{t}}\right)\operatorname{E}_{\mathcal{P}}\left(\tilde{\mathscr{p}}_{\mathrm{t}}\right)=
		\varepsilon \, \frac{a_{\st{0}}\,\dot{\varsigma}_{\st{2}}}{2\,A}
		+O(\varepsilon^{2})
	\end{align}
	where
	\begin{align}
		\varsigma_{\st{2}}=\mu_{\st{2}}^{({2})}-(\mu_{\st{2}}^{({1})})^{2}
		\nonumber
	\end{align}
	and
	\begin{align}
		\dot{\varsigma}_{\st{2}}&=-2\,\int_{\mathbb{R}}\mathrm{d}\nq \,\rho_{\st{2}} \,\nq \,\partial_{\nq}\left(\sigma_{\st{2}}+\alpha\ln\rho_{\st{2}}\right )-2\,\dot{\mu}_{\st{2}}^{({1})}\,\mu_{\st{2}}^{({1})}
		\nonumber\\
		&=\partial_{\st{2}}\left(\mu_{\st{2}}^{({2})}-(\mu_{\st{2}}^{({1})})^{2}\right)\,\equiv\,\partial_{\st{2}}\varsigma_{\st{2}}\,.
		\label{eq:Kfun}
	\end{align}
	\item[Position variance] we obtain its expression by evaluating the difference between
	\begin{align}
		&	\operatorname{E}_{\mathcal{P}}\tilde{\mathscr{q}}_{\nt}^{2}=\mu_{\st{2}}^{({2})}
		+2\,\varepsilon^{2}\,g\,\int_{\mathbb{R}}\mathrm{d}\nq  \,\nq \,\nf_{\st{0},\st{2}}^{(1:1)}(\nq)  -2\, \varepsilon^{2}\,(1+g) \int_{\mathbb{R}}\mathrm{d}\nq  \,\nq\,\left(\dfrac{\st{0}}{\tf}\int_0^{\tf}\mathrm{d}\ns\, -\int_0^{\st{0}}\mathrm{d}\ns\, \right)\nf_{\ns,\st{2}}^{(1:1)}(\nq)
		\nonumber
	\end{align}
	and the squared mean value
	\begin{align}
			\left(\operatorname{E}_{\mathcal{P}}\tilde{\mathscr{q}}_{t}\right)^{2}&=(\mu_{\st{2}}^{({1})})^{2}
		+	2\,\varepsilon^{2}\,g\, \frac{a_{\st{0}}-b_{\st{0}}}{A-B}\,\mu_{\st{2}}^{({1})}\dot{\mu}_{\st{2}}^{(1)}\nonumber \\&-2\,\frac{\varepsilon^{2}\,(1+g)\,\mu_{\st{2}}^{({1})}\,\dot{\mu}_{\st{2}}^{(1)}}{A-B} \left( \frac{\st{0}}{\ttf}\int_{0}^{\tf}\mathrm{d}\ns\,  -\int_{0}^{\st{0}}\mathrm{d}\ns\,   \right)(a_{\ns}-b_{\ns})
		+O(\varepsilon^{3})\,.
		\nonumber
	\end{align}
	After some algebra, we find that the expression of the variance reduces to
	\begin{align}
		&	\operatorname{E}_{\mathcal{P}}\tilde{\mathscr{q}}_{\nt}^{2}-\left(\operatorname{E}_{\mathcal{P}}\tilde{\mathscr{q}}_{\nt}\right)^{2}=
		\varsigma_{\st{2}}
		+\varepsilon^{2}\,g\,\frac{a_{\st{0}}}{A}\,\dot{\varsigma}_{\st{2}}
		- \varepsilon^{2}\,\frac{1+g}{A}\dot{\varsigma}_{\st{2}}
		\left(\frac{\st{0}}{\ttf}\int_0^{\ttf}\mathrm{d}\ns -\int_0^{\st{0}}\mathrm{d}\ns\right)a_{\ns}+O(\varepsilon^{3})
	\end{align}
	with $\varsigma_{\st{2}} $ and $\dot{\varsigma}_{\st{2}} $ defined by (\ref{eq:Kfun}).
	\item[Momentum variance] the  expectation value of the squared momentum conditioned on the position process is
	\begin{align}
			\operatorname{E}\left(\tilde{\mathscr{p}}_{\nt}^{2}\,\big{|}\, \tilde{\mathscr{q}}_{\nt}=\nq\right)\rho_{\st{2}}(\nq)
			=\int_{\mathbb{R}}\mathrm{d}\np\, \np^2 \,\nf_{\nt}(\nq,\np) 
			= 	\rho_{\st{2}}(\nq)
		+ \varepsilon^2 \,\left(\nf_{\st{0},\st{2}}^{(0:2)}(\nq)+2\,\nf_{\st{0},\st{2}}^{(2:2)}(\nq) \right)+O(\varepsilon^3)\,.    
		\nonumber
	\end{align}
	with
	\begin{equation}
		\label{eq:f22}
		\nf_{\st{0},\st{2}}^{(2:2)}=\frac{\left(\nf_{\st{0},\st{2}}^{(1:1)}\right)^2}{2\,\rho_{\st{2}}}-\rho_{\st{2}} \int_0^{\st{0}}\mathrm{d}\ns\, e^{-2(\st{0}-\ns)} \partial_{\nq} \left(\frac{\nf_{\st{0},\st{2}}^{(1:1)}}{\rho_{\st{2}}}\right)\,.
	\end{equation}
	After some tedious algebra we arrive at
	\begin{align}
	\operatorname{E}\left(\tilde{\mathscr{p}}_{\nt}^{2}\right)-(\operatorname{E}\left(\tilde{\mathscr{p}}_{\nt}\right))^{2}=
		1&-\varepsilon^{2}\, \frac{a_{\st{0}}^{2}}{A^{2}}\,(\dot{\mu}_{\st{2}}^{({1})})^{2}+2 \,\varepsilon^{2}\int_0^{\st{0}}\mathrm{d}\ns\, e^{-2(\st{0}-\ns)}	\frac{a_{\ns}}{A}\int_{\mathbb{R}}\mathrm{d}\nq\,\rho_{\st{2}} \,\partial_{\nq}^{2}(
	\sigma_{\st{2}}+\alpha \ln \rho_{\st{2}})
		\nonumber\\
	&	+\varepsilon^{2}\,\frac{a_{\st{0}}^{2}}{A^{2}}\int_{\mathbb{R}}\mathrm{d}\nq\,
	\rho_{\st{2}}\big{(}\partial_{\nq}(
	\sigma_{\st{2}}+\alpha \ln \rho_{\st{2}})\big{)}^{2}
		+O(\varepsilon^{3})\,.
	\end{align}
	We notice that the variance satisfies the boundary conditions in consequence of the identity
	\begin{align}
		\int_0^{\ttf}\mathrm{d}\ns\, e^{-2(\st{0}-\ns)}	a_{\ns}=0
		\nonumber
	\end{align}
	which follows from (\ref{eq:f11bc}).
\end{description}

\subsubsection{Optimal control potential}
\label{section:optimal_drift}

Similarly, the cell problem yields the leading order expression for the gradient of the optimal control potential 
\begin{equation}\begin{split}
    		    \partial_{\nq}  \nU_{\nt} (\nq)&=-\partial_{\nq}\ln\rho_{\st{2}}(\nq)
		 -\frac{\dot{a}_{\st{0}}+a_{\st{0}}}{A}\,\partial_{\nq}\mathscr{c}_{\st{2}}(\nq)\\
	&\qquad-\frac{(B\,\dot{a}_{\st{0}}-A \,\dot{b}_{\st{0}})+(B\,a_{\st{0}}-A \,b_{\st{0}})}{A\,(A-B)}\,\dot{\mu}_{\st{2}}^{({1})} +O(\varepsilon)
 \label{eq:optdrift}
 \end{split}
\end{equation}
with $$\dot{a}_{\st{0}}=\partial_{\st{0}}\,a_{\st{0}}$$
and
\begin{align}
	\mathscr{c}_{\st{2}}(\nq)=-\sigma_{\st{2}}(\nq)-\alpha \ln \rho_{\st{2}}(\nq)
	\label{optimal_drift:cp}
\end{align}
the current potential of the cell problem. In other words, the gradient of (\ref{optimal_drift:cp}) is the current velocity \cite{NelE2001} which allows us to represent (\ref{eq:sol_o2_f}) as a mass conservation equation for any strictly positive $\alpha$. 

By definition, the current velocity vanishes when the system is in a Maxwell-Boltzmann equilibrium state. Hence, finite time transitions at minimum cost are not between Maxwell-Boltzmann equilibrium states, as we see from the explicit expression of the drift at the end times
\begin{align}
& \partial_{\nq}  \nU_{0} (\nq)=-\partial_{\nq}\ln\rho_{0}(\nq)-\frac{\dot{a}_{0}}{A}\,\partial_{\nq} \mathscr{c}_{0}(\nq)-\frac{B\,\dot{a}_{0}-A \,\dot{b}_{0}}{A\,(A-B)}\,\dot{\mu}_{0}^{({1})} +O(\varepsilon)
	\nonumber
\end{align}
and
\begin{align}
	& \partial_{\nq}  \nU_{\varepsilon^{2}\ttf} (\nq)=-\partial_{\nq}\ln\rho_{\varepsilon^{2}\ttf}(\nq)-\frac{\dot{a}_{\varepsilon^{2}\ttf}}{A}\,\partial_{\nq}	\mathscr{c}_{\varepsilon^{2}\ttf}(\nq)
	-\frac{B\,\dot{a}_{\varepsilon^{2}\ttf}-A \,\dot{b}_{\varepsilon^{2}\ttf}}{A\,(A-B)}\,\dot{\mu}_{\varepsilon^{2}\ttf}^{({1})} +O(\varepsilon)\,.
	\nonumber
\end{align}
From the physics point of view, this means transitions minimizing thermodynamic cost functionals have non-vanishing current velocity at the start and end of the protocol.  Mathematically, this is unsurprising because the boundary conditions associated to the optimal control problem do not impose any conditions on the terminal values of the control potentials. %

For all practical purposes, the shape of potential corresponding to the boundary equilibrium states can be matched at zero cost, through an instantaneous change of the control.

\subsubsection{Minimum cost}
\label{sec:mcd}

We evaluate the expression for the minimum cost using the duality relation (\ref{oc:dual}).

\begin{description}[style=unboxed,leftmargin=0cm]
	\item\hyperref[itm:a]{KL}:  The projection onto Hermite polynomials couches  (\ref{oc:dual}) into the form
	\begin{align}
	\operatorname{K}(\mathcal{P}\mathrel{\Vert}\mathcal{Q})=\int_{\mathbb{R}}\mathrm{d}\nq\,\left(\nf_{0}^{(0)} v_{0}^{(0)}-\nf_{\ttf}^{(0)} v_{\ttf}^{(0)}\right)\,.
		\nonumber
	\end{align}
	At leading order, multiscale perturbation theory yields the approximation
	\begin{align}
		\operatorname{K}(\mathcal{P}\mathrel{\Vert}\mathcal{Q})=\int_{\mathbb{R}}\mathrm{d}\nq\,\left(\nf_{0,0}^{(0:0)} v_{\ttf,0}^{(0:0)}-\nf_{0,\varepsilon^{2}\ttf}^{(0:0)} v_{\ttf,\varepsilon^{2}\ttf}^{(0:0)}\right)+O(\varepsilon^{2})\,.
		\nonumber
	\end{align}
	This is because the non-perturbative boundary conditions only allow contributions that are proportional to $ \nf_{0,0}^{(0:n)}$. 
 In addition, we subtract secular terms in the value function
	expansion by requiring
	\begin{align}
		v_{\ttf,\st{2}}^{(0:2)}=v_{0,\st{2}}^{(0:2)}\,.
		\nonumber
	\end{align}
	Thus, in our multiscale framework,  the value of $v_{\ttf,\varepsilon^{2}\ttf}^{(0:2)}$ can be only determined by  higher order cell problems.
	
	To gain insight into the predicted features of the minimum, we couch the optimum value of the divergence into the form
	\begin{align}
		\operatorname{K}(\mathcal{P}\mathrel{\Vert}\mathcal{Q})=-\int_{0}^{\varepsilon^{2}\ttf}\mathrm{d}\st{2}\int_{\mathbb{R}}\mathrm{d}\nq\,
		\partial_{\st{2}} \left(\nf_{0,\st{2}}^{(0:0)} \,v_{\ttf,\st{2}}^{(0:0)}\right)+O(\varepsilon^{2})\,.
		\nonumber
	\end{align}
	The above representation allows us to express the divergence in terms of the cell problem density (\ref{eq:rho}) and the identity
	\begin{align}
	&	v_{\ttf,\st{2}}^{(0:0)}=\frac{\sigma_{\st{2}}+(\alpha-A) \ln \rho_{\st{2}} }{2\,A}-\frac{\nU_{\star}}{2} 
	-\frac{B\,\nq\,\dot{\mu}_{\st{2}}^{({1})}}{2\,A\,(A-B)}
		-\frac{B}{4\,A\,(A-B)}\int_{0}^{\st{2}}\mathrm{d}\ns\, (\dot{\mu}_{\ns}^{({1})})^{2}
		\nonumber
	\end{align}
	stemming from (\ref{eq:zeta}) and (\ref{eq:sigma}) in section~\ref{sec:maobo}. 
	Indeed,  straightforward algebra yields
	\begin{equation}
\begin{split}
	\operatorname{K}(\mathcal{P}\mathrel{\Vert}\mathcal{Q})
	&=\int_{0}^{\varepsilon^{2}\ttf}\mathrm{d}\ns\int_{\mathbb{R}}\mathrm{d}\nq\,\rho_{\ns}\,\dfrac{\big{(}\partial_{\nq} (\sigma_{t_2}-\alpha\nU_{\star})\big{)}^2	}{4\,A}\\
 &\qquad+\frac{A-\alpha}{2\,A}\int_{\mathbb{R}}\mathrm{d}\nq \,\left(\rho_{\varepsilon^{2}\ttf}\,\ln\frac{\rho_{\varepsilon^{2}\ttf}}{\rho_{\star}}-\rho_{0}\,\ln\frac{\rho_{0}}{\rho_{\star}}\right)	
\\
&\qquad+\frac{B}{4\,A\,(A-B)}\int_{0}^{\varepsilon^{2}\ttf}\mathrm{d}\ns\, (\dot{\mu}_{\ns}^{({1})})^{2}+O(\varepsilon^{2})
\end{split}
		\label{eq:KLopt}
	\end{equation}
	where
	\begin{align}
&		\ln\rho_{\star}=-\nU_{\star}-\ln \int_{\mathbb{R}}\mathrm{d}\nq\,e^{-\nU_{\star}}
		\nonumber
	\end{align}
	and
		\begin{align}
				A-B=(1+g)\left(1-\frac{2 \tanh\frac{\ttf}{2}}{\ttf}\right)
		\nonumber
	\end{align}
	which is positive definite when $\ttf\,>\,2$.
	
In  (\ref{eq:KLopt}), all terms but the first vanish in the limit of infinite scale separation $\ttf$ tending to infinity. 
Further elementary considerations shed more light on the sign of the corrections. Recalling (\ref{optimal_drift:cp}) and the properties of the current velocity, we obtain the identity
\begin{equation}
	\begin{split}
\int_{\mathbb{R}}\mathrm{d}\nq\,\rho_{\st{2}} \big{(}\partial_{\nq} (\sigma_{t_2}-\alpha\nU_{\star})\big{)}^2
	&=\int_{\mathbb{R}}\mathrm{d}\nq\,\rho_{\st{2}}\,(\partial_{\nq}\mathscr{c}_{t_2})^{2}+\alpha^{2}\int_{\mathbb{R}}\mathrm{d}\nq\,\rho_{\st{2}}\left(\partial_{\nq} \ln \frac{\rho_{\st{2}}}{\rho_{\star}}\right)^{2}\\&+2\,\alpha \,\partial_{\nt}\int_{\mathbb{R}}\mathrm{d}q\,\rho_{\st{2}}\ln \frac{\rho_{\st{2}}}{\rho_{\star}}\,.
	\end{split}
	\label{mc:dec}
\end{equation} 
We then re-write (\ref{eq:KLopt}) as
\begin{align}
				\begin{split}
				\operatorname{K}(\mathcal{P}\mathrel{\Vert}\mathcal{Q})&=
				\frac{1}{4\, \alpha}\int_{0}^{\varepsilon^{2}\ttf}\mathrm{d}\st{2}\,\int_{\mathbb{R}}\mathrm{d}\nq\,\rho_{\st{2}} \big{(}\partial_{\nq} (\sigma_{\st{2}}-\alpha\nU_{\star})\big{)}^2\\
				&-\frac{B}{4\,A\,(A-B)}\int_{0}^{\varepsilon^{2}\tf}\mathrm{d}\st{2}\,\left(\int_{\mathbb{R}}\mathrm{d}\nq\,\rho_{\st{2}}(\partial_{\nq}\mathscr{c}_{\st{2}})^{2}-(\dot{\mu}_{\st{2}}^{({1})})^{2}\right)
				\nonumber\\[3pt]
				&+\frac{\alpha-(A-B)}{4\,\alpha\,(A-B)}	\int_{0}^{\varepsilon^{2}\tf}\mathrm{d}\st{2}\,\int_{\mathbb{R}}\mathrm{d}\nq\,\rho_{\st{2}}(\partial_{\nq}\mathscr{c}_{\st{2}})^{2}\\
				&+\alpha\,\frac{\alpha-A}{4\,A}\int_{0}^{\varepsilon^{2}\tf}\mathrm{d}\st{2}\,\int_{\mathbb{R}}\mathrm{d}\nq\,\rho_{\st{2}}\left(\partial_{\nq} \ln \frac{\rho_{\st{2}}}{\rho_{\star}}\right)^{2}+O(\varepsilon^{2})\,.
			\end{split}		
	\nonumber
\end{align}
The identity
\begin{align}
	\dot{\mu}^{({1})}=\int_{\mathbb{R}}\mathrm{d}\nq\,\rho_{\st{2}}(\partial_{\nq}\mathscr{c}_{\st{2}})
	\nonumber
\end{align}
and the Cauchy-Schwarz inequality then ensure that all corrections are positive for $\ttf\,>\,2$. Thus for any $\ttf$ sufficiently large to ensure a separation of time scales, we arrive at the inequality
\begin{align}
	\operatorname{K}(\mathcal{P}\mathrel{\Vert}\mathcal{Q})\,\geq\,\frac{1}{4\,\alpha}\int_{0}^{\varepsilon^{2}\ttf}\mathrm{d}\st{2}\,\int_{\mathbb{R}}\mathrm{d}\nq\,\rho_{\st{2}}\big{(}\partial_{\nq} (\sigma_{\st{2}}-\alpha\nU_{\star})\big{)}^2
	\nonumber
\end{align}
whence we read the multiscale perturbation theory prediction of the Talagrand-Otto-Villani constant $\mathcal{C}_{\text{TOV}}$ in (\ref{FG:TOV}).
To do so, we focus on entropic transport and set $\nU_{\star} $ to zero. Next, we recall that  any solution of  the
cell problem (\ref{eq:cellFP})-(\ref{eq:cellB})  enjoys the lower bound \cite{LehJ2013}
\begin{align}
	\int_{0}^{\varepsilon^{2}\ttf}\mathrm{d}\st{2}\int_{\mathbb{R}}\mathrm{d}\nq\,\rho_{\st{2}}\big{(}\partial_{\nq} \sigma_{\st{2}}\big{)}^2\,\geq\,\frac{\operatorname{E}_{\widetilde{\mathcal{P}}}\left|\tilde{\mathscr{q}}_{\varepsilon^{2}\ttf}-\tilde{\mathscr{q}}_{0}\right|^{2}}{\varepsilon^{2}\ttf}
	\label{mc:bound}
\end{align}
where $ \widetilde{\mathcal{P}}$ is the measure generated by the overdamped Schr\"odinger bridge in $[0,\varepsilon^{2}\ttf]$ 
associated to the stochastic differential equation 
\begin{align}
	\mathrm{d}\tilde{\mathscr{q}}_{\st{2}}=-(\partial \sigma_{\st{2}})(\tilde{\mathscr{q}}_{\st{2}})\,\mathrm{d}\st{2}+\sqrt{2\,\alpha}\,\mathrm{d}\mathscr{w}_{\st{2}}\,.
	\label{mc:sde}
\end{align}
The inequality (\ref{mc:bound}) is a consequence of the law of iterated expectation (see e.g. \cite{PavG2014} pag.~310). Indeed, it ensures that
\begin{align}
\int_{0}^{\varepsilon^{2}\ttf}\mathrm{d}\st{2}\int_{\mathbb{R}}\mathrm{d}\nq\,\rho_{\st{2}}\big{(}\partial_{\nq} \sigma_{\st{2}}\big{)}^2
&	\,\equiv\,\int_{0}^{\varepsilon^{2}\ttf}\mathrm{d}\st{2}\operatorname{E}_{\widetilde{\mathcal{P}}}\big{(}(\partial \sigma_{\st{2}})(\tilde{\mathscr{q}}_{\st{2}})\big{)}^{2}
	\nonumber\\
&	=\int_{0}^{\varepsilon^{2}\ttf}\mathrm{d}\st{2}\operatorname{E}_{\widetilde{\mathcal{P}}}\left(\operatorname{E}_{\widetilde{\mathcal{P}}}\left(\big{(}(\partial \sigma_{\st{2}})(\tilde{\mathscr{q}}_{\st{2}})\big{)}^{2}\,\big{|}\,\tilde{\mathscr{q}}_{0}\right)\right)
\nonumber\\
&\,\geq\,\int_{0}^{\varepsilon^{2}\ttf}\mathrm{d}\st{2}\operatorname{E}_{\widetilde{\mathcal{P}}}\left(\Big{(}\operatorname{E}_{\widetilde{\mathcal{P}}}\left((\partial \sigma_{\st{2}})(\tilde{\mathscr{q}}_{\st{2}})\,\big{|}\,\tilde{\mathscr{q}}_{0}\right)\Big{)}^{2}\right)\,.
	\nonumber
\end{align}
We now invert the order of integration and apply the Benamou-Brenier argument \cite{BeBr2000} to the stochastic paths generated by (\ref{mc:sde})
and find
\begin{align}
&\int_{0}^{\varepsilon^{2}\ttf}\mathrm{d}\st{2}\int_{\mathbb{R}}\mathrm{d}\nq\,\rho_{\st{2}}\big{(}\partial_{\nq} \sigma_{\st{2}}\big{)}^2\,\geq\,\frac{	\operatorname{E}_{\widetilde{\mathcal{P}}}\left(\Big{(}\operatorname{E}_{\widetilde{\mathcal{P}}}\left( \tilde{\mathscr{q}}_{\varepsilon^{2}\ttf}-\tilde{\mathscr{q}}_{0} -\sqrt{2\,\alpha}\,\tilde{\mathscr{w}}_{\varepsilon^{2}\ttf} \big{|}\tilde{\mathscr{q}}_{0}\right)\Big{)}^{2}\right)}{\varepsilon^{2}\,\ttf}\,.
	\nonumber
\end{align}
The inequality now follows because $ \tilde{\mathscr{w}}$ is the Wiener process with respect to the measure $\widetilde{\mathcal{P}} $
and as such has zero conditional expectation with respect to $\tilde{\mathscr{q}}_{0} $. In appendix~\ref{app:PI} we present a path integral derivation of the same result.

The upshot is that for entropic transport we get a Talagrand-Otto-Villani type inequality
\begin{align}
	\operatorname{K}(\mathcal{P}\mathrel{\Vert}\mathcal{Q})\,\geq\,\frac{1}{4\,\alpha}\frac{\operatorname{E}_{\widetilde{\mathcal{P}}}\left|\tilde{\mathscr{q}}_{\varepsilon^{2}\ttf}-\tilde{\mathscr{q}}_{0}\right|^{2}}{\varepsilon^{2}\ttf}\,.
	\nonumber
\end{align}
In dimensional units, the same result reads
\begin{align}
		\operatorname{K}(\mathcal{P}\mathrel{\Vert}\mathcal{Q})\,\geq\,\frac{1}{4\,\alpha}\frac{\beta\,m\operatorname{E}_{\widetilde{\mathcal{P}}}\left|\mathscr{q}_{\tf}-\mathscr{q}_{0}\right|^{2}}{\tau\,\tf}\,.
	\nonumber
\end{align}

\item\hyperref[itm:b]{EP}: Upon contrasting (\ref{model:ep}) with (\ref{oc:BP}),  the exact expression of the minimum mean entropy production reads
	\begin{align}
		\mathcal{E}=\int_{\mathbb{R}}\mathrm{d}\nq\,
		\left(\nf_{0}^{(0)} \big{(}v_{0}^{(0)}+\ln \nf_{0}^{(0)}\big{)}-\nf_{\tf}^{(0)} \big{(}v_{\tf}^{(0)}+\ln \nf_{\tf}^{(0)}\big{)}\right)\,.
		\nonumber
	\end{align}
	The multiscale approximation then is
		\begin{align}
		\mathcal{E}=-\int_{0}^{\varepsilon^{2}\ttf}\hspace{-0.4cm}\mathrm{d}\st{2}\int_{\mathbb{R}}\mathrm{d}\nq\,
		\partial_{\st{2}} \left(\nf_{0,\st{2}}^{(0:0)} (v_{\ttf,\st{2}}^{(0:0)}+\ln \nf_{0,\st{2}}^{(0:0)})\right)+O(\varepsilon^{2})
		\nonumber
	\end{align}
	where, by (\ref{eq:zeta}) and (\ref{eq:sigma}), the identity
	\begin{align}
	&	v_{\ttf,\st{2}}^{(0:0)}+\ln \nf_{0,\st{2}}^{(0:0)}=\frac{2\,\sigma_{\st{2}}}{A}+\frac{2\,B\,\nq\,\dot{\mu}^{({1})}}{A(A-B)}
		+\frac{B\,(\dot{\mu}^{({1})})^{2}\,\st{2}}{A\,(A-B)}
		\nonumber
	\end{align}
	with
	\begin{align}
		\dot{\mu}^{({1})}=\frac{\mu_{\varepsilon^{2}\ttf}^{({1})}-\mu_{0}^{({1})}}{\varepsilon^{2}\,\ttf}
		\nonumber
	\end{align}
	holds true. After some algebra, we arrive at 
	\begin{equation}
		\begin{split}
	\mathcal{E}&=\dfrac{1}{1+g}\int_{0}^{\varepsilon^{2}\ttf}\mathrm{d}\st{2}\,\int_{\mathbb{R}}\mathrm{d}\nq\,\rho_{\st{2}} (\partial_{\nq}\sigma_{\st{2}})^{2}	+\dfrac{1+g-A}{(1+g)\,A}\int_{0}^{\varepsilon^{2}\tf}\hspace{-0.4cm}\mathrm{d}\st{2}\,\left(\int_{\mathbb{R}}\mathrm{d}\nq\,\rho_{\st{2}} (\partial_{\nq}\sigma_{\st{2}})^{2}-(\dot{\mu}^{({1})})^{2}\right)
		\\[6pt]
		&+	\dfrac{1+g-(A-B)}{4\,(1+g)\,(A-B)}\,(\dot{\mu}^{({1})})^{2}\,\varepsilon^{2}\,\ttf+O(\varepsilon^{2})
		\end{split}
		\label{eq:epopt}
	\end{equation}
	with
	\begin{align}
		A-B=(1+g)\left(1-\frac{2 \tanh\frac{\omega\,\ttf}{2}}{\omega\,\ttf}\right)\,.
		\nonumber
	\end{align}
	All addends in (\ref{eq:epopt}) are positive. Furthermore, the last two vanish both in the limit of infinite scale separation and upon recalling the definition (\ref{eq:omega}) of $\omega$ when the coupling constant $g$ is vanishing
	\begin{align}
		\lim_{g\searrow 0}\mathcal{E}=\int_{0}^{\varepsilon^{2}\ttf}\mathrm{d}\st{2}\,\int_{\mathbb{R}}\mathrm{d}\nq\,\rho_{\st{2}} \,(\partial_{\nq}\sigma_{\st{2}})^{2}\,.
		\nonumber
	\end{align}
	We also emphasize for case~\hyperref[itm:b]{EP}, the field $\sigma$ satisfies the compressible Euler equation. As a consequence, we can directly apply to (\ref{eq:epopt}) the Benamou-Brenier inequality \cite{BeBr2000} and straightforwardly recover the bound (\ref{FG:ep}).
	\end{description}	 

\subsubsection{Accuracy of the multiscale approximation}
\label{sec:BBGKY}

Infinite hierarchies of equations such as (\ref{eq:eqhermite}) appear in the study of Liouville's and Boltzmann equations  \cite{WyBa1987,BobA2006}. Many numerical methods resort to a phenomenological truncation of the hierarchy. The multiscale method provides a controlled truncation at the level of second order equations.
In fact, all cumulants up to second order can be reconstructed from an effective first order system embodied by the cell problem. 

In Fig.~\ref{fig:scheme}, we summarize how the secular term cancellation (or, equivalently, solvability) conditions allow us to re-order contributions of the regular perturbative expansion within the hierachy. The upshot is that the predictions for cumulants and total cost obtained from the solution of the cell problem have different accuracies in $\varepsilon$. 

\subsection{Order-by-order solution}
\label{sec:maobo}
 In this Section, we solve the hierarchy of equations~\eqref{eq:eqhermite} in a multiscale perturbative series in powers of $\varepsilon$. 
To this goal, we insert Eqs.~\eqref{eq:expeps1}~-~\eqref{eq:expeps3} into Eq.~\eqref{eq:eqhermite}, and identify equations of distinct order in the power expansion, taking into account the time differentiation, which acts on the multiscale dependence of the probability density and value function according to (\ref{ms:td}). The derivation of the results is briefly outlined in words below.

 At order zero in $\varepsilon$, the equations for the density and value function give rise to two decoupled infinite systems of first order differential equations in the fast time $\st{0}$. These systems are trivially integrable with respect to the fast time $\st{0}$, implicitly keeping all information about the boundary condition in the unresolved dependence of the integration constants upon the slow times. 
 
 Remarkably, at order $\varepsilon^{1}$ the boundary (\ref{eq:bchermite}) and stationary conditions reduce the non-trivial contribution of the two infinite hierarchies of equations to a system of two first order differential equations in the fast time for $ \nf^{(1:1)}$ and $v^{(1:1)}$. Dependence upon higher order coefficients of the expansion in Hermite polynomials enters these equations in the form of functions of the slow time $\st{2}$ that must be determined at order $\varepsilon^{2}$ in the regular perturbative expansion. As no secular term appears at this order we can assume within accuracy independence of the solution of the extremal equations from $\st{1}$. 
 
 At order $\varepsilon^{2}$, we can determine all unknown quantities inherited from lower orders in the regular perturbative expansion by imposing the cancellation of secular terms. This fixes the dynamical dependence upon the slow time $\st{2}$ in the form of a cell problem. We enforce the correct boundary conditions in terms of  $ \nf^{(0:2)}$,  $ \nf^{(2:2)}$ and $v^{(0:2)}$, $v^{(2:2)}$. Finally, if we set all the $ \nf^{(n:0)}$,  $\nf^{(n:1)}$  that are not sustained by the drift and all the $ v^{(n:0)}$,  $ v^{(n:1)}$ that are not needed to control the non-vanishing contributions to the density to zero,  it is self-consistent to set 
 \begin{align}
 	\nf^{(1:2)}= v^{(1:2)}=0\,.
 	\nonumber
 \end{align}
 Fig.~\ref{fig:scheme} is a stylized summary of the procedure. Additional details are provided in Appendix~\ref{app:oboe}.

In principle, it is possible to extend the analysis to orders higher than $\varepsilon^{2}$, as done in \cite{WyBa1987}. The appearance of spatial derivatives of higher order than the second may, however, call for the introduction of appropriate variables to perform partial resummations \cite{BobA2006}. We return to this point in section~\ref{sec:outlook}

\begin{figure*}
    \centering
    \includegraphics[width=\linewidth]{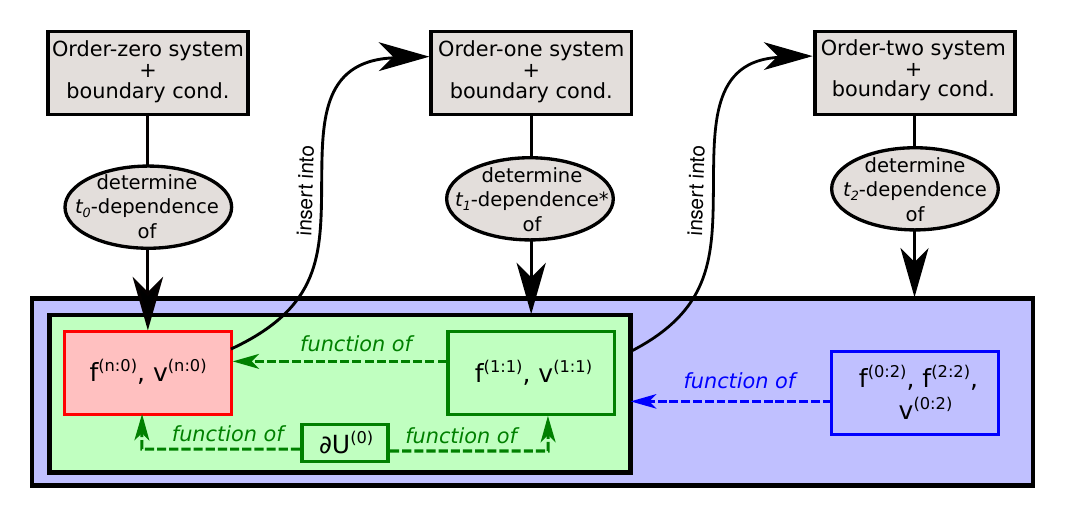}
    \caption{Scheme of the multiscale approach presented in the text. The logical order of the calculation is represented by the black solid arrows, going through the sequence of solutions of the differential systems at the different orders. The quantities computed with this strategy are reported in the coloured boxes, where different colours correspond to different steps of the order-by-order multiscale calculation. Dashed arrows show the functional dependencies of the computed quantities. \newline
    $^\star$~The calculation actually shows that there is no dependence on the $\st{1}$-time scale. }
    \label{fig:scheme}
\end{figure*}

\subsubsection{Boundary conditions}

The boundary conditions (\ref{model:ini}), (\ref{model:fin}) are by hypothesis independent of the Stokes time and therefore remain the same once expressed in non-dimensional units. 
%
%
Consequently, all $\nf_{\slt{0}}^{(n:i)}$'s with $n\,\geq\, 1$ vanish at the boundaries, so that
\begin{subequations}
\label{eq:bcf}
    \begin{eqnarray}
        \nf_{0,\slt{1}}^{(n:i)}(\nq)&=&\nf_{0}^{(0)}(\nq)\,\delta_{n,0}\,\delta_{i,0}\,\\[6pt]
        \nf_{\tf,\slt{1}}^{(n:i)}(\nq)&=&\nf_{\tf}^{(0)}(\nq)\,\delta_{n,0}\,\delta_{i,0}\,,
    \end{eqnarray}
\end{subequations}
where $\delta_{i,j}$ is a Kronecker delta. Without loss of generality, we set $\ti=0$.

The non-perturbative boundary behavior is not assigned a priori but is determined by that of the probability density. However, in multiscale perturbation theory, we have the freedom to choose how partial resummations to cancel secular terms is performed \cite{VerF2005}. We have reasoned that contributions to the cost can only come from the same time scale as those where the control varies, which gives the following resonance subtraction condition
\begin{equation}
\label{eq:bcv}
    v_{\tf, \slt{1}}^{(0:i)}(\nq)-v_{0, \slt{1}}^{(0:i)}(\nq)=\cbr{v_{\tf}^{(0)}(\nq)-v_{0}^{(0)}(\nq)}\delta_{i,0}\,.
\end{equation}

\subsubsection{Solution of the problem at order zero}
\label{sec:sys0}

The calculation starts at order zero of the $\varepsilon$-expansion.  
Equation~\eqref{eq:eq1hermite} can be written at order zero in $\varepsilon$ as:
$$
\cbr{\partial_{\st{0}}+n}\,\nf_{\st{0},\slt{1}}^{(n:0)}=0\,,
$$
which implies 
\begin{equation}    \nf_{\slt{0}}^{(n:0)}=c_{\slt{1}}e^{-n \st{0}}\,.
\end{equation}
Here $c_{\slt{1}}$ is fixed by imposing the initial condition at time $\st{0}=0$:
$$
\nf_{\slt{0}}^{(n:0)}\Big|_{\st{0}=0}=\delta_{n,0}\,\nf_{0,\slt{1}}^{(0:0)}\,,
$$
following from Eq.~\eqref{eq:bcf}. This observation leads to 
\begin{equation}
    \nf_{\slt{0}}^{(n:0)}=\delta_{n,0}\,\nf_{0,\slt{1}}^{(0:0)} \label{eq:sol_o0_f}\,.
\end{equation}

Solving the value function equation~\eqref{eq:eq2hermite} at order zero in $\varepsilon$ gives
\begin{equation}    
v_{\slt{0}}^{(n:0)}-\,v_{\ttf,\slt{1}}^{(n:0)}e^{\st{0}-\ttf}\\
=\begin{cases}
    0\quad&\syma\\[0.3cm]
    \delta_{n,2}\,(1-e^{2(\st{0}-\ttf)})/2\,.\quad&\symb
\end{cases}
\end{equation}
This time we have no boundary conditions to impose. However, it follows from Eq.~\eqref{eq:eq3hermite} that
\begin{equation}
v_{\slt{0}}^{(1:0)}=0\,,
\end{equation}
hence
\begin{equation}
\begin{aligned}
v_{\slt{0}}^{(n:0)}&-(1-\delta_{n,1})\,v_{\ttf,\slt{1}}^{(n:0)}\,e^{\st{0}-\ttf}=\begin{cases}
    0\quad&\syma\\[0.3cm]
    \delta_{n,2}(1-e^{2(\st{0}-\ttf)})/2\,.\quad&\symb
\end{cases}\label{eq:sol_o0_v}
\end{aligned}
\end{equation}
The $\st{0}$-dependence of the probability density and the value function is completely determined at order zero. We have no way to enforce the boundary condition at $t=\ttf$ for $\nf_{\slt{0}}^{(n:0)}$ at this stage: we will need to impose it on a slower time scale, in this way exploiting the additional freedom provided by the multiscale approach.

\subsubsection{Solution of the problem at order one}
\label{sec:sys1}

By expanding Eq.~\eqref{eq:eq1hermite} at order one in $\varepsilon$, one gets
$$
\begin{aligned}
\label{eq:exp_herm_f1}
    \partial_{\st{0}} \nf_{\slt{0}} ^{(n:1)}&+\partial_{\st{1}}\nf_{\slt{0}}^{(n:0)}
    +n \,\nf_{\slt{0}}^{(n:1)} + \cbr{n+1}\partial_{\nq} \nf_{\slt{0}}^{(n+1:0)}+\cbr{\partial_{\nq} +\cbr{\partial_{\nq} \nU_{\slt{0}}^{(0)}}}\nf_{\slt{0}}^{(n-1:0)}=0\,.
\end{aligned}
$$
The boundary conditions for the probability density force all terms of order higher than zero in $\varepsilon$ vanish at $t=0$ and $t=\ttf$. This is a consequence of our assumption that the protocol starts and ends in equilibrium states, which cannot depend on the relaxation time scale $\varepsilon$. They must coincide with the stationary states of the overdamped limit $\varepsilon \to 0$.
The $n=0$ case of Eq.~\eqref{eq:exp_herm_f1}, by recalling Eq.~\eqref{eq:sol_o0_f}, implies therefore that $\nf_{\slt{0}} ^{(0:0)}$ is independent of $\st{1}$, hence
\begin{equation}
\nf^{(0:0)}_{\slt{0}}=\nf_{0,\slt{1}}^{(0:0)}=\nf_{0\vv\slt{2}}^{(0:0)}\,,
\end{equation}
where we have introduced the notation
\begin{equation}
\label{eq:notationvv}
\nf_{\st{0}\vv\slt{2}}\equiv \nf_{\st{0},0,\slt{2}}\,.
\end{equation}
Similarly, the equations with $n \ge 2$ lead to
\begin{equation}
 \nf^{(n:1)}_{\slt{0}}=0\,,\quad \quad n\ge2\,.   
\end{equation}
The case $n=1$ is less trivial and brings about the relation
\begin{equation}
\label{eq:fp_appf}
    \partial_{\st{0}} \nf_{\slt{0}}^{(1:1)}+\nf_{\slt{0}}^{(1:1)}+\partial_{\nq} \nf_{0\vv\slt{2}}^{(0:0)}+\nf_{0\vv\slt{2}}^{(0:0)}\partial_{\nq} \nU^{(0)}_{\slt{0}}=0\,.
\end{equation}
Similarly, the value function equation~\eqref{eq:eq2hermite} at order one, for the case $n=1$, gives
\begin{equation}
\label{eq:vf_appf}
\begin{split}
&	    \cbr{\partial_{\st{0}}-1}v^{(1:1)}_{\slt{0}}=-\partial_{\nq} v^{(0:0)}_{\slt{0}}-2\cbr{\partial_{\nq} +\bcbr{\partial_{\nq} \nU_{\slt{0}}^{(0)}}}v^{(2:0)}_{\slt{0}}\,.
\end{split}
\end{equation}
Once complemented with a condition for the drift $\partial_{\nq} U^{(0)}_{\slt{0}}$, Eqs.~\eqref{eq:fp_appf} and~\eqref{eq:vf_appf} form a closed system of differential equations. The missing relation can be obtained from the stationarity condition~\eqref{eq:eq3hermite}, which at order one in $\varepsilon$ reads
\begin{equation}
\label{eq:u_appf}
\begin{split}
  g\, \nf_{0\vv\slt{2}}^{(0:0)} \partial_{\nq} v^{(0:0)}_{\slt{0}}+\nf_{0\vv\slt{2}}^{(0:0)} v^{(1:1)}_{\slt{0}}+2 \,\nf_{\slt{0}}^{(1:1)} v^{(2:0)}_{\slt{0}}= \begin{cases}
    \dfrac{1+g}{2}\,\nf_{\st{0}\,\vv\slt{2}}^{(0:0)} \,\partial_{\nq}  \nU^{(0)}_{\slt{0}}\quad &\syma \\[0.3cm]
    2\,g \,\nf_{\st{0}\vv\slt{2}}^{(0:0)}\, \partial_{\nq}  \nU^{(0)}_{\slt{0}}+g\,\partial_{\nq} \nf^{(0:0)}_{\st{0},\slt{2} }\,.\quad &\symb 
\end{cases}  
\end{split}
\end{equation}
Eq.~\eqref{eq:u_appf} provides an expression for the drift, which can be inserted into Eq.~\eqref{eq:fp_appf} to obtain a relation for $v^{(1:1)}_{\slt{0}}$ (see Eq.~\eqref{eq:v11} in Appendix~\ref{app:oboe}).
The system is then solved by differentiating the resulting equation with respect to $\st{0}$, and eliminating $\partial_{\st{0}}v^{(1:1)}_{\slt{0}}$ through~\eqref{eq:vf_appf} and $v^{(1:1)}_{\slt{0}}$ through~\eqref{eq:v11}. 
A second-order ODE for $\nf_{\slt{0}}^{(1:1)}$ is found:
\begin{equation}
\label{eq:diff2order}
\partial_{\st{0}}^2 \nf_{\slt{0}}^{(1:1)}-\omega^2 \, \nf_{\slt{0}}^{(1:1)} = F_{\slt{0}}\,,  
\end{equation}
with $\omega$ as defined in (\ref{eq:omega}). The dependence of $F_{\slt{0}}(\nq)$ on $\st{0}$ is known; for its explicit expression, see Eq.~\eqref{eq:rhsF}. The equation can be solved by recalling that the Green function for the second order differential equation~\eqref{eq:diff2order} is
\begin{equation}
\label{eq:green}
    G_{t,s}=J_{t,s}+J_{s,t}
\end{equation}
with
$$
J_{t,s}=-\theta\,(t-s)\,\dfrac{\sinh\cbr{\omega(\ttf-t)}\,\sinh\cbr{\omega s}}{\omega \sinh\cbr{\omega \ttf}}\,,
$$
with $\theta(\cdot)$ being the Heaviside step-function. By introducing the notation
\begin{equation}
\label{eq:gk}
G^{(k)}_t=\int_{0}^{\ttf}\mathrm{d}s \, G_{t,s}\, e^{-k(\ttf-s)}\,,
\end{equation}
one obtains for $\nf_{\slt{0}}^{(1:1)}$ the relation
\begin{equation} 
\label{eq:sys1:f11}
\begin{aligned}
     \nf_{\st{0}\vv\slt{2}}^{(1:1)} &= \omega^2 \,G^{(0)}_{\st{0}}\,\nf_{0\vv\slt{2}}^{(0:0)} \,\partial_{\nq} \zeta_{\slt{2}}+\begin{cases}
         \dfrac{4}{1+g}\,\partial_{\nq} \cbr{v^{(2:0)}_{\ttf\vv\slt{2}} \,\nf_{0\vv\slt{2}}^{(0:0)}}\, G^{(2)}_{\st{0}} \quad&\syma\\[0.3cm]
          \dfrac{1}{g}\,\partial_{\nq} \cbr{v^{(2:0)}_{\ttf\vv\slt{2}} \,\nf_{0\vv\slt{2}}^{(0:0)}-\dfrac{\nf_{0\vv\slt{2}}^{(0:0)}}{2}} G^{(2)}_{\st{0}} \quad&\symb
     \end{cases}   
\end{aligned}
\end{equation}

where

\begin{equation}
\label{eq:zeta}
    \zeta_{\st{2}}(\nq)=\begin{cases}
        2\, v_{\ttf\vv\slt{2}}^{(0:0)}(\nq)+\nU_{\star}(\nq)+\ln \nf_{0\vv\slt{2}}^{(0:0)}(\nq) \quad&\syma\\[0.3cm]
        \dfrac{ 1}{2}\cbr{v_{\ttf\vv\slt{2}}^{(0:0)}(\nq)+\ln \nf_{0\vv\slt{2}}^{(0:0)}}(\nq)\,,\quad&\symb
    \end{cases}
\end{equation}
and a notation analogous to~\eqref{eq:notationvv} is adopted also for the value function.
The last step of the solution of the order $\varepsilon^{1}$ consists in writing the optimal control potential as a function of (\ref{eq:sys1:f11}). From Eq.~\eqref{eq:fp_appf} we find
\begin{equation}
\label{eq:fp_appf2}
	\partial_{\nq}  \nU_{\slt{0}}^{(0)}(\nq)=-\frac{\partial_{\nq}\nf_{0\vv\slt{2}}^{(0,0)}(\nq)+\partial_{\st{0}}\nf_{\st{0}\vv\slt{2}}^{(1:1)}(\nq)+\nf_{\st{0}\vv\slt{2}}^{(1:1)}(\nq)}{\nf_{0\vv\slt{2}}^{(0,0)}(\nq)}\,.
\end{equation}

Since there are no equations for $\partial_{\st{1}}\nf_{\slt{0}}^{(1:1)}$ nor $\partial_{\st{1}}v^{(1:1)}_{\slt{0}}$, i.e. no secular terms are found on the time scale $\st{1}$, we can assume the solution to be independent of $\st{1}$. Once $\nf_{\slt{0}}^{(1:1)}$ is known, an explicit expression for $v^{(1:1)}_{\slt{0}}$ can also be found from~\eqref{eq:vf_appf} (see Eq.~\eqref{eq:v11} in Appendix~\ref{app:oboe}).
From the above equation it is possible to derive the expression~\eqref{eq:optdrift} for the optimal drift, by using the expressions of $\nf_{\st{0}\vv\slt{2}}^{(0:0)}$ and $\nf_{\st{0}\vv\slt{2}}^{(1:1)}$ that will be found in the next subsection.

\subsubsection{Solution of the problem at order two}
\label{sec:sys2}
The order-two expansion of~\eqref{eq:eq1hermite} provides the following relations
\begin{subequations}
\label{eq:exp_herm_f2}
    \begin{eqnarray}  
\partial_{\st{0}}\nf_{\slt{0}}^{(0:2)}+\partial_{\st{2}}\nf_{0\vv\slt{2}}^{(0:0)}&=&-\partial_{\nq} \nf_{\st{0}\vv\slt{2}}^{(1:1)}
     +g\,\partial_{\nq}^{2} \nf_{0\vv\slt{2}}^{(0:0)} + g \,\partial_{\nq}\cbr{\nf_{\st{0}\vv\slt{2}}^{(0:0)} \,\partial_{\nq} \nU^{(0)}_{\st{0}\vv\slt{2}}}  \quad 
\label{eq:exp_herm_f2_a}\\[6pt]   
\partial_{\st{0}}\nf_{\slt{0}}^{(1:2)}+ \nf_{\slt{0}}^{(1:2)} &=& - \partial_{\nq} \nf_{\slt{0}\vv\slt{2}}^{(0:1)}-\nf_{\st{0}\vv\slt{2}}^{(0:0)}\,\partial_{\nq} \nU_{\slt{0}}^{(1)}  \quad
\label{eq:exp_herm_f2_b}\\[6pt]  
\partial_{\st{0}}\nf_{\slt{0}}^{(2:2)}+2\, \nf_{\slt{0}}^{(2:2)} &=& - \partial_{\nq} \nf_{\st{0}\vv\slt{2}}^{(1:1)}-\nf_{\st{0}\vv\slt{2}}^{(1:1)}\,\partial_{\nq} \nU_{\st{0}\vv\slt{2}}^{(0)}   \quad
\label{eq:exp_herm_f2_c}\\[6pt]   
\partial_{\st{0}}\nf_{\slt{0}}^{(n:2)}+n \,\nf_{\slt{0}}^{(n:2)}&=&0,\qquad n>2\,. \quad
    \label{eq:exp_herm_f2_d}\end{eqnarray}
\end{subequations}

The last equation~\eqref{eq:exp_herm_f2_d} ensures that all terms $\nf_{\slt{0}}^{(n:2)}$ with $n>2$ vanish once equilibrium boundary conditions are taken into account. Equation~\eqref{eq:exp_herm_f2_b} provides a relation for $\nf_{\slt{0}}^{(1:2)}$ that requires knowledge of $\partial_{\nq} \nU^{(1)}_{\slt{0}}$. If we expand the stationary condition (\ref{eq:eq3hermite}) to second order in $\varepsilon$ and assume that all $v_{\slt{0}}^{(n:0)} $, $v_{\slt{0}}^{(n:1)} $
that are not needed to control the non-vanishing $\nf_{\slt{0}}^{(n:2)}$'s can be set to zero, we get
\begin{align}
	\partial_{\nq} \nU^{(1)}_{\slt{0}}= \begin{cases}\frac{2}{g+1}\left(v_{\slt{0}}^{(1:2)}+\frac{2\, \nf_{\slt{0}}^{(1:2)}\,v_{\slt{0}}^{(2:0)}}{\nf_{\slt{0}}^{(0:0)}}\right)
		\quad &\syma \\[0.5cm]
	\frac{\omega^{2}-1}{2}\left(v_{\slt{0}}^{(1:2)}+\frac{2\, \nf_{\slt{0}}^{(1:2)}\,v_{\slt{0}}^{(2:0)}}{\nf_{\slt{0}}^{(0:0)}}\right) \,.	\quad &\symb
	\end{cases} 
	\nonumber
\end{align}
We insert this result into (\ref{eq:exp_herm_f2_b}) and the corresponding equation for $v_{\slt{0}}^{(1:2)} $, and after straightforward, albeit tedious, algebra we arrive at
 \begin{align}
 	\partial_{\st{0}}\nf_{\slt{0}}^{(1:2)}-\omega^{2}\,\nf_{\slt{0}}^{(1:2)}=0\,.
 	\nonumber
 \end{align}
Taking into account the boundary conditions, we get
\begin{align}
\nf_{0,\slt{1}}^{(1:2)}=\nf_{\ttf,\slt{1}}^{(1:2)}=0
	\nonumber
\end{align}
and we conclude that for any $\st{0}$,
\begin{align}
	\nf_{\st{0},\slt{1}}^{(1:2)}=0\,.
	\nonumber
\end{align}
The same applies to $v_{\st{0},\slt{1}}^{(1:2)}$.

Let us focus first on Eq.~\eqref{eq:exp_herm_f2_c}.
Integrating over $\st{0}$, one has
$$
\nf^{(2:2)}_{\slt{0}}=-\int_{0}^{\st{0}}\mathrm{d}\ns\, e^{-2(\st{0}-\ns)}\cbr{\partial_{\nq}\nf_{\ns\vv\slt{2}}^{(1:1)}+\nf_{\ns\vv\slt{2}}^{(1:1)}\partial_{\nq} \nU_{\ns\vv\slt{2}}^{(0)}}\,.
$$
By substituting the expression of the drift obtained from Eq.~\eqref{eq:fp_appf2} and integrating the term proportional to $\partial_s\nf_{\ns\vv\slt{2}}^{(1:1)}$ by parts, we find
$$
\nf^{(2:2)}_{\st{0}\vv\slt{2}}=\dfrac{\cbr{\nf_{\st{0}\vv\slt{2}}^{(1:1)}}^{2}}{2\,\nf_{0\vv\slt{2}}^{(0:0)}}-\nf_{0\vv\slt{2}}^{(0:0)}\int_{0}^{\st{0}}\mathrm{d}\ns\, e^{-2(\st{0}-\ns)}\partial_{\nq} \cbr{\dfrac{\nf_{\ns\vv\slt{2}}^{(1:1)}}{\nf_{0\vv\slt{2}}^{(0:0)}}}\,,
$$
which is Eq.~\eqref{eq:f22}. This relation implies, recalling the boundary conditions, that
\begin{equation}
\label{eq:f11bc}
    \int_{0}^{\ttf}\mathrm{d}\ns\, e^{-2\,(\ttf-\ns)}\,\partial_{\nq} \cbr{\dfrac{\nf_{\ns\vv\slt{2}}^{(1:1)}}{\nf_{0\vv\slt{2}}^{(0:0)}}}=0\,.
\end{equation}
By substituting Eq.~\eqref{eq:sys1:f11}, an equation for the term $\partial_{\nq}\cbr{v^{(2:0)}_{\ttf\vv\slt{2}} \,\nf_{0\vv\slt{2}}^{(0:0)}}$ can be derived (see Eq.~\eqref{eq:v20f00} in Appendix~\ref{app:oboe}). Once plugged back into Eq.~\eqref{eq:sys1:f11} itself, it yields
\begin{equation}
    \label{eq:f11_appg}    \nf_{\st{0}\vv\slt{2}}^{(1:1)}(\nq)=-\nf_{0\vv\slt{2}}^{(0:0)}\left( a_{\st{0}}\, (\partial\zeta_{\slt{2}})(\nq)-b_{\st{0}}\,\kappa_{\slt{2}} \right)\,.
\end{equation}

Here we introduce the functions whose explicit expression we gave in (\ref{eq:a&b})
\begin{align}
&	a_{\st{0}}=-\omega^2\,\left(G^{(0)}_{\st{0}}-\frac{G^{(0:2)}}{G^{(2:2)}}\,G^{(2)}_{\st{0}}\right)
\nonumber\\
&b_{\st{0}}=\omega^2\,\frac{G^{(0:2)}}{G^{(2:2)}}\,G^{(2)}_{\st{0}}\,.
	\nonumber
\end{align}
By (\ref{eq:gk}) the two functions $a_{\st{0}}$ and $b_{\st{0}}$ are non homogeneous solution of the unstable oscillator equation weighed by constant coefficient also depending upon integrals over the Green function
\begin{equation}
    G^{(k:l)}=\int_{0}^{\ttf}\mathrm{d}s\, e^{-l(\ttf-s)}\,G^{(k)}_s\,.
    \label{eq:Gkl}
\end{equation}
In (\ref{eq:f11_appg}) we also introduce
\begin{equation}
    \kappa_{\slt{2}}=\int_{\mathbb{R}}\mathrm{d}\nq\, \nf_{\st{0}\vv\slt{2}}^{(0:0)}(\nq)\, (\partial\zeta_{\slt{2}})(\nq)\,,
\end{equation}
where we use the function $\zeta_{\slt{2}}$  defined in Eq.~\eqref{eq:zeta}. Equation~\eqref{eq:f11_appg} will be crucial in the following, as it allows to write a closed system of differential equations for $\nf_{0\vv\slt{2}}^{(0:0)}$ and $v^{(0:0)}_{\ttf\vv\slt{2}}$, which can be reshaped as in Eqs.~\eqref{eq:sol_o2}.

Taking into account the boundary conditions and Eq.~\eqref{eq:fp_appf2}, Eq.~\eqref{eq:exp_herm_f2_a} can be integrated over $\st{0}$ to give
\begin{equation}
\label{eq:f00bc}
    \partial_{\st{2}}\nf_{0\vv\slt{2}}^{(0:0)}+\frac{g+1}{\ttf}\int_{0}^{\ttf}\mathrm{d}\ns\, \partial_{\nq}\nf_{\ns\vv\slt{2}}^{(1:1)}=0\,.
\end{equation}
If we now substitute Eq.~\eqref{eq:f11_appg} we get
\begin{equation}
\label{eq:solf-int}
    \partial_{\st{2}}\nf_{0\vv\slt{2}}^{(0:0)}=A\,\partial_{\nq}\cbr{\nf_{0\vv\slt{2}}^{(0:0)}\bcbr{\partial_{\nq}\zeta_{\slt{2}}}}-B\kappa_{\slt{2}}\partial_{\nq}\nf_{0\vv\slt{2}}^{(0:0)}
\end{equation}
where
\begin{subequations}
\label{eq:AB_constants}
\begin{eqnarray}
    A&=&\frac{g+1}{\ttf}\int_{0}^{\ttf}\mathrm{d}\ns \,a_{\ns}
    =-\dfrac{\omega^2\,(1+g)}{\ttf}\cbr{G^{(0:0)}-\dfrac{\cbr{G^{(0:2)}}^2}{G^{(2:2)}}}\label{eq:A_constant}\\[5pt]
    B&=&\frac{g+1}{\ttf}\int_{0}^{\ttf}\mathrm{d}s \,b_{\ns}=\dfrac{\omega^2\,(1+g)}{\ttf}\,\dfrac{\cbr{G^{(0:2)}}^2}{G^{(2:2)}} \,.\label{eq:B_constant}
\end{eqnarray}    
\end{subequations}
The above relations lead to Eq.~\eqref{eq:A&B}.

We now need to find an equation for $\zeta_{\slt{2}}$ in order to close the differential system and find the $\st{2}$-dependence of $\nf_{0\vv\slt{2}}^{(0:0)}$. To this aim, we consider the case $n=0$ for the expansion of Eq.~\eqref{eq:eq2hermite} at order two in $\varepsilon$. It reads
\begin{equation}
\label{eq:v02}
\begin{split}
\partial_{\st{0}}v_{\st{0}\vv\slt{2}}^{(0:2)}+
&\partial_{\st{2}}v_{\ttf\vv\slt{2}}^{(0:0)}+\left(\partial_{\nq} -(\partial_{\nq} \nU_{\st{0}\vv\slt{2}}^{(0)})\right)\left(v_{\st{0}\vv\slt{2}}^{(1:1)}+g\,\partial_{\nq} v_{\ttf\vv\slt{2}}^{(0:0)}\right)
\\&
\qquad=\begin{cases}
    -\dfrac{g+1}{4}\cbr{\partial_{\nq}\bcbr{\nU_{\st{0}\vv\slt{2}}^{(0)}- \nU_{\star}}}^2\quad&\syma\\[0.3cm]
    -g \cbr{\bcbr{\partial_{\nq} \nU^{(0)}_{\st{0}\vv\slt{2}}}^2-\partial_{\nq}^2\nU_{\st{0}\vv\slt{2}}}\,.\quad&\symb
\end{cases}
\end{split}
\end{equation}
We integrate the above equation over $\st{0}$. By substituting~\eqref{eq:fp_appf2} and making repeated use of Eqs.~\eqref{eq:diff2order},~\eqref{eq:f11_appg} and~\eqref{eq:f11bc} (see Appendix~\ref{app:oboe} for details), one finds
\begin{equation}
\label{eq:solv-int}
\begin{aligned}
    \partial_{\nt_2}\zeta_{\slt{2}}&=\frac{A-B}{2}(\partial_{\nq}\zeta_{\slt{2}})^2+\frac{B}{2}\cbr{\partial_{\nq}\zeta_{\slt{2}}-\kappa_{\slt{2}}}^2+\frac{\alpha^2}{A} \cbr{W_{\star}+ \frac{\partial_{\nq}^2\nf_{0\vv\slt{2}}^{(0:0)}}{\bcbr{\nf_{0\vv\slt{2}}^{(0:0)}}^2} -\frac{1}{2}\cbr{\frac{\partial_{\nq}\nf_{0\vv\slt{2}}^{(0:0)}}{\nf_{0\vv\slt{2}}^{(0:0)}}}^2}\,,
    \end{aligned}
\end{equation} which is the closure equation for $\zeta_{\slt{2}}$. where the constant $\alpha$ is defined by Eq.~\eqref{eq:alpha}. 

The differential system for $\nf_{0\vv\slt{2}}^{(0:0)}$ and $\zeta_{\slt{2}}$ can be rewritten in a much more convenient form by introducing the auxiliary field
\begin{equation}
\label{eq:sigma}
\begin{split}
	    \sigma_{\slt{2}}(\nq)&=A \,\zeta_{\slt{2}}(\nq)-\alpha \ln \nf_{0\vv\slt{2}}^{(0:0)}(\nq) -B\,\cbr{\nq\,\kappa_{\slt{2}}+\frac{A-B}{2}\int_{0}^{\st{2}}\mathrm{d}\ns\, \kappa^2_{\ns,\slt{3}}}\,.
\end{split}
\end{equation}
Indeed, taking into account Eq.~\eqref{eq:derk}, it is easy to verify that Eqs.~\eqref{eq:solf-int} and~\eqref{eq:solv-int} are amenable to the form~\eqref{eq:sol_o2}.
Let us stress that, in terms of the field $\sigma$, Eq.~\eqref{eq:solf-int} becomes
\begin{align}
	\kappa_{\slt{2}}=\frac{1}{A-B}\int_{\mathbb{R}}\mathrm{d}\nq\, \nf_{\st{0}\vv\slt{2}}^{(0:0)}(\nq) \,(\partial\sigma_{\slt{2}})(\nq)\,.
	\label{eq:ksigma}
\end{align}
Upon inserting this identity, ~\eqref{eq:sigma}, and~\eqref{eq:mus}, in \eqref{eq:f11_appg}, we recover the expression \eqref{eq:sol_o1_f}. 

Finally, by plugging Eq.~\eqref{eq:f00bc} and Eq.~\eqref{eq:fp_appf2} in Eq.~\eqref{eq:exp_herm_f2_a} one gets:
$$
\partial_{\st{0}} \nf_{\slt{0}}^{(0:2)} -\dfrac{1+g}{\ttf}\int_{0}^{\ttf}\mathrm{d}\ns\, \partial_{\nq} \nf_{\ns\vv\slt{2}}^{(1:1)}=-(1+g)\,\partial_{\nq}\nf_{\slt{0}}^{(1:1)}-g\partial_{\st{0}}\partial_{\nq} \nf_{\slt{0}}^{(1:1)}\,.
$$
Integrating over $\st{0}$ leads to Eq.~\eqref{eq:f02}.

\section{Analytic results for the Gaussian case}
\label{sec:gauss_results}

As discussed in Section~\ref{sec:results} and shown analytically in Section~\ref{sec:maobo}, in order to find the explicit solution of the optimal problem, one first needs to address the differential system~\eqref{eq:sol_o2}. In most cases, the solution can only be 
found numerically: this is discussed in the next Section. However, if the assigned initial and final conditions are Gaussian probability density functions (meaning that the particle is subject to harmonic confinement), the solution can be found analytically. 

To do this, we plug a Gaussian ansatz for the density and a parabolic one for $\sigma_{\st{2}}$, namely
\begin{subequations}
\label{eq:ansatzgauss}
\begin{eqnarray}
    \rho_{\st{2}}&=&\frac{1}{\sqrt{2\, \pi \,\varsigma_{\st{2}}}}\exp\left(-\frac{(\nq-\mu_{\st{2}}^{(1)})^2}{2 \,\varsigma_{\st{2}}}\right)\\
    \sigma_{\st{2}}&=&\sigma^{(0)}_{\st{2}}+\sigma^{(1)}_{\st{2}}\nq+\sigma^{(2)}_{\st{2}}\nq^2
\end{eqnarray}
\end{subequations} 
where $\mu^{(1)}$ and $\mu^{(2)}$ are consistent with  Eq.~\eqref{eq:mus}, into Eqs.~\eqref{eq:sol_o2}. 

Next, we solve for the coefficients, taking into account the boundary conditions. The derivation is straightforward and not carried out here. For both  cases~\hyperref[itm:a]{KL} and~\hyperref[itm:b]{EP}, and $\nU_{\star}=0$, the explicit expressions for the relevant coefficients appearing in Eqs.~\eqref{eq:ansatzgauss} are
\begin{align}
    \mu_{\st{2}}^{(1)}&=\mu^{(1)}_{0}+\frac{\st{2}}{\varepsilon^2 \ttf}\cbr{\mu^{(1)}_{\varepsilon^2 \ttf}-\mu^{(1)}_{0}}
    \nonumber\\
    \varsigma_{\st{2}}&=\frac{(\st{2}-\varepsilon^2 \ttf)^{2}\,\varsigma_{0} +\st{2} \left(2 \, (\varepsilon^2 \,\ttf-\st{2})\,\lambda_{\varepsilon^2 \ttf}+\st{2}\, \varsigma_{\varepsilon^2 \ttf}\right)}{\varepsilon^4 \ttf^2}
    \nonumber\\
    \sigma_{\st{2}}^{(1)}&=\frac{\mu^{(1)}_{0} \left(\varsigma_{\varepsilon^2 \ttf}-\varepsilon^{2}\ttf\,\alpha+\lambda_{\varepsilon^{2}\ttf}\right)-\mu^{(1)}_{\varepsilon^2 \ttf} \left(\varsigma_{0}+\varepsilon^{2}\ttf\,\alpha+\lambda_{\varepsilon^{2}\ttf}\right)}{\varepsilon^4\, \ttf^2\,\varsigma_{\st{2}}}
    \nonumber
\end{align}
where
\begin{align}
	\lambda_{\varepsilon^2 \ttf}=\sqrt{\varepsilon^{4}\,\ttf^{2}\,\alpha^{2}\,+\varsigma_{0} \,\varsigma_{\varepsilon^2 \ttf}}
	\nonumber
\end{align}
and for $\dot{\varsigma}_{\st{2}}=\partial_{\st{2}}\varsigma_{\st{2}}$
\begin{equation}
	\label{gauss_results:s2}
	\sigma_{\st{2}}^{(2)}=\frac{2\,\alpha-\dot{\varsigma}_{\st{2}}}{2\,\varsigma_{\st{2}}}
\end{equation}
Knowing these coefficients allows us to compute the cumulants discussed in Section~\ref{section:cumulants} for the general case.

It is worth noticing that these results result in a remarkably simple expression for mean entropy production at $g=0$
\begin{align}
		\mathcal{E}=\frac{\left(\mu^{(1)}_{\varepsilon^2 \ttf}-\mu^{(1)}_{0}\right)^{2}+\left (\sqrt{\varsigma_{\varepsilon^{2}\ttf}}-\sqrt{\varsigma_{0}}\right )^{2}}{2\,\varepsilon^2 \ttf}\,.
	\nonumber
\end{align}
When 
\begin{align}
	\nU_{*}=\nU_{*}^{(1)}\nq+\frac{1}{2}\nU_{*}^{(1)}\nq^{2}
	\nonumber
\end{align}
(\ref{gauss_results:s2}) remains valid, whereas it is possible to close the hierachy with a second order equation for the variance of the position process
\begin{align}
	2\,\varsigma_{\st{2}}\,\ddot{\varsigma}_{\st{2}}-\left(\dot{\varsigma}_{\st{2}}\right)^{2}-4\,\alpha^{2}\left(\nU_{\star}^{(2)}\,\varsigma_{\st{2}}-1\right)=0
	\label{gauss_results:var}
\end{align}
The general solution of this equation takes the form
\begin{align}
	\varsigma_{\st{2}}=\frac{ c_{1}\,e^{2\,\alpha\,\nU_{\star}^{(2)}\,\st{2}}+c_{2}\,e^{-2\,\alpha\,\nU_{\star}^{(2)}\,\st{2}}+c_{3}}{\nU_{\star}^{(2)}}
	\label{gauss_results:varsol}
\end{align}
with the constants $c_{i}$, $i=1,2,3$, related by the algebraic equation
\begin{align}
	4\,c_{1}\,c_{2}+1-c_{3}^{2}=0
	\nonumber
\end{align}
Unfortunately resolving the $c_{i}$'s in terms of generic boundary conditions leads to somewhat cumbersome expressions. In section~\ref{sec:exco} we consider a special case 
of particular relevance. 

\section{Numerically assisted applications}
\label{sec:num}

In this section, we apply numerical methods to the multiscale expansion to analyze the underdamped dynamics, both in the case of Gaussian boundary conditions and in more complex boundary conditions, in particular, those modelling Landauer's one bit of memory erasure. 

In the Gaussian case, we have a system of differential equations specifying the non-perturbative solution, and we can therefore use numerical integration to solve the associated boundary value problems, from which we can obtain the first and second order phase space cumulants. We then use the perturbative approach to compute the same values, which show good agreement: see Fig.~\ref{fig:kl_gauss} for case \hyperref[itm:a]{KL} and Fig.~\ref{fig:ep_gauss} for case \hyperref[itm:b]{EP}. Additionally, we take a look at expansion and compression with Gaussian boundary conditions in Section~\ref{sec:exco}.

Furthermore, the perturbative approach can be used to make predictions for the cumulants when no analytic solution is available. We demonstrate this using boundary conditions modelling Landauer's one bit of memory erasure, as illustrated in Fig.~\ref{fig:bvp_landauer}. This requires numerically solving the cell problem \eqref{eq:cell}, from which we obtain the optimal control protocol and the marginal distribution of the position in the overdamped dynamics. We can then compute leading order corrections to approximate the quantities in the underdamped dynamics.


\subsection{Gaussian Case}
In cases~\hyperref[itm:a]{KL} and \hyperref[itm:b]{EP}, when the boundary conditions are assigned as Gaussian random variables, we have two boundary value problems for the first and second order cumulants. For case~\hyperref[itm:a]{KL}, we compute approximate solutions to the systems \eqref{G:c2} and \eqref{G:c1}, and, for case~\hyperref[itm:b]{EP}, we make the amendments as described in Section~\ref{section:epgauss}. 

The perturbative approach follows Section \ref{sec:gauss_results}, and instead we have only one boundary value problem. The dependant quantities: momentum mean, momentum variance and the position-momentum cross correlation, as well as the higher order corrections to the position mean and variance can then be computed. 

\begin{figure*}
\centering
\includegraphics[width=\textwidth]{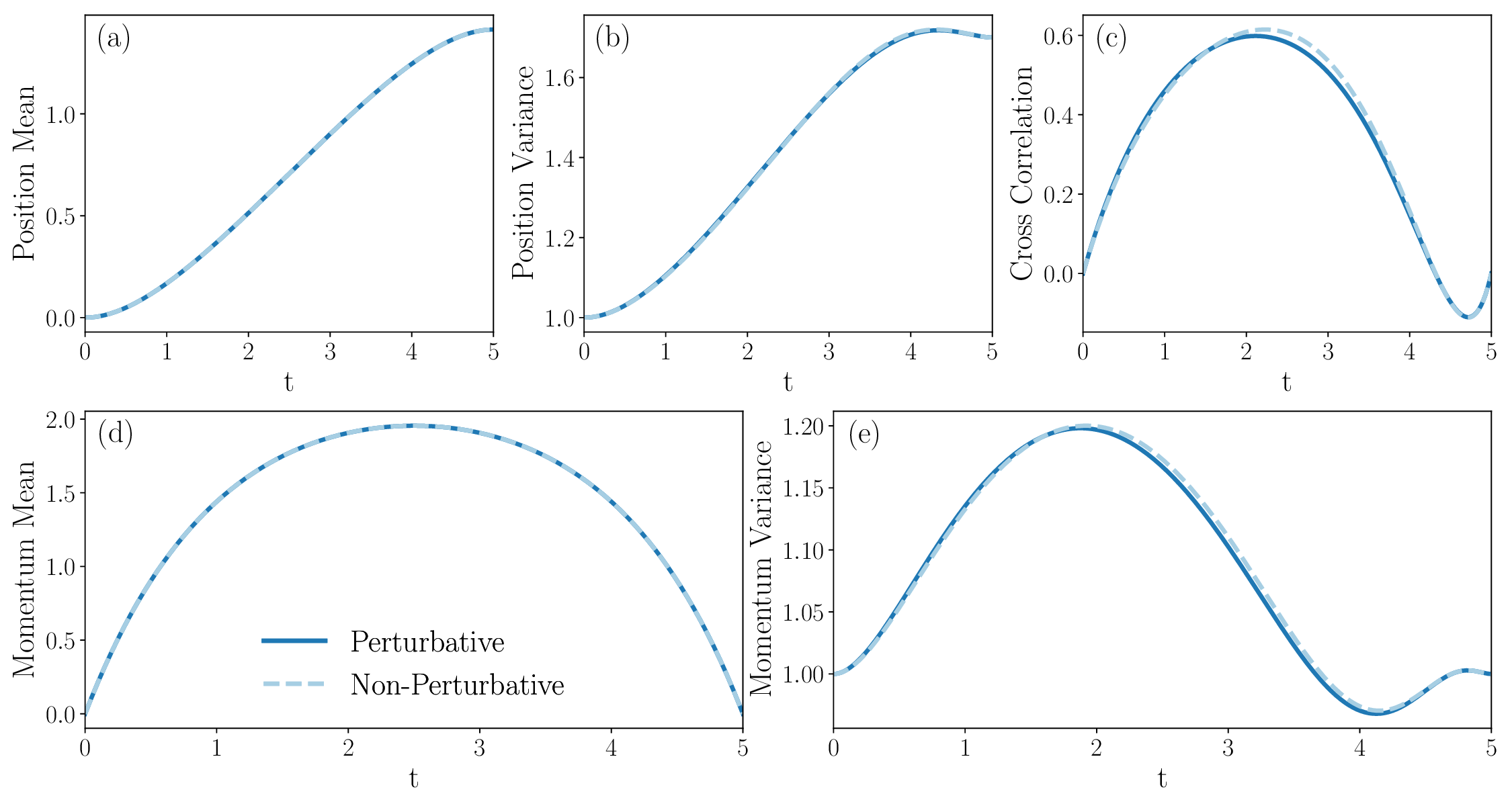}
\caption{Position mean \textbf{(a)} and variance \textbf{(b)}; position-momentum cross correlation \textbf{(c)}; and momentum mean \textbf{(d)} and variance \textbf{(e)} for the underdamped problem minimising the Kullback-Leibler divergence (Case \hyperref[itm:a]{KL}) from a free diffusion with assigned Gaussian initial and final conditions. We plot the expressions of cumulants up to second order predicted by our perturbative approach, solid dark blue line, and contrast them with the numeric solution of the corresponding non-perturbative equation of subsection~\ref{sec:G:KL} shown with a dashed light blue line. 
To plot the perturbative expressions we use the exact solution of the cell problem given in section~\ref{sec:gauss_results} and then we determine the full perturbative  prediction for each of the cumulants using the expressions given in section~\ref{section:cumulants}.  
We impose Gaussian boundary conditions through the first and second order cumulants of the position and momentum (tildes denoting non-dimensional coordinates): at initial time $\nt = 0$, we set the position and momentum variance $\tilde{\styleD{Q}}_{0}=\tilde{\styleD{P}}_{0}=1$, the position-momentum cross correlation $\tilde{\styleD{C}}_{0} = 0$; and the position and momentum means $\operatorname{E}_{\mathcal{P}}\tilde{\bm{\mathscr{q}}}_{\tti} =\operatorname{E}_{\mathcal{P}}\tilde{\bm{\mathscr{p}}}_{\tti} =0$. At final time $\nt = \ttf$: the momentum variance $\tilde{\styleD{P}}_{\ttf}=1$; the position-momentum cross correlation $\tilde{\styleD{C}}_{\ttf} = 0$; the position variance $\tilde{\styleD{Q}}_{\ttf}=1.7$;  the momentum mean $ \operatorname{E}_{\mathcal{P}}\tilde{\bm{\mathscr{p}}}_{\ttf} =0$; and the position mean $\operatorname{E}_{\mathcal{P}}\tilde{\bm{\mathscr{q}}}_{\ttf} = \sqrt{2}$. We use $\ttf=5$, $\varepsilon=0.2$ and $g=0$.\newline
Numerical integration is performed by a fourth order co-location method in the DifferentialEquations.jl library \cite{diffeqsjl}.}
 \label{fig:kl_gauss}
\end{figure*}

\begin{figure*}[!ht]
\centering
\begin{subfigure}{\textwidth}
\includegraphics[width=\textwidth]{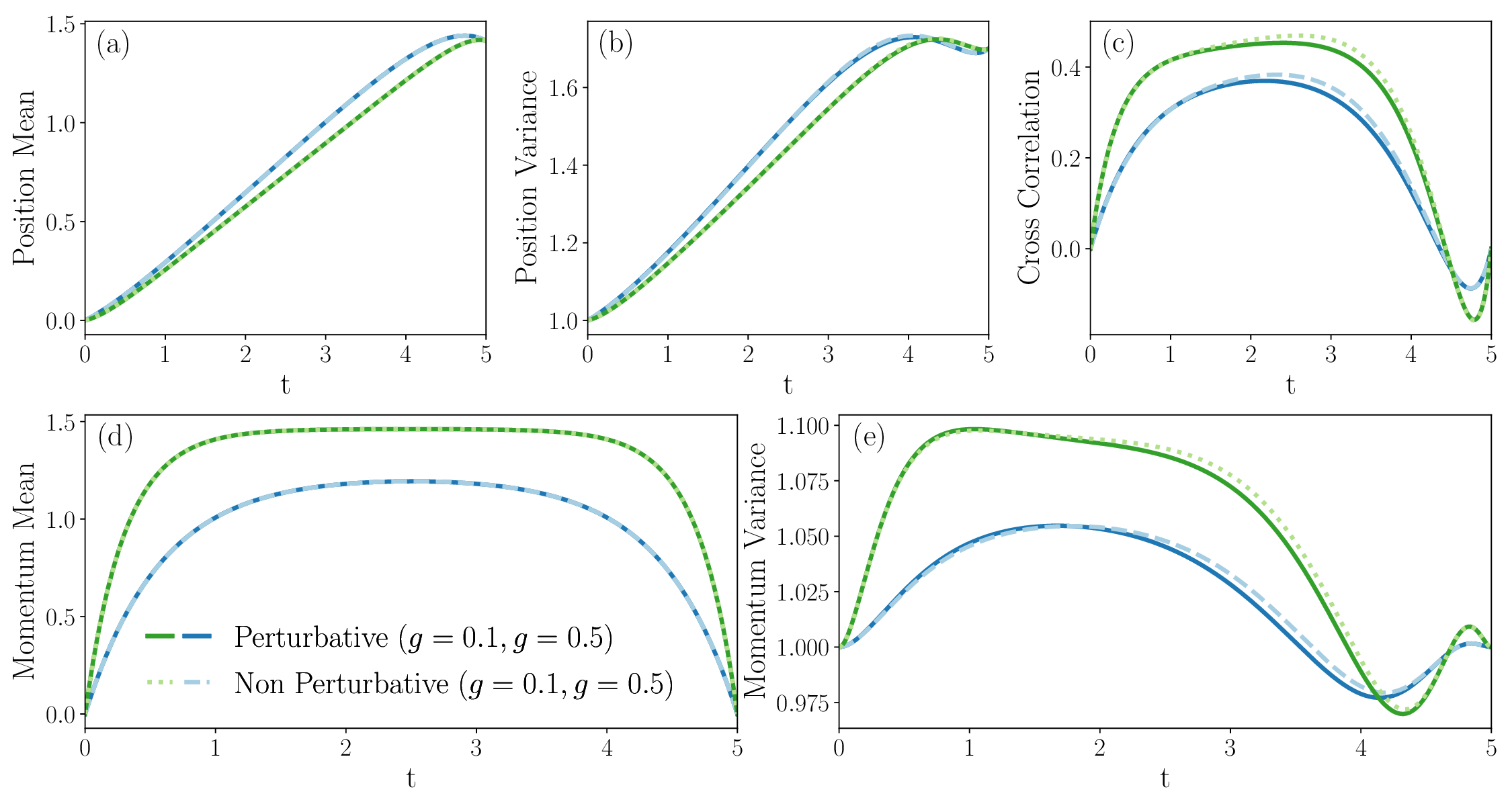}
\centering
\end{subfigure}
\caption{Position mean \textbf{(a)} and variance \textbf{(b)}; position-momentum cross correlation \textbf{(c)}; and momentum mean \textbf{(d)} and variance \textbf{(e)} for the underdamped problem minimising the entropy production~\hyperref[itm:b]{EP} from a free diffusion with assigned Gaussian initial and final conditions. We use $\ttf = 5$, $\varepsilon = 0.2$, and compare the values computed by the perturbative approach (solid lines) with the numeric solution of the non-perturbative system at $g=0.5$ (blue, dashed) and $g=0.1$ (green, dotted), and $\omega = \sqrt{(1+g)/g}$. We impose Gaussian boundary conditions through the first and second order cumulants of the position and momentum. We use the same values for the boundary conditions and the same numerical integration method as in Fig.~\ref{fig:kl_gauss}.
We compute the perturbative predictions in the same way as explained in the caption of Fig.~\ref{fig:kl_gauss}. The equations of the non-perturbative system are those of section~\ref{section:epgauss}.
}
 \label{fig:ep_gauss}
\end{figure*}

The respective boundary value problems are integrated numerically using the DifferentialEquations.jl \cite{diffeqsjl} library in the Julia programming language. The results of the perturbative and non-perturbative integrations for case~\hyperref[itm:a]{KL} are in Fig.~\ref{fig:kl_gauss} and for case~\hyperref[itm:b]{EP} are in Fig.~\ref{fig:ep_gauss}. In both case~\hyperref[itm:a]{KL} and case~\hyperref[itm:b]{EP}, we see that the perturbative expansion gives a very good approximation of the true solution.

 \subsubsection{Asymmetry in optimal approaches to equilibrium}
 \label{sec:exco}

\begin{figure*}[t]
\hspace*{-1.2cm}
     \includegraphics[width=1.2\linewidth]{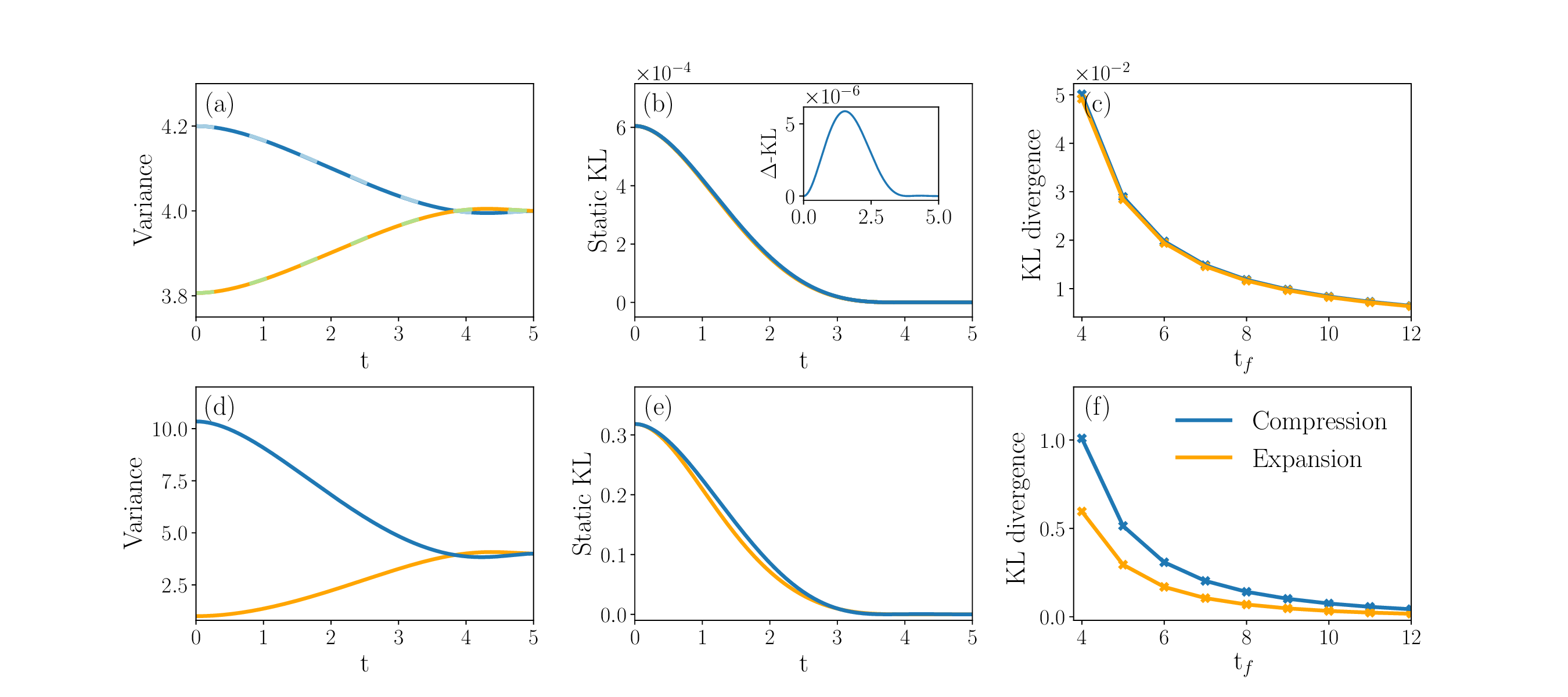}
     \vspace{-0.5cm}
    \caption{Thermal kinematics in the underdamped dynamics. We compare a compression (blue) and expansion (orange) process starting from thermodynamically equidistant states from the final state.  
     We fix the final position variance as $\styleD{Q}_{\tf} = 4$, the corresponding reference potential $\styleD{U}_{\star} = 1/4$, and $\beta = 1$ in all panels. Panels \textbf{(a)-(c)} show the picture where the initial variances are chosen to be close together. We use $\styleD{Q}^{(c)}_{0} = 4.20$ for compression  and $\styleD{Q}^{(e)}_{0} \approx 3.8065$ for expansion. For \textbf{(a)-(b)}, we use $\ttf=5$ and $\varepsilon =0.1$. Panel~\textbf{(a)} shows the variance of the position, with the solid lines computed non-perturbatively and the overlayed dashed lines computed using the linearized approach outlined in Section~\ref{sec:exco}. Panel~\textbf{(b)} shows the static Kullback-Leibler divergence~\eqref{exco:KL}, with the inset axes showing the difference between the compression and expansion process. Panel \textbf{(c)} shows the difference in the dynamic Kullback-Leibler divergence as a function of the time horizon $\ttf$, with $\varepsilon=0.5$. \newline
     Panels \textbf{(d)-(e)} illustrate the case when the initial variances are chosen to be further apart. We use $\styleD{Q}^{(c)}_{0} \approx 10.3467$ for compression  and $\styleD{Q}^{(e)}_{0} = 1$ for expansion. We use $\tf=5$ and $\varepsilon=0.1$. Panel \textbf{(d)} shows the variances and \textbf{(e)} shows the static Kullback-Leibler divergence as functions of the time interval. Panel \textbf{(f)} shows the dynamic Kullback-Leibler divergence as a function of the time horizon $\ttf$, with $\varepsilon =0.5$. First order cumulants do not play a role in the analysis and are equal to $0$ in all panels. All numerical integration is performed by DifferentialEquations.jl \cite{diffeqsjl}, as in Fig~\eqref{fig:kl_gauss}.}\label{exco:Fig}
    \end{figure*}
 Very recently, \cite{LaGo2020,IbDiLaGoRi2024} highlighted the existence of a cooling versus heating asymmetry in the relaxation to a thermal equilibrium from hotter and colder states that are \textquotedblleft thermodynamically equidistant\textquotedblright. Although not strictly  a distance, the Kullback-Leibler divergence from the thermal state may be used to identify the dual processes \cite{LaGo2020}. 
 We show that a similar asymmetry also occurs in optimally controlled isothermal compressions versus expansions of a small system. 
 
 To this goal we make the following observations. Choosing a reference potential $U_{\star}$ in (\ref{model:KL}) equal to the potential in the final condition
  (\ref{model:fin}) forces the current velocity specified by the optimal protocol to be as small as possible at the end of the control horizon. In this sense, the optimal control problem models a relaxation to a thermal equilibrium in finite time. Well-established laboratory techniques \cite{BeArPeCiDiLu2012,MaPeGuTrCi2016,MaRoDiPePaRi2016} use the fact that the optical potential generated by a laser to trap a colloidal nanoparticle is effectively Gaussian. We combine these two observations to compare the compression versus the expansion of a nanosystem in an isothermal environment when the initial data are thermodynamically equidistant from the final equilibrium state. Mathematically, this means that the position marginals of the boundary conditions (\ref{model:ini}), (\ref{model:fin}) are centered Gaussians that differ only in the variance. In such a case, the only non-trivial optimal control equations are (\ref{G:c2}) and (\ref{G:v2}). Our aim is to compare a compression and an expansion process starting from \textquotedblleft dual\textquotedblright\ initial states. Duality is with respect the Kullback-Leibler divergence from the end state whose value is initially the same for the two opposite processes. In the notation of  section~\ref{sec:G} the Kullback-Leibler divergence for $d=1$ reads
  \begin{align}
  	\operatorname{K}(\tilde{\styleC{f}}_{t}\mathrel{\Vert} \tilde{\styleC{f}}_{\tf})=\frac{1}{2}\left(\frac{\mathscr{Q}_{t}}{\mathscr{Q}_{\tf}} -1-\ln \frac{\mathscr{Q}_{t}}{\mathscr{Q}_{\tf}}\right)
  	\label{exco:KL}
  \end{align}
  We fix the terminal condition
  \begin{align}
  	&\mathscr{Q}_{\tf}^{(i)}=(\beta\,\mathscr{U}_{\star})^{-1}=:\mathscr{Q}_{\star},\qquad i=e,c
  	\nonumber
  \end{align}
and compare the evolution of probability densities specified by initial conditions at $\ti=0$
\begin{align}
	\mathscr{Q}_{0}^{(e)}\,<\,\mathscr{Q}_{\star}\,<\,\mathscr{Q}_{0}^{(c)}
	\nonumber
\end{align}
such that the initial position marginals have equal Kullback-Leibler divergence from the final state 
\begin{align}
	\operatorname{K}(\tilde{\styleC{f}}_{0}^{(e)}\mathrel{\Vert} \tilde{\styleC{f}}_{\star})=\operatorname{K}(\tilde{\styleC{f}}_{0}^{(c)}\mathrel{\Vert} \tilde{\styleC{f}}_{\star})
	\nonumber
\end{align}
The dynamic Schr\"odinger bridge with boundary conditions $(\mathscr{Q}_{0}^{(e)},\mathscr{Q}_{\star})$ / $(\mathscr{Q}_{0}^{(e)},\mathscr{Q}_{\star})$ provides a model of optimal expansion/compression of the system towards the equilibrium state characterized by $ \mathscr{Q}_{\star}$.

The multiscale prediction for the position variance is   
\begin{align}
	\dfrac{\mathscr{Q}_{t}}{\ell^{2}}=\varsigma_{\frac{\varepsilon^{2}\,t}{\tau}}
	- \varepsilon^{2}\,\frac{\dot{\varsigma}_{\frac{\varepsilon^{2}\,t}{\tau}}}{A}
	\left(\frac{t}{\tf}\int_{0}^{\frac{\tf}{\tau}}\mathrm{d}s -\int_{0}^{\frac{t}{\tau}}\mathrm{d}s\right)a_{s}+O(\varepsilon^{3})
	\nonumber
\end{align}
where for the sake of simplicity we set $g=0$. To relate non-dimensional quantities to their dimensional counterparts, we explicitly write the Stokes time $\tau$ and the typical length-scale $\ell$ of the transition.
We suppose that the variance of the non-dimensional cell problem at the beginning of the control horizon is
\begin{align}
	&\varsigma_{0}=\frac{\mathsf{v}}{\nU_{2}},\hspace{0.5cm}
	\mbox{with}\,\nU_{2}=\frac{\beta\,\mathscr{U}_{\star} }{\ell^{2}}
	\nonumber
\end{align}
How much the non-dimensional constant $\mathsf{v}$ differs from unity controls the thermodynamic distance from the final state.
In such a case we find that the coefficients $c_{i}$'s in (\ref{gauss_results:varsol}) are
\begin{align}
&	c_{1}=y\,c_{2}
\nonumber\\
&c_{2}=\frac{2 \,e^{2 \,\alpha\,\varepsilon^{2}  \,\ttf \,\nU_2} \left(y\, e^{4\, \alpha \,\varepsilon^{2}  \, \ttf \nU_2}+1\right)}{\left(y \,e^{4\, \alpha \,\varepsilon^{2}  \, \ttf \nU_2}-1\right){}^2}
\nonumber\\
& c_{3}=-\frac{1+6 \,e^{4 \,\alpha\,\varepsilon^{2}  \,\ttf \,\nU_2}\,y+ e^{8 \,\alpha\,\varepsilon^{2}  \,\ttf \,\nU_2} y^{2}}{\left(y \,e^{4\, \alpha \,\varepsilon^{2}  \, \ttf \nU_2}-1\right){}^2}
	\nonumber
\end{align}
  with 
 \begin{align}
 	y&=\frac{2 \cosh \left(2 \,\alpha  \,\varepsilon^{2}  \,\ttf \,\nU_2\right)-3}{\left(\mathsf{v}+1\right) e^{4\, \alpha  \,\varepsilon^{2}  \,\ttf \,\nU_2}-2\, e^{2\, \alpha  \,\varepsilon^{2}  \,\ttf \,\nU_2}}
 \nonumber\\
&\qquad +\frac{\mathsf{v}-2 \sqrt{2} \sinh \left(\alpha  \,\varepsilon^{2}  \,\ttf \,\nU_2\right) \sqrt{2\, \mathsf{v}+\cosh \left(2\, \alpha  \,\varepsilon^{2}  \,\ttf \,\nU_2\right)-1}}{\left(\mathsf{v}+1\right) e^{4\, \alpha  \,\varepsilon^{2}  \,\ttf \,\nU_2}-2\, e^{2\, \alpha  \,\varepsilon^{2}  \,\ttf \,\nU_2}}
 	\nonumber
 \end{align} 
   The above expressions are exact and provide a useful benchmark for exact numerical integration of the cumulant hierarchy (see Fig~\ref{exco:Fig}). 
   
   For transitions describing small deformations of the position marginal of the system, it is however expedient to resort to simpler approximated expressions. We obtain these by linearizing (\ref{gauss_results:var}) around the final condition of the transition. In other words, we look for a solution of the form 
   \begin{align}
   	\varsigma_{\st{2}}=\frac{1}{\nU_{2}}+\varsigma_{\st{2}}^{\prime}+\dots
   	\nonumber
   \end{align} 
   with dots corresponding to higher order terms in the non-linearity. We obtain
   \begin{align}
   \varsigma_{\st{2}}^{\prime}=	-\frac{\left(\mathsf{v}-1\right) e^{-2 \,\alpha \, \st{2}\, \nU_2} \left(e^{4\, \alpha  \, \st{2}\,  \nU_2}-e^{4\, \alpha  \, \varepsilon^{2}\,\ttf\,  \nU_2}\right)}{\nU_2 \left(e^{4\, \alpha  \, \varepsilon^{2}\,\ttf\,  \nU_2}-1\right)}
   	\nonumber
   \end{align} 
  This expression allows us to analytically compare the behavior of the divergence from a common end state of system undergoing an expansion and a compression. We see that if we choose
  \begin{align}
  	\mathsf{v}^{(e)}=1-\eta
  	\nonumber
  \end{align}
  for $\eta\sim O(10^{-1})$ then within $O(10^{-4})$ accuracy the initial data for the dual compression process is
  \begin{align}
  		\mathsf{v}^{(c)}=1+\eta +\frac{2 \eta ^2}{3}+\frac{4 \eta ^3}{9}+\frac{44 \eta ^4}{135} +O(\eta^{5})
  	\nonumber
  \end{align}
 A straightforward calculation then shows that  within leading order accuracy 
    \begin{align}
    &	\operatorname{K}(\tilde{\styleC{f}}_{t}^{(c)}\| \tilde{\styleC{f}}_{\star})-\operatorname{K}(\tilde{\styleC{f}}_{t}^{(e)}\| \tilde{\styleC{f}}_{\star})
    \,\geq\, 0\,,\qquad \forall\,0\,\leq\,t\,\leq\,\ttf
    	\nonumber
    \end{align}
The result holds analytically for small deformations of the potential $\eta \ll 1$ and close to the overdamped limit $\varepsilon\,\ll\,1$.  

Another thermodynamic indicator encoding similar information is the cost of the dynamic Schr\"odinger bridge (\ref{model:KL}). This quantity is a global indicator of the transition that can be studied versus the duration of the horizon.  
Consistently with the analytic perturbative result the evaluation of  (\ref{model:KL}) shows that the divergence from equilibrium is larger for compression processes. The difference between compression and expansion tends to zero as the duration of the horizon tends to infinity, thus indicating symmetry restoration for adiabatic processes. 

Our findings are summarized in Figure~\ref{exco:Fig}. Our analysis is in line with the findings of \cite{LaGo2020}. If we interpret the divergence from equilibrium at any fixed time as an indirect quantifier of the speed with which the system ultimately thermalizes, our analytic and numerical results confirm that expansion is faster than compression for Gaussian models .  
    
\subsection{Landauer's erasure problem}
\begin{figure}[h]
     \includegraphics[width=\textwidth]{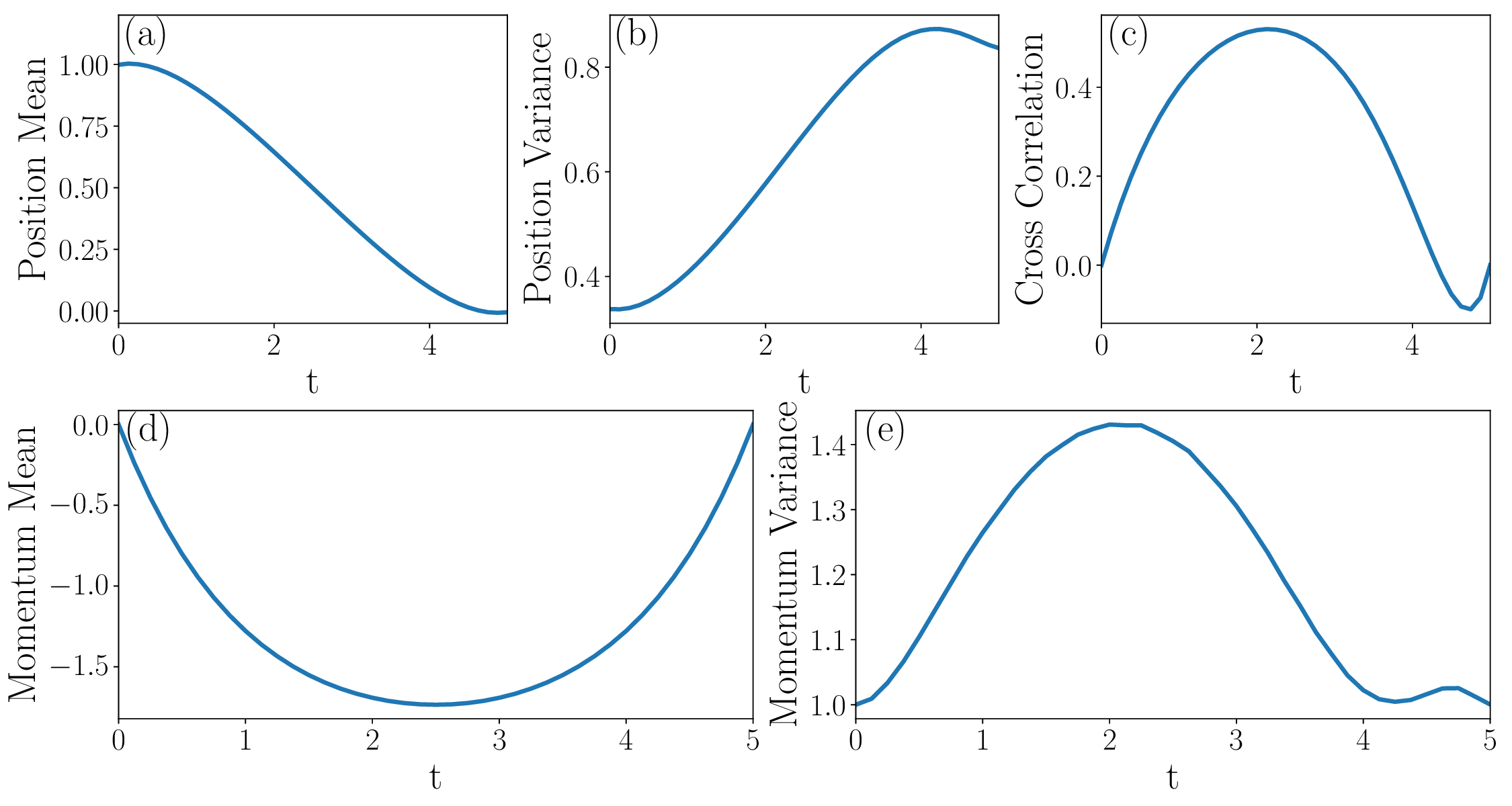}
    \caption{Predictions for the position mean \textbf{(a)} and variance \textbf{(b)}; position-momentum cross correlation \textbf{(c)}; momentum variance \textbf{(d)} and mean \textbf{(e)} in the underdamped dynamics minimizing the Kullback-Leibler divergence (Case \hyperref[itm:a]{KL}) from a free diffusion computed using the perturbative expansion. The boundary conditions are assigned on the marginal density of the position:  
    the initial state at $\nt= 0$ is a single peaked distribution centered at $x_{\mathfrak{o}} = 1$ (see Eq.~\eqref{num:initial_state}) and the final is a double peaked distribution with peaks at $-1$ and $1$ (see Eq.~\eqref{num:final_state}). We set $\ttf = 5$, $\varepsilon =0.2$, $g=0$,  $\omega = 1$  and use $\alpha = \sqrt{(1+g)\,A}\approx 0.64$, where $A$ is as defined in \eqref{eq:A_constant}. 
    The functions $\rho_{\st{2}}(\nq)$ and $\sigma_{\st{2}}(\nq)$ are approximated numerically for values of  $\st{2}\in[0,\varepsilon^2\, \ttf]$ as the solution of Eq.~\eqref{eq:cell} by a forward-backward iteration. We perform a total of $15$ forward and backward passes of the iteration. At each step, the factors $\phi$ and $\hat\phi$ are computed by means of Eq.~\eqref{num:mcphis} and normalized for numerical stability. We use a total of $8000$ sample points for $\nq$ in the interval $[-6,6]$ and evolve $50\ 000$ independent Monte Carlo trajectories started from each $\nq$ by an Euler-Maruyama discretization of the SDE \eqref{num:auxsde} with step-size $h=0.005$. The forward-backward iteration is initialized by $\hat\phi_{\ttf}$ with a vector of ones. We recover the functions $\rho$ and $\sigma$ from $\phi$ and $\hat\phi$ using Eqns. \eqref{num:rhosigma}; these are then smoothed using a convolution with a box filter with window size $\delta = 0.096$. The predictions for the underdamped moments are then computed using expressions found in Section~\eqref{section:cumulants}.}
    \label{fig:kl_cumulants}
\end{figure}

We model the Landauer's one bit of memory erasure \cite{LeOrPoSn2019} as a Schrodinger bridge problem between an initial state single-peaked distribution and final state as a double-peaked distribution, as illustrated in Figure~\ref{fig:bvp_landauer}. 
We can make predictions for the first and second order cumulants of the position and momentum distributions from the perturbative expansion, by computing the numerical solution to the cell problem \eqref{eq:cell} and hence the appropriate corrections. We focus only on case~\hyperref[itm:a]{KL}. 

We assign the initial and final state of the position marginal distribution
\begin{align*}
\rho_{\varepsilon^2\tti}(\nq) &= \int_{\mathbb{R}} \mathrm{d} \np\ \mathtt{p}_{\tti}(\nq,\np) =: P_{\iota}(\nq) \\
\rho_{\varepsilon^2 \ttf}(\nq) &= \int_{\mathbb{R}} \mathrm{d} \np\ \mathtt{p}_{\ttf}(\nq,\np) =: P_{\styleB{f}}(\nq)
\end{align*}
where $P_{\iota}$ and $P_{\styleB{f}}$ denote the assigned initial and final distributions, and here take the explicit forms
\begin{equation}
\begin{split}
    \label{num:initial_state}
  P_{\iota}(\nq)& =\dfrac{1}{Z_{\iota}} \exp\Bigl(-\, (\nq-x_{\mathfrak{o}})^4\Bigr)
 \end{split}
\end{equation}
\begin{equation}
\begin{split}
    \label{num:final_state}
 P_{\styleB{f}}(\nq) & =\dfrac{1}{Z_{\styleB{f}}}\exp\Bigl(-\, (\nq^2-x_{\mathfrak{o}}^{2})^2\Bigr)
\end{split}
\end{equation}
with $Z_{\iota},\ Z_{\styleB{f}}$ normalizing constants. The initial condition is a single peaked distribution centered at $x_{\mathfrak{o}}$, and final condition is a double peaked distribution, with peaks at $x_{\mathfrak{o}}$ and $-x_{\mathfrak{o}}$.

We look at the case of $U_{\star}=0$. The cell problem \eqref{eq:sol_o2} can be approximated numerically using a forward-backward iteration. This specifically means computing the numerical solution of two coupled non-linear partial differential equations to obtain the functions $\rho$ and $\sigma$ of the slow time $\st{2} = \varepsilon^2 \nt$. 

We adopt the methodology of \cite{CaHa2022}, beginning with the Hopf-Cole transform
\[\hat\phi_{\st{2}}(\nq) = \rho_{\st{2}}(\nq)\exp\left({\dfrac{\sigma_{\st{2}}(\nq)}{2\,\alpha}}\right),\quad\phi_{\st{2}}(\nq)  = \exp\left(-\dfrac{\sigma_{\st{2}}(\nq)}{2\,\alpha}\right)\]
yielding a pair of Fokker-Planck equations
    \begin{subequations}
    \label{num:fpk}
\begin{eqnarray}
\label{num:pde:phi}
    \partial_{\st{2}} \phi_{\st{2}}(\nq) + \alpha\,\partial^2_{\nq} \phi_{\st{2}}(\nq) &= 0 \\ 
\label{num:pde:phihat}
\partial_{\st{2}} \hat\phi_{\st{2}}(\nq) - \alpha\,\partial_{\nq}^2 \hat\phi_{\st{2}}(\nq)        &= 0 
\end{eqnarray}
\end{subequations}
with coupled boundary conditions
      \begin{subequations}
\label{num:boundary}
\begin{align}
    \label{num:phi_bc}
    \phi_{\varepsilon^2\ttf}(\nq)  &= P_{\styleB{f}}(\nq)\,/\,\hat\phi_{\varepsilon^2\ttf}(\nq) \\
\label{num:phihat_bc}
    \hat\phi_0(\nq) &= P_{\iota}(\nq)\,/\, \phi_0(\nq)
\end{align}
\end{subequations}
In this form, the cell problem can be solved using the forward-backward iteration, an adaptation of Algorithm 1 of \cite{CaHa2022}. We make a slight simplification, in that we perform the numerical integration of equations \eqref{num:fpk} by a Monte Carlo method, computing 
\begin{subequations}
\label{num:mcphis}
    \begin{align}
            \label{num:mcphi_hat1}
        \hat\phi_{\varepsilon^2 \ttf}(\nq) &= \operatorname{E}\Bigl(\hat\phi_0\left(\mathscr{q}_{\varepsilon^2 \ttf}\right)\,\Big{|}\, \mathscr{q}_{0}=\nq\Bigr) \\
        \label{num:mcphi0}
       \phi_{0}(\nq) &= \operatorname{E}\Bigl(\phi_{\varepsilon^2\ttf}\left(\mathscr{q}_{0}\right)\,\Big{|}\, \mathscr{q}_{\varepsilon^2 \ttf}=\nq\Bigr)
    \end{align}
\end{subequations}
using the forward and backward evolution respectively of the underlying auxiliary (Ito) stochastic process 
\begin{equation}
    \label{num:auxsde}
\mathrm{d}\mathscr{q}_{\st{2}} = \sqrt{2\,\alpha}\, \mathrm{d}\mathscr{w}_{\st{2}}\,,\end{equation}
where $\{\mathscr{w}_{\st{2}}\}_{\st{2}\geq 0}$ denotes a standard Wiener process. The values of $\mathscr{q}_{\st{2}}$ are approximated with discretized trajectories of \eqref{num:auxsde} by the Euler-Maruyama scheme.

The forward-backward iteration goes as follows: We begin by sampling a set of values for $\nq$ from an interval on which both the initial and final assigned distributions $P_{\iota}$ and $P_{\styleB{f}}$ are compactly supported. We initialize the forward-backward iteration by taking a set of (positive) values for $\hat\phi_{\varepsilon^2\ttf}$, which are then used to compute the boundary condition \eqref{num:phi_bc} for equation \eqref{num:pde:phi}. We integrate equation \eqref{num:pde:phi} using the expression \eqref{num:mcphi0} to obtain $\phi_0$ and recompute $\hat\phi_0$ using \eqref{num:phihat_bc}. By integrating \eqref{num:pde:phihat} using \eqref{num:mcphi_hat1} up to $\ttf$, we once again obtain $\hat\phi_{\varepsilon^2\ttf}$. This procedure is then repeated until convergence; we verify that the boundary condition relations \eqref{num:boundary} are satisfied, and the mean-squared difference between two iterations of $\phi_{\ttf}$ and  $\hat\phi_{\ttf}$ is less than a specified tolerance. We can then recover the values of $\rho_{\st{2}}$ and $\sigma_{\st{2}}$ by the relations
\begin{equation}
\label{num:rhosigma}
        \begin{split}
        \rho_{\st{2}}(\nq) &= \hat\phi_{\st{2}}(\nq)\, \phi_{\st{2}}(\nq) \\
        \sigma_{\st{2}}(\nq) &= - 2\,\alpha\,\log\left(\phi_{\st{2}}(\nq)\right)\,.
    \end{split}
\end{equation}

\begin{figure*}
\hspace*{-1.5cm}
     \includegraphics[scale=0.35]{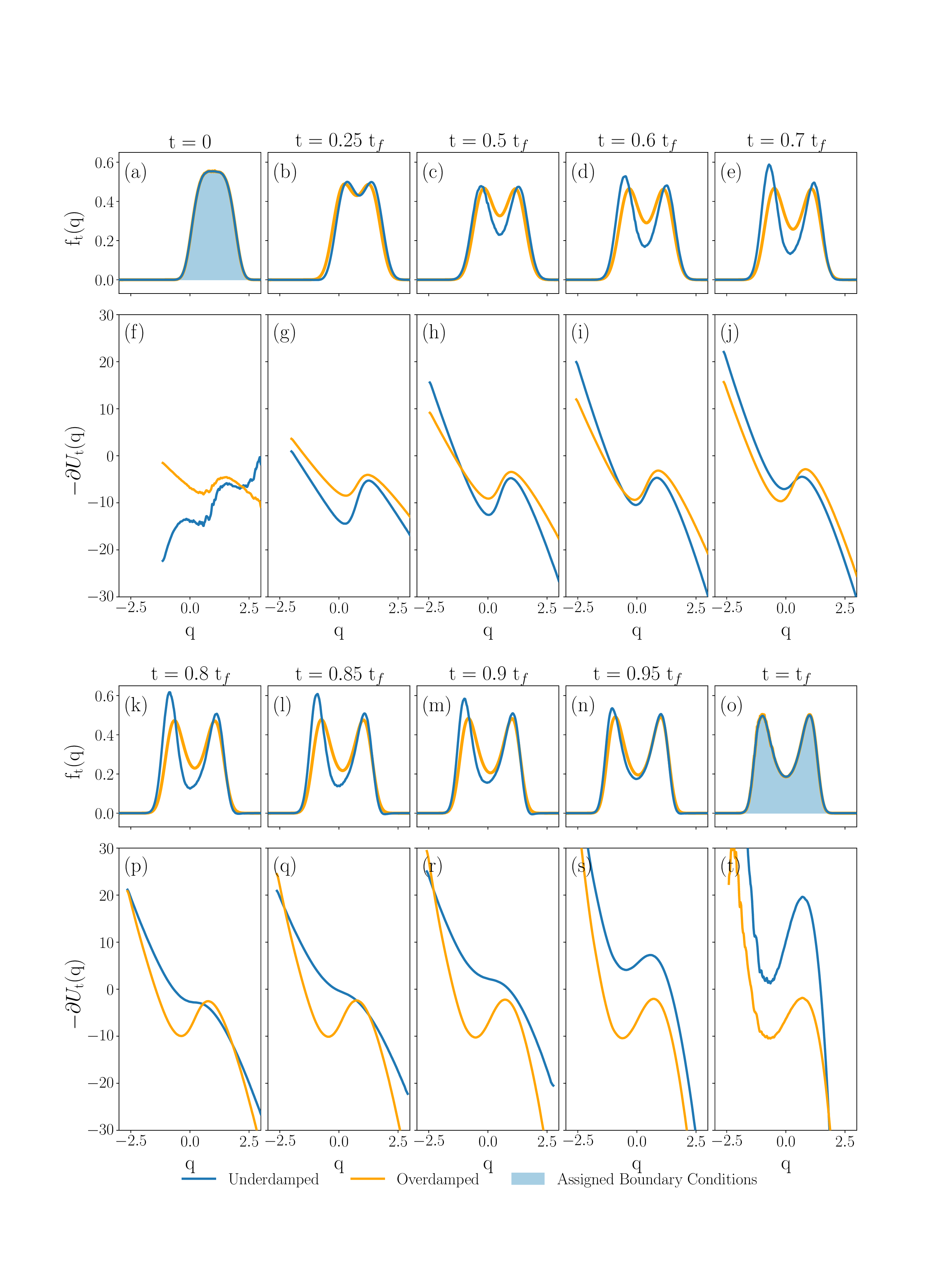}\vspace*{-1cm}
   \caption{Predictions for the marginal density of the position \textbf{(a)}-\textbf{(e)} and \textbf{(k)} -\textbf{(o)} and the gradient of the optimal control protocol \textbf{(f)}-\textbf{(j)} and \textbf{(p)} -\textbf{(t)} in the overdamped (orange) and underdamped (blue) dynamics minimizing the Kullback-Leibler divergence (Case \hyperref[itm:a]{KL}) from a free diffusion, $U_{\star}=0$. We show the distribution $\rho$ and the gradient of the optimal control protocol $-\partial_q \sigma$ \eqref{num:rhosigma} for the overdamped dynamics, and compute the corrections needed to obtain the corresponding quantities $\nf_{\nt}$ and $-\partial_{\nq} U_{\nt}$ for the underdamped dynamics. The shaded region in panels \textbf{(a)} and \textbf{(o)} show the assigned boundary conditions \eqref{num:initial_state} and \eqref{num:final_state} respectively: the initial state at $\nt= 0$ is a single peaked distribution centered at $x_{\mathfrak{o}} = 1$ and the final is a double peaked distribution with peaks at $-1$ and $1$.
   We set $\ttf = 5$, $\varepsilon =0.2$, $g=0$, $\omega = 1$  and $\alpha = \sqrt{(1+g)\,A}\approx 0.64$, where $A$ is as defined in \eqref{eq:A_constant}. 
    The functions $\rho_{\st{2}}(\nq)$ and $\sigma_{\st{2}}(\nq)$ are computed as in Fig.~\eqref{fig:kl_cumulants}, and predictions for the underdamped are computed using the expressions in \ref{sec:results}. All distributions are normalized.}
    \label{fig:kl_drift}
 \end{figure*}

\begin{figure}[!ht]
\centering
\hspace*{0.5cm}
         \includegraphics[width=0.9\textwidth]{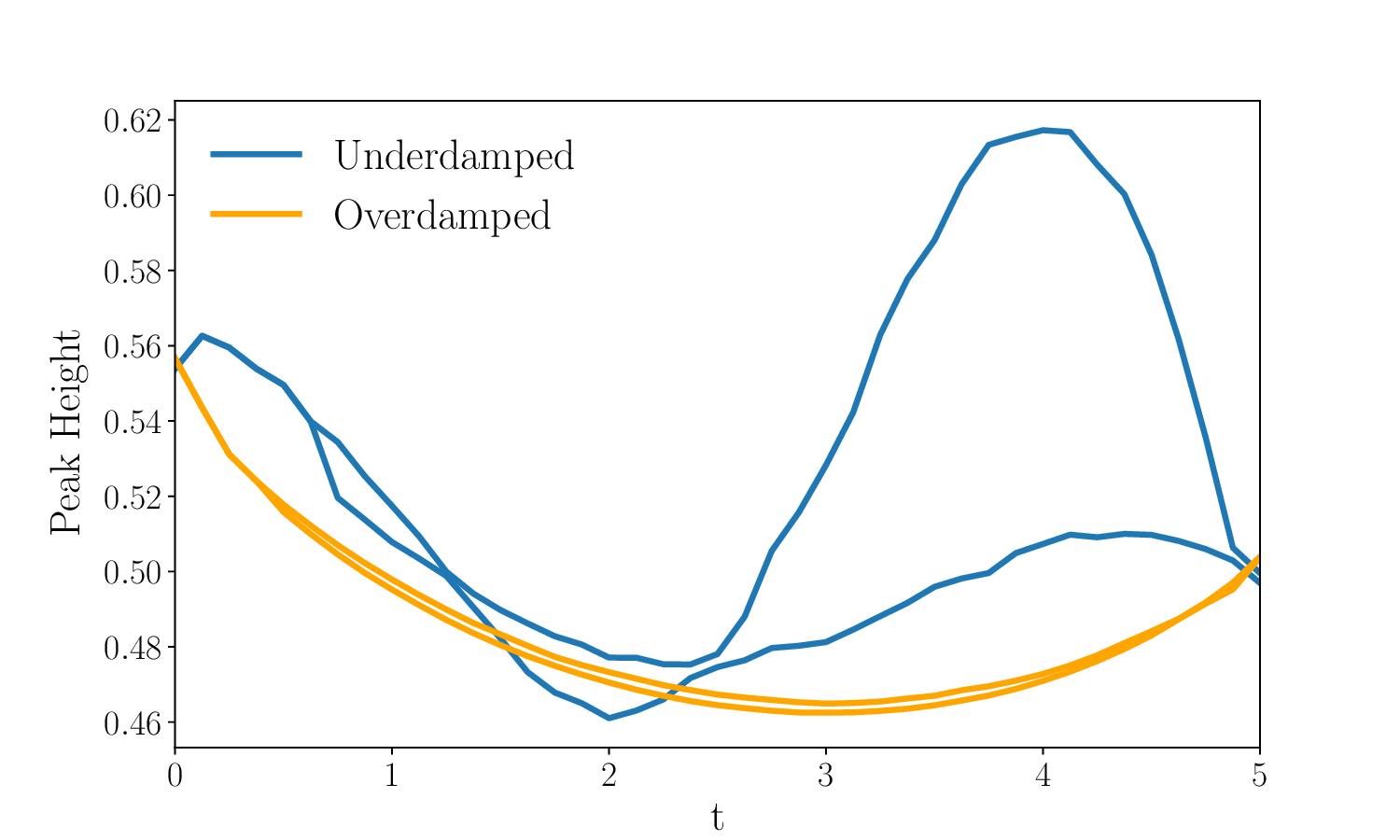}
    \caption{Predictions for the heights of the peaks of the marginal distribution of the position in the overdamped (orange) and underdamped (blue) dynamics. Both peaks in the overdamped remain at roughly equal height, while the underdamped diverge. The distribution in the overdamped and the corrections to obtain the underdamped are computed as in Fig.~\ref{fig:kl_drift}, using $\mathrm{t}_{\styleB{f}} = 5$, $\varepsilon = 0.2$, and $g=0$.
    \label{fig:kl_peak_heights}   }
\end{figure}

The optimal control protocol in the overdamped case is $\sigma$. From here, we use the relevant equations in Sections~\ref{section:cumulants} and \ref{section:optimal_drift} to make predictions for the first and second order cumulants of the position and momentum in the underdamped dynamics, which are shown in Figure~\ref{fig:kl_cumulants}. The predicted marginal distribution of the position and the gradient of the optimal control protocol is shown in Figure~\ref{fig:kl_drift}. Figure~\ref{fig:kl_peak_heights} contrasts the heights of the peaks of the marginal distribution of the position in the underdamped and overdamped dynamics over the time interval.

\section{Conclusions and outlook}
\label{sec:outlook}

In this paper, we address the problem of finding optimal control protocols analytically for finite time stochastic thermodynamic transitions described by underdamped dynamics. To such end, we introduce a multiscale expansion whose order parameter vanishes in the overdamped limit. Within second order accuracy, we are able to find corrections for the linear and quadratic moments of the process. When the boundary conditions are Gaussian, our results are in excellent agreement with the solutions found by non-perturbative numerical methods. 

We expect our theoretical predictions to provide a necessary benchmark for design and interpretation of experiments on nanomachine thermodynamics.  In particular, this is the case for statistical indicators of the momentum process, whose dynamical properties are a distinctive trait of the underdamped regime. Our predictions for the momentum variance and the position-momentum cross correlation are in qualitative agreement with the very recent experimental observations in related laboratory setups \cite{RaGuGuOdTr2023}. 

We envisage several directions to extend the present work. In our view, the most urgent and possibly relevant for applications is devising efficient numerical algorithms to determine regular extremals for general (non-Gaussian) boundary conditions. The non-local nature of the equations determining the regular extremals 
 hamper the direct application of proximal algorithms~\cite{CaHa2020,CaHa2022} and Monte Carlo methods. We address the problem of generalizing these methods to the underdamped case in a forthcoming companion contribution \cite{SaBaMG2024}. Here, we also compute inertial corrections to the numerical solution of the overdamped problem~\cite{BrFrHeLoMaMoSo2003} for minimal entropy production in Landauer's problem \cite{AuGaMeMoMG2012}.

A second main result of the present work is the proof that the optimal control for transitions between Gaussian states solve a Lyapunov equation in any number of dimensions.
This is a strong indication of the existence of regular extremals in phase spaces of any number of dimensions: in view of \cite{WyBa1987c}, the extension of the multiscale method is very cumbersome, but otherwise conceptually straightforward. A more subtle issue is instead the computation of corrections of orders higher than two, which are prone to instabilities already at third order. Ideas motivated by normal form theory \cite{BobA2006} offer a promising way to overcome this difficulty. Yet, the application to optimal control on a finite time horizon is still an open challenge.

From the physics perspective, the multiscale expansion appears best suited to deal with nanoscale dynamics when inertial effects are present, but are small in comparison to thermal fluctuations. A possible alternative approach is the underdamped expansion (see e.g. Chapter~6 of \cite{PavG2014}). This technique could be used to extract complementary information to that obtained here.

In terms of applications, our results are relevant for all physical contexts where random fluctuations and inertial effects cannot be disregarded.
This is the case, for example, in bit manipulation in electronic devices. 
Information bits are encoded using bi-stable states governed by double-well potentials. Inertia is required to improve the efficiency of most logic operations~\cite{lopez2016sub}.

Our results find natural applications also in biophysics. The control of biological systems such as bacteria suspensions and swarms is nowadays accessible to experimentation through several techniques~\cite{dileonardo2015, peng2016,  cavagna2023characterization, Pellicciotta_2023}. 
This has generated increasing interest in the theoretical challenge of applying control theory to active matter models, i.e. out-of-equilibrium dynamics showing complex phenomena inspired by biology~\cite{Shankar_2022, baldovin2023, davis2023active, frim2023shortcut}. So far, however, only overdamped dynamics have been considered. While this does describe the behaviour of microscopic biological systems at high Reynolds numbers (e.g., bacteria in liquid suspensions) fairly, it is well known that inertial effects do play a fundamental role in some classes of such systems~\cite{Manacorda2017, Scholz2018, Lwen2019InertialEO}. A meaningful description of the collective behaviour of flocks and swarms requires taking into account inertial effects that allow efficient propagation of information within the system~\cite{Attanasi_2014, Cavagna_2014, Cavagna_2017}. Any approach to the control of these models should therefore be carried out in the underdamped regime: even if our results cannot be straightforwardly applied to collective dynamics, they may provide a promising starting point for the development of control theory in this context.

\subsection*{Acknowledgments}
	
The authors are pleased to acknowledge discussions with Luca Peliti and Paolo Erdman. 
JS was supported by the Centre of Excellence in Randomness and Structures of the Academy of Finland 
 and by a University of Helsinki funded doctoral researcher position, Doctoral Programme in Mathematics and Statistics.
 MB was supported by ERC Advanced Grant RG.BIO (Contract No. 785932).

\section*{Declarations}

\subsection*{Data availability statement}
Data sets generated during the current study are available from the corresponding author on reasonable request. 

\subsection*{Conflict of interest}
The authors have no relevant financial or non-financial interests to disclose.

\appendix

\section*{Appendices}


\section{Derivation of the cost functionals}
\label{app:KL}

The physics-style derivation of (\ref{model:KL}) and (\ref{model:ep}) proceeds by constructing finite dimensional approximation
on families of time lattices $\ti \,\leq\,t_{0}\,\leq\,\dots\,\leq\,t_{N+1}=\tf $ with mesh size
\begin{align}
	h=\dfrac{\tf-\ti}{N+2}\,.
	\nonumber
\end{align}
The one-step approximation of the transition probability density of (\ref{model:sde}) in the pre-point prescription is
\begin{align}
\operatorname{T}_{t_{i+1},t_{i}}^{(h)}(\bm{x}_{i+1}\mid\bm{x}_{i})=
	\dfrac{\exp\big{(}-\operatorname{A}_{t_{i+1},t_{i}}^{(h)}(\bm{x}_{i+1}\mid\bm{x}_{i})\big{)}}{Z_{i}}
	\label{app:KL:Ito}
\end{align}
where $\bm{x}_{i}=\bm{q}_{i}\oplus \bm{p}_{i}$ and 
\begin{align}
	\operatorname{A}_{t_{i+1},t_{i}}^{(h)}(\bm{x}_{i+1}\mid\bm{x}_{i})&=\dfrac{\beta\,m}{4\,g\,\tau\,h}\left\|\bm{q}_{i+1}-\bm{q}_{i}-
\left(\dfrac{\bm{p}_{i}}{m}-\dfrac{g\,\tau}{m}(\bm{\partial}U_{t_{i}})(\bm{q}_{i})\right)\,h\right\|^{2}
\nonumber\\[3pt]
&\qquad+\dfrac{\beta\,\tau}{4\,m\,h}\left\|\bm{p}_{i+1}-\bm{p}_{i}-
	\left(\dfrac{\bm{p}_{i}}{\tau}+(\bm{\partial}U_{t_{i}})(\bm{q}_{i})\right)\,h\right\|^{2}\,,
	\label{app:KL:Ito2}
\end{align}
while $Z_{i}$ is a normalization constant irrelevant for the present considerations.
Within accuracy (\ref{app:KL:Ito}) satisfies the Chapman-Kolmogorov equation \cite{PavG2014}. Hence, we obtain the transition probability over any finite time 
interval by means of the limit
\begin{align}
&	\operatorname{T}_{t,\tilde{t}}\,(\bm{x}\mid\bm{\tilde{x}})=
\nonumber\\
&\lim_{\substack{h \downarrow 0 Nh=t-\tilde{t} }}\,\prod_{i=1}^{N}\,\int\mathrm{d}^{2\,d }\bm{z}_{i}\,\operatorname{T}_{s_{i+1},s_{i}}^{(h)}(\bm{z}_{i+1}\mid\bm{z}_{i})\,\operatorname{T}_{s_{1},s_{0}}^{(h)}(\bm{z}_{1}\mid\bm{z}_{0})
	\label{app:KL:tp}
\end{align} 
where we hold fixed in the limit $\tilde{t}=s_{0}\,\leq\,s_{N+1}=t$ and $\bm{z}_{0}=\bm{\tilde{x}}$, $ \bm{z}_{N+1}=\bm{x}$. 
For any admissible potential, (\ref{app:KL:tp}) satisfies by hypothesis the bridge boundary conditions
\begin{align}
	\operatorname{f}_{\tf}(\bm{x})=\int_{\mathbb{R}^{2\,d}}\mathrm{d}^{2\,d}\bm{y}\,\operatorname{T}_{\ttf,\ti}(\bm{x}\mid\bm{y})\operatorname{f}_{\ti}(\bm{y})
	\nonumber
\end{align}
with $ \operatorname{f}_{\ti}$, $\operatorname{f}_{\tf}$ respectively assigned by (\ref{model:ini}) and  (\ref{model:fin}). 

\subsection{Case~\hyperref[itm:a]{KL}}

Proceeding in a similar fashion, the finite dimensional approximation of (\ref{model:KL}) is by definition
\begin{align}
		\operatorname{K}(\mathds{P}^{N} \,||\,\mathds{Q}_{\star}^{N} )&=\int_{\mathbb{R}^{2 d (N+2)}}\hspace{-0.3cm}\mathrm{d}^{2 d}\bm{x}_{0}\,\mathrm{d}^{2 d}\bm{x}_{N+1}\prod_{j=1}^{N} \mathrm{d}^{2 d}\bm{x}_{j}\, \operatorname{f}_{\ti}(\bm{x}_{0})
	\nonumber\\
	&\qquad\times\,\operatorname{T}_{t_{j+1},t_{j}}^{(h)}(\bm{x}_{j+1}\mid\bm{x}_{j})\ln \prod_{k=1}^{N}  \dfrac{\operatorname{T}_{t_{k+1},t_{k}}^{(h)}(\bm{x}_{k+1}\mid\bm{x}_{k})}{\operatorname{T}_{\star~t_{k+1},t_{k}}^{(h)}(\bm{x}_{k+1}\mid\bm{x}_{k})}\,,
	\label{app:KL:def}
\end{align}
where $\operatorname{T}_{\star}^{(h)}$ is defined with respect to the reference potential $U_{\star}$.
Using the properties of the logarithm and the normalization of the transition probability, the definition reduces to the sum 
\begin{align}
			\operatorname{K}(\mathds{P}^{N} \,||\,\mathds{Q}_{\star}^{N} )&=\sum_{i=1}^{N}\,\int_{\mathbb{R}^{2 d (i+2)}}\hspace{-0.3cm}\mathrm{d}^{2 d}\bm{x}_{0}\,\mathrm{d}^{2 d}\bm{x}_{i+1}\,\prod_{j=1}^{i} \mathrm{d}^{2 d}\bm{x}_{j}
	\nonumber\\
	&\qquad\times\, \operatorname{f}_{\ti}(\bm{x}_{0}) \operatorname{T}_{t_{j+1},t_{j}}^{(h)}(\bm{x}_{j+1}\mid\bm{x}_{j})\ln \dfrac{\operatorname{T}_{t_{j+1},t_{j}}^{(h)}(\bm{x}_{j+1}\mid\bm{x}_{j})}{\operatorname{T}_{\star~t_{j+1},t_{j}}^{(h)}(\bm{x}_{j+1}\mid\bm{x}_{j})}
	\label{app:KL:fd}
\end{align}
Next, we observe that 
\begin{align}
	\ln \frac{\operatorname{T}_{t_{i+1},t_{i}}^{(h)}(\bm{x}_{i+1}\mid\bm{x}_{i})}{\operatorname{T}_{\star~t_{i+1},t_{i}}^{(h)}(\bm{x}_{i+1}\mid\bm{x}_{i})}&=
		\frac{\beta\,\tau\,(1+g)\,h}{4\,m}\Big{(}(\bm{\partial}U_{\star})^{2}(\bm{q}_{i})-(\bm{\partial}U_{t_{i}})^{2}(\bm{q}_{i})\Big{)}
	\nonumber\\[3pt]
&\qquad	+\frac{\beta}{2}\left(\bm{q}_{i+1}-\bm{q}_{i}-h\,\dfrac{\bm{p}_{i}}{m}\right)\cdot\Big{(}(\bm{\partial}U_{\star})(\bm{q}_{i})-(\bm{\partial}U_{t_{i}})(\bm{q}_{i})\Big{)}
\nonumber\\[3pt]
&\qquad+\frac{\beta\,\tau}{2\,m} \left(\bm{p}_{i+1}-\bm{p}_{i}+
h\,\dfrac{\bm{p}_{i}}{\tau}\right)\cdot\Big{(}(\bm{\partial}U_{\star})(\bm{q}_{i})-(\bm{\partial}U_{t_{i}})(\bm{q}_{i})\Big{)}
	\nonumber
\end{align}
 The outermost integrals in (\ref{app:KL:fd}) over $\bm{q}_{i+1},\bm{p}_{i+1}$ are Gaussian and  equal to 
\begin{align}
&	\int \mathrm{d}^{2d}\bm{x}_{i+1}  \operatorname{T}_{t_{i+1},t_{i}}^{(h)}(\bm{x}_{i+1}\mid\bm{x}_{i})
\begin{cases}
	\bm{q}_{i+1}
\\
\bm{p}_{i+1}
\end{cases}		=
\begin{cases}
&\bm{q}_{i}+\left(\dfrac{\bm{p}_{i}}{m}-\dfrac{g\,\tau}{m}(\bm{\partial}U_{t_{i}})(\bm{q}_{i})\right )\,h
\\[0.3cm]
&	\bm{p}_{i}-
	\left(\dfrac{\bm{p}_{i}}{\tau}+(\bm{\partial}U_{t_{i}})(\bm{q}_{i})\right)\,h
\end{cases}		
	\nonumber
\end{align}
We thus arrive at
\begin{align}
		\int \mathrm{d}^{2 d}\bm{x}_{i+1} \operatorname{T}_{t_{i+1},t_{i}}^{(h)}&(\bm{x}_{i+1}\mid\bm{x}_{i})\ln \dfrac{\operatorname{T}_{t_{i+1},t_{i}}^{(h)}(\bm{x}_{i+1}\mid\bm{x}_{i})}{\operatorname{T}_{t_{{1}},t_{{0}}}^{(h)}(\bm{x}_{i+1}\mid\bm{x}_{i})}
	\nonumber\\
	&\qquad=\dfrac{\beta\,\tau\,(1+g)\,h}{4\,m}
	\left \|(\bm{\partial}U_{\star})(\bm{q}_{i})-(\bm{\partial}U_{t_{i}})(\bm{q}_{i})\right \|^{2}
	\nonumber
\end{align}
Inserting this result into (\ref{app:KL:fd}) and passing to the continuum limit recovers (\ref{model:KL}).

\subsection{Case~\hyperref[itm:b]{EP}}

The starting point is (\ref{app:KL:def}) where we replace $\operatorname{T}_{\star}^{(h)}$ with the transition
probability generated by the \emph{backward} stochastic differential equations
\begin{align}
   &\mathrm{d}^{\flat}\bm{\mathscr{q}}_{t}=\left (\dfrac{\bm{\mathscr{p}}_{t}}{m}
	+\dfrac{g\,\tau}{m}(\bm{\partial}U_{t})(\bm{\mathscr{q}}_{t})\right)\mathrm{d}t
	+\sqrt{ \dfrac{2\,\tau\,g}{m\,\beta}}\,\mathrm{d}^{\flat}\bm{\mathscr{w}}^{(1)}_{t}	
	\nonumber\\
	&
	\mathrm{d}^{\flat}\bm{\mathscr{p}}_{t}=\left(\dfrac{\bm{\mathscr{p}}_{t}}{\tau}
	-(\bm{\partial}U_{t})(\bm{\mathscr{q}}_{t})\right)\mathrm{d}t
	+\sqrt{ \dfrac{2\,m}{\tau\,\beta}}\,\mathrm{d}^{\flat}\bm{\mathscr{w}}^{(2)}_{t}\,,	
	\nonumber
\end{align}
where the label $\flat$ recalls that the evolution proceeds backwards
The one-step approximation of the transition probability density on the lattice using the adapted post-point prescription yields
\begin{align}
	\operatorname{T}_{t_{i},t_{i+1}}^{\flat(h)}(\bm{x}_{i}\mid\bm{x}_{i+1})=
	\dfrac{\exp\big{(}-A_{t_{i},t_{i+1}}^{\flat(h)}(\bm{x}_{i}\mid\bm{x}_{i+1})\big{)}}{Z_{i}^{\flat}}
	\label{app:KL:post}
\end{align}
with
\begin{align}
	A_{t_{i},t_{i+1}}^{\flat(h)}(\bm{x}_{i}\mid\bm{x}_{i+1})&=\dfrac{\beta\,m}{4\,g\,\tau\,h}\left\|\bm{q}_{i}-\bm{q}_{i+1}+
	\left(\dfrac{\bm{p}_{i+1}}{m}+\dfrac{g\,\tau}{m}(\bm{\partial}U_{t_{i+1}})(\bm{q}_{i+1})\right)\,h\right\|^{2}
	\nonumber\\
&\qquad+\dfrac{\beta\,\tau}{4\,m\,h}\left\|\bm{p}_{i}-\bm{p}_{i+1}+
\left(\dfrac{\bm{p}_{i+1}}{\tau}-(\bm{\partial}U_{t_{i+1}})(\bm{q}_{i+1})\right)\,h\right\|^{2}\,.
\nonumber
\end{align}
We recover (\ref{model:ep}) by contrasting ratios of (\ref{app:KL:Ito}) and (\ref{app:KL:post}) over the same time intervals and by
identifying the sum of two finite dimensional approximations of stochastic integrals over the same integrand but evaluated in the pre-point 
and post-point prescription as twice the same integral in the Stratonovich prescription \cite{PavG2014}. We refer to \cite{MaReMo2000} for the
details of the calculation or e.g. to \cite{MGPe2023} a derivation directly in the continuum limit using stochastic calculus and Girsanov formula.

\section{Consistency of the definition of mean entropy production with stochastic thermodynamics}
\label{app:st}

The calculation follows the same steps as \cite{PMG2014}. Let 
\begin{align}
	H_{t}(\bm{x})=\frac{\left\|\bm{p}\right\|^{2}}{2\,m}+U_{t}(\bm{q})
	\nonumber
\end{align}
the kinetic plus potential Hamiltonian specified by the control potential in (\ref{model:ep}). We define the work done on the system  during a time interval $[0,t]$ as
\begin{align}
	\mathscr{W}_{t}=\int_{0}^{t}\mathrm{d}s\,(\partial_{s}H_{s})(\bm{\mathscr{x}}_{s})
	\,\equiv\,\int_{0}^{t}\mathrm{d}s\,(\partial_{s}U_{s})(\bm{\mathscr{x}}_{s})
	\nonumber
\end{align}
Here, $\mathscr{\bm{x}}_{t}$ ia a realization of  (\ref{model:sde}). The partial derivative only affects the explicit time dependence of the control potential. Correspondingly, we identify the heat released by the system during the same realization with the Stratonovich 
stochastic integral
\begin{align}
	\mathscr{Q}_{t}=
	-\int_{0}^{t}\mathrm{d}\bm{\mathscr{x}}_{s}{\cdot}(\bm{\partial}H_{s})(\bm{\mathscr{x}}_{s})\,,
	\nonumber
\end{align}
so that for any path satisfying (\ref{model:sde}) the identity
\begin{align}
	H_{t}=\mathscr{W}_{t}-\mathscr{Q}_{t}
	\label{st:H}
\end{align}
ensures the validity of  the first law of thermodynamics. The dynamics (\ref{mc:sde}) describes an open system in contact with an environment at constant
temperature $\beta^{-1}$. Hence the average heat released by the system in $[0,t]$ also specifies the change of entropy in the environment
\begin{align}
	\Delta \mathscr{S}_{t}^{(e)}=\beta\,\operatorname{E}\mathscr{Q}_{t}
	\label{st:heat}
\end{align}
The total entropy change is the sum of this quantity and the change of the Gibbs-Shannon entropy of the system \cite{AuGaMeMoMG2012}
\begin{align}
	\Delta \mathscr{S}_{t}=\Delta \mathscr{S}_{t}^{(e)}+\operatorname{E}\ln\dfrac{\styleC{p}_{0}(\bm{\mathscr{x}}_{0})}{\styleC{p}_{t}(\bm{\mathscr{x}}_{t})}
	\label{st:th_mep}
\end{align}
In order to justify referring to (\ref{model:ep}) as mean entropy production, we need to show how it is related to (\ref{st:heat}). 
Indeed, from the properties of the Stratonovich integral
\begin{align}
	\beta\,\operatorname{E}\mathscr{Q}_{t}= -\beta\operatorname{E}\int_{0}^{t}\mathrm{d}s\,\bm{v}_{s}(\bm{\mathscr{x}}_{s})\cdot (\bm{\partial}H_{s})(\bm{\mathscr{x}}_{s})
	\nonumber
\end{align}
On the right hand side we introduce the current velocity 
\begin{align}
\bm{v}_{t}(\bm{x})	=\mathsf{J}\cdot (\bm{\partial}H_{t})(\bm{x})
	- \frac{m}{\tau}\mathsf{S}_{g}\cdot\left((\bm{\partial} H_{t})(\bm{x})+\frac{1}{\beta}(\bm{\partial}\ln \styleC{p}_{t})(\bm{x})\right)
	\nonumber
\end{align}
which is most conveniently written in terms of the $2\,d \,\times\,2\, d$ real matrices 
\begin{align}
	\mathsf{J}=
	\begin{bmatrix}
		\mathsf{0} & \mathsf{1}_{d}
		\\
		-\mathsf{1}_{d} & \mathsf{0}
	\end{bmatrix}
	\hspace{1.0cm}\&\hspace{1.0cm}
	\mathsf{S}_{g}=
	\begin{bmatrix}
		\frac{g\,\tau^{2}}{m^{2}}\,\mathsf{1}_{d} & \mathsf{0}
		\\
		\mathsf{0}  & \mathsf{1}_{d}
	\end{bmatrix}
	\nonumber
\end{align}
Probability conservation and antisymmetry of $\mathsf{J}$ and imply
\begin{align}
	\operatorname{E}\int_{0}^{t}\mathrm{d}s\,\partial_{s}\ln \styleC{p}_{s}(\bm{\mathscr{x}}_{s})=
	\operatorname{E}\int_{0}^{t}\mathrm{d}s\, (\bm{\partial}H_{s})(\bm{\mathscr{x}}_{s})\mathsf{J}\cdot(\bm{\partial}\ln \styleC{p}_{s})(\bm{\mathscr{x}}_{s})=0
	\nonumber
\end{align}
and thus allow us to arrive at
\begin{align}
	\beta\,\operatorname{E}\mathscr{Q}_{t}=\operatorname{E}\ln\dfrac{\styleC{p}_{t}(\bm{\mathscr{x}}_{t})}{\styleC{p}_{0}(\bm{\mathscr{x}}_{0})}
+	\frac{m}{\tau}\operatorname{E}\int_{0}^{t}\mathrm{d}s\,
	\left \|\sqrt{\mathsf{S}_{g}}\cdot \left( (\bm{\partial}H_{s})(\bm{\mathscr{x}}_{s})+\frac{1}{\beta}(\bm{\partial}\ln \styleC{p}_{s})(\bm{\mathscr{x}}_{s})\right)\right \|^{2}
	\nonumber
\end{align}
From this expression we readily verify that the thermodynamic mean entropy production (\ref{st:th_mep}) is positive definite. 
Next, we unfold the quadratic form in the integrand
\begin{align}
&	\Delta \mathscr{S}_{t}=\frac{m\,\beta}{\tau}\operatorname{E}\int_{0}^{t}\mathrm{d}s\,
\| \sqrt{\mathsf{S}_{g}}\cdot(\bm{\partial}H_{s})(\bm{\mathscr{x}}_{s})\|^{2}
\nonumber\\
&+\frac{m}{\tau}\operatorname{E}\int_{0}^{t}\mathrm{d}s\,
	\left(2\,(\bm{\partial}H)(\bm{\mathscr{x}}_{s})\cdot (\mathsf{S}_{g}\bm{\partial}\ln \styleC{p}_{s})(\bm{\mathscr{x}}_{s})
	+\frac{1}{\beta}\|\sqrt{\mathsf{S}_{g}}\cdot (\bm{\partial}\ln \styleC{p}_{s})(\bm{\mathscr{x}}_{s})\|^{2}
	\right)
	\nonumber
\end{align}
Under our working the hypothesis that confining mechanical potentials produce probability density decreasing at infinity sufficiently fast, an integration by parts yields
\begin{align}
	0\leq \operatorname{E}\|\sqrt{\mathsf{S}_{g}}\cdot (\bm{\partial}\ln \styleC{p}_{s})(\bm{\mathscr{x}}_{s})\|^{2}
	=-\operatorname{E}\operatorname{Tr}\mathsf{S}_{g} (\partial_{\bm{\mathscr{x}}_{t}}\otimes\partial_{\bm{\mathscr{x}}_{t}}\ln\styleC{p}_{s} )(\bm{\mathscr{x}}_{s})
	\nonumber
\end{align}
We are therefore in the position to apply the chain of identities 
\begin{align}
&	\operatorname{E}\ln\dfrac{\styleC{p}_{t}(\bm{\mathscr{x}}_{t})}{\styleC{p}_{0}(\bm{\mathscr{x}}_{0})}
	=\int_{0}^{t}\mathrm{d}s\operatorname{E}(\partial_{s}+\mathfrak{L}_{\bm{\mathscr{x}}_{s}})\ln \styleC{p}_{s}(\bm{\mathscr{x}}_{s})
	\nonumber\\
&	=\frac{m}{\tau}\int_{0}^{t}\mathrm{d}s
	\operatorname{E}\left(- (\bm{\partial}H_{s})(\bm{\mathscr{x}}_{s})\cdot(\mathsf{S}_{g}\bm{\partial}\ln \styleC{p}_{s})(\bm{\mathscr{x}}_{s})
	+\frac{1}{\beta}\operatorname{Tr}\mathsf{S}_{g} (\bm{\partial}\otimes\bm{\partial}\ln \styleC{p}_{s})(\bm{\mathscr{x}}_{s})
	\right)
	\nonumber
\end{align}
to couch (\ref{st:th_mep}) into the form
\begin{align}
&	\Delta \mathscr{S}_{t}=-\operatorname{E}\ln\dfrac{\styleC{p}_{t}(\bm{\mathscr{x}}_{t})}{\styleC{p}_{0}(\bm{\mathscr{x}}_{0})}
\nonumber\\
&	+\frac{m}{\tau}\operatorname{E}\int_{0}^{t}\mathrm{d}s\,\left(\beta
\| \sqrt{\mathsf{S}_{g}}\cdot(\bm{\partial}H_{s})(\bm{\mathscr{x}}_{s})\|^{2}
+(\bm{\partial}H_{s})(\bm{\mathscr{x}}_{s})\cdot (\mathsf{S}_{g}\bm{\partial}\ln \styleC{p}_{s})(\bm{\mathscr{x}}_{s})
\right)
	\nonumber
\end{align}
The last step is to make use of the kinetic plus potential form of the Hamiltonian. Straightforward algebra and an integration  by parts 
yield
\begin{align}
	\Delta \mathscr{S}_{t}=\mathcal{E}
	\nonumber
\end{align}
Hence, the Kullback-Leibler divergence (\ref{model:ep}) coincides with the thermodynamic entropy production. The above calculation can also be conceptualized as a special case of the general theory expounded in \cite{ChGa2008}.

\section{Proof of the mean entropy production lower bound}
\label{app:ep}

It is well known (see e.g. \cite{ChGa2008,PMG2014}) that (\ref{model:ep}) can be couched into the explicitly positive form
\begin{align}
		\mathcal{E}&=\operatorname{E}_{\mathcal{P}}\int_{\ti}^{\tf}\mathrm{d}t\,\dfrac{m\,\beta}{\tau}
	\left \|\dfrac{\bm{\mathscr{p}}_{t}}{m}+\dfrac{1}{\beta}\partial_{\bm{\mathscr{p}}_{t}}\ln \styleC{f}_{t}(\bm{\mathscr{q}}_{t},\bm{\mathscr{p}}_{t})\right \|^{2}
	\nonumber\\
	&\qquad+	\operatorname{E}_{\mathcal{P}}\int_{\ti}^{\tf}\mathrm{d}t\,\dfrac{\beta\,g\,\tau}{m}
	\left \|(\bm{\partial}U_{t})(\bm{\mathscr{q}}_{t})+\dfrac{1}{\beta}\partial_{\bm{\mathscr{q}}_{t}}\ln \styleC{f}_{t}(\bm{\mathscr{q}}_{t},\bm{\mathscr{p}}_{t})\right \|^{2}
	\label{app:ep:def}
\end{align}
We add and subtract 
\begin{align}
	\bm{k}_{t}(\bm{q})=\int_{\mathbb{R}^{d}}\mathrm{d}^{d}\bm{p}\,\dfrac{\styleC{f}_{t}(\bm{q},\bm{p})}{\tilde{\styleC{f}}_{t}(\bm{q})}\, \dfrac{\bm{p}}{m}
	\label{app:ep:k}
\end{align}
in the squared norm of the first integrand in (\ref{app:ep:def})  and
\begin{align}
	\bm{h}_{t}(\bm{q})=(\bm{\partial}U_{t})(\bm{q})+\dfrac{1}{\beta}
	\,\bm{\partial}_{\bm{q}}\ln \tilde{\styleC{f}}_{t}(\bm{q})
	\nonumber
\end{align}
to the second one. In both expressions we introduce the position marginal density 
\begin{align}
	\tilde{\styleC{f}}_{t}(\bm{q})=\int_{\mathbb{R}^{d}}\mathrm{d}^{d}\bm{p}\,\styleC{f}_{t}(\bm{q},\bm{p})
	\label{app:ep:mp}
\end{align}
Upon expanding the norm squared into inner products, and taking advantage of the cancellation of the mixed term, we get
\begin{align}
  \mathcal{E}\,\geq\,
	\dfrac{m\,\beta}{\tau}\operatorname{E}_{\mathcal{P}}\int_{\ti}^{\tf}\mathrm{d}t\,\left(\dfrac{g\,\tau^{2}}{m^{2}}\left\|\bm{h}_{t}(\bm{\mathscr{q}}_{t})
	\right\|^{2}
	+\left\|\bm{k}(\bm{\mathscr{q}}_{t})\right\|^{2}
	\right)
\end{align}
 For any $g>0$ and $a$, $b$ arbitrary real numbers, the inequality
\begin{align}
(g\, a+b)^2\leq\,(1+g)\,(g\,a^2+b^2)
	\label{app:lb:ineq}
\end{align}
holds true. The upshot is
\begin{align}
	 \mathcal{E}\,\geq\,\dfrac{m\,\beta}{\tau\,(1+g)}\operatorname{E}_{\mathcal{P}}\int_{\ti}^{\tf}\mathrm{d}t\,\left\|\bm{\tilde{v}}_{t}(\bm{\mathscr{q}}_{t})\right\|^{2}
	\nonumber
\end{align}
The vector field appearing on the right hand-side of the inequality
\begin{align}
	\bm{\tilde{v}}_{t}(\bm{q})=\bm{k}_{t}(\bm{q})-\dfrac{g\,\tau}{m}\,\bm{h}_{t}(\bm{q})
	\label{app:ep:cv}
\end{align}
is exactly the current velocity transporting the position marginal distribution:
\begin{align}
	\partial_{t}\tilde{\styleC{f}}_{t}(\bm{q})+\bm{\partial}_{\bm{q}}\cdot\Big{(}\bm{\tilde{v}}_{t}(\bm{q})\tilde{\styleC{f}}_{t}(\bm{q})\Big{)}=0
	\label{app:ep:mpeq}
\end{align}
We are therefore in the position to apply the Benamou-Brenier inequality \cite{BeBr2000}
\begin{align}
	&	\operatorname{E}_{\mathcal{P}}\int_{\ti}^{\tf}\mathrm{d}t\left\|\bm{\tilde{v}}_{t}(\bm{\mathscr{q}}_{t})\right\|^{2}\,\geq\,\operatorname{E}_{\mathcal{P}}\dfrac{\left\|\int_{\ti}^{\tf}\mathrm{d}t\,\bm{\tilde{v}}_{t}(\bm{\mathscr{q}}_{t})\right\|^{2}}{\tf-\ti}
	=\operatorname{E}_{\mathcal{P}}\dfrac{\left\|\bm{\mathscr{q}}_{\ti}-\bm{\mathscr{q}}_{\ti}\right\|^{2}}{\tf-\ti}
	\label{app:lb:BB}
\end{align}
whence we finally recover (\ref{FG:ep}).

\section{Details of the expansion in Hermite polynomials}
\label{app:hermite}
The Hermite polynomials are defined as
\begin{equation} 
	\begin{aligned}
    H_n(\np)=(-1)^n \,e^{\np^2/2} \dfrac{\mathrm{d}^n}{\mathrm{d}\np^n}\,e^{-\np^2/2}\,,
\end{aligned} 
\end{equation} 
so that
$$H_0(\np)=1,\quad H_1(\np)=p,\quad H_2(\np)=\np^{2}-1,\quad \ldots$$
and so on. They fulfill the following orthonormality condition:
\begin{equation} \begin{aligned}
    \av{H_n,H_m}= \int_{\mathbb{R}}\mathrm{d}\np\, \dfrac{e^{-p^2/2}}{\sqrt{2 \pi}}\,H_n(\np) H_m(\np) = n!\, \delta_{n,m}\,.
\end{aligned} \end{equation} 

Let us notice that from the definition of Hermite polynomials it follows:
\begin{equation} 
	\begin{aligned}
    \partial_{\np} H_n(\np)=n\, H_{n-1}(\np)
\end{aligned} 
\end{equation} 
and
\begin{equation} 
	\begin{aligned}
    p H_n(\np)=H_{n+1}(\np)+n\, H_{n-1}(\np)\,.
\end{aligned} 
\end{equation} 
The above identities, together with decomposition~\eqref{eq:pdfhermite}, can be used to write
\begin{equation} 
\begin{aligned}
	p\,\partial_{\nq}\nf_{t}&=p\, \dfrac{e^{-\frac{\np^{2}}{2}}}{\sqrt{ 2\,\pi}} \,\partial_{\nq}\sum_{n}\,\nf_{\nt}^{(n)}\,H_{n}\\
 &=\dfrac{e^{-\frac{\np^{2}}{2}}}{\sqrt{ 2\,\pi}}\,\sum_{n}\,\partial_{\nq}\nf_{\nt}^{(n)}\,(H_{n+1}+n\,H_{n-1})\\
 &=\dfrac{e^{-\frac{\np^{2}}{2}}}{\sqrt{ 2\,\pi}}\left(\sum_{n\geq 1}\,\partial_{\nq}\nf_{\nt}^{(n-1)}\,H_{n}+\sum_{n}\,(n+1)\,\partial_{\nq}\nf_{\nt}^{(n+1)}H_{n}\right)\,.
\end{aligned}    
\end{equation} 

Similarly, one has
\begin{equation} \begin{aligned}
    \partial_{\np} \nf_{t}= -\dfrac{e^{-\frac{\np^{2}}{2}}}{\sqrt{ 2\,\pi}} \sum_{n} \,\nf_{\nt}^{(n)} H_{n+1}    
\end{aligned} \end{equation} 	
and
\begin{equation} \begin{aligned}
    \partial_{\np}^2 \nf_{t} =\dfrac{e^{-\frac{\np^{2}}{2}}}{\sqrt{ 2\,\pi}} \,\sum_{n}\, \nf_{\nt}^{(n)} H_{n+2}\,.	
\end{aligned} \end{equation}

Substituting the above relations into the Fokker-Planck equation~\eqref{eq:fp_dl} and projecting onto $H_n$ we get
\begin{equation} \begin{aligned}
	\partial_{t}\nf_{\nt}^{(n)}+n\,\nf_{\nt}^{(n)}+\varepsilon\,(n+1)\partial_{\nq}\nf_{\nt}^{(n+1)}+\varepsilon\,\partial_{\nq}\nf_{\nt}^{(n-1)}+\varepsilon\,(\partial_{\nq}\nU_{t})\,\nf_{\nt}^{(n-1)}=0	
\end{aligned} \end{equation} 
which can be recast into Eq.~\eqref{eq:eq1hermite}.

A similar approach can be followed for the value function. Recalling Eq.~\eqref{eq:valuehermite} one gets

\begin{equation*} 
	\begin{aligned}
	\cbr{\partial_{t} - n}v_{\nt}^{(n)} &=-\varepsilon\,(n+1) \bcbr{\partial_{\nq}-(\partial_{\nq}\nU_{t})}v_{\nt}^{(n+1)} -\varepsilon\, \partial_{\nq} v_{\nt}^{(n-1)}-\delta_{n,0}\, \dfrac{\varepsilon^{2}}{4}\,\Big{(}\partial_{\nq}\,(\nU_{\star}-\nU_{t})\Big{)}^2\,,
\end{aligned} 
\end{equation*} 
hence Eq.~\eqref{eq:eq2hermite} follows. Eq.~\eqref{eq:eq3hermite} comes from an analogous expansion of Eq.~\eqref{eq:control2_dl}.

\section{Path integral proof of the inequality (\ref{mc:bound})}
\label{app:PI}

We start from the discretized stochastic differential equation 
\begin{align}
	\mathscr{q}_{i+1}-\mathscr{q}_{i}=b_{i}(\mathscr{q}_{i})\,h+\sqrt{2\,\alpha}\,\eta_{i+1}\,\sqrt{h}
	\nonumber
\end{align}
where the label $i$ runs over bins of the time discretization. We take uniform mesh $h$. $b$ is a sufficiently regular drift, and the $ \eta_{i}$'s are independent
identically distributed centered Gaussian random variables with unit variance. 
We set out to compute
\begin{align}
&	\operatorname{E} \left(\int_{0}^{\tf}\mathrm{d}t\, b(\xi_{t})\right)^{2} = \lim_{\substack{h \downarrow 0\\ Nh =\tf}}\int \frac{\prod_{k=0}^{N+1}\mathrm{d}x_{k}}{Z_{h}} \operatorname{p}(x_{0})\,\prod_{j=0}^{N}\,\operatorname{T}_{j}(x_{j+1}\mid x_{j})\left(\sum_{k=0}^{N}\,b(x_{k})\,h\right)^{2} 
	\nonumber
\end{align}
with
\begin{align}
	\operatorname{T}_{i}(x_{i+1}\mid x_{i})=\exp\left(- \frac{(x_{i+1}-x_{i}-b(x_{i})\,h)^{2}}{4\,\alpha\,h}\right)
	\nonumber
\end{align}
and $ Z_{h}$ a mesh dependent normalization constant. We emphasize the use of the pre-point prescription in our construction of finite dimensional approximations of the path integral. We perform the change of variables
\begin{align}
	&	y_{i}=x_{i}-x_{i-1}-b(x_{i-1})\,h && i\,\geq\,1
	\nonumber\\
	&x_{0}=y_{0}
	\nonumber
\end{align}
in consequence whereof the chain of identities
\begin{align}
	\sum_{i=0}^{N}b(x_{i})\,h &= \sum_{i=0}^{N}(x_{i+1}-x_{i}-y_{i+1})
\nonumber\\
&=x_{N+1}(y_{N+1},\dots,y_{0})-y_{0}-\sum_{i=0}^{N}\,y_{i+1}
	\nonumber
\end{align}
holds true. As the change of variables in the pre-point prescription has unit Jacobian, we arrive at
\begin{align*}
	\operatorname{E} \left(\int_{0}^{\tf}\mathrm{d}t \,b(\xi_{t})\right)^{2}&=\lim_{\substack{h \downarrow 0\\ N h =\tf}}\int\, \prod_{k=0}^{N+1}\mathrm{d}y_{k}\, \frac{\operatorname{p}(y_{0})}{Z_h}\\
 &\qquad\times\,e^{-\sum_{i=1}^{N+1}  \frac{ y_{k}^{2}}{4\alpha h}}\left(x_{N+1}(y_{N+1},\dots,y_{0})-y_{0}-\sum_{i=0}^{N}\,y_{i+1}\right)^{2}
\nonumber\\
&\geq\,\lim_{\substack{h \downarrow 0\\ N h =\tf}} \int \mathrm{d}y_{0} \operatorname{p}(y_{0})\,\Big{(} X_{N+1}(y_{0})-y_{0}\Big{)}^{2}
	\nonumber
\end{align*}
by the Cauchy inequality, with
\begin{align}
	X_{N+1}(y_{0})=\int\,\prod_{k=1}^{N+1}\,\mathrm{d}y_{k}\,e^{- \frac{ y_{k}^{2}}{4\alpha h}}\,x_{N+1}(y_{N+1},\dots,y_{0})\,.
	\nonumber
\end{align}
This inequality holds for any drift and therefore also for the one implementing the optimal bridge.

\section{Further details on the order-by-order multiscale expansion}
\label{app:oboe}

In this Appendix, we present additional details about the order-by-order solution of the multiscale problem presented in Section~\ref{sec:maobo}. While it is not essential to follow the logic of our method, these intermediate steps may be a useful reference for the reader interested in the detailed derivation of the results.

In Section~\ref{sec:sys1}, we describe how to obtain a second order differential equation in $\st{0}$ for $\nf_{\slt{0}}^{(1:1)}$, namely Eq.~\eqref{eq:diff2order}.
The first step is to find a relation for $v^{(1:1)}_{\slt{0}}$ by plugging Eq.~\eqref{eq:u_appf} in Eq.~\eqref{eq:fp_appf}. We get
    \begin{equation}
    \label{eq:v11}
       v^{(1:1)}_{\slt{0}}=\begin{cases}
           -g\, \partial_{\nq} v^{(0:0)}_{\ttf,\slt{1}} -\dfrac{1+g}{2}\,\cbr{\partial_{\nq}\cbr{U_{\star}+\ln \nf_{0\vv\slt{2}}^{(0:0)}} +\dfrac{\partial_{\st{0}}\nf_{\st{0}\vv\slt{2}}^{(1:1)}}{\nf_{0\vv\slt{2}}^{(0:0)}}+\cbr{1+\dfrac{4\, v^{(2:0)}_{\st{0},\slt{1}}}{1+g}}\dfrac{\nf_{\st{0}\vv\slt{2}}^{(1:1)}}{\nf_{0\vv\slt{2}}^{(0:0)}} } \quad&\syma\\[0.3cm]
            -g\,\partial_{\nq}\cbr{\ln \nf_{0\vv\slt{2}}^{(0:0)}+  v^{(0:0)}_{\ttf,\slt{1}}} - \dfrac{2\,g}{ \nf_{0\vv\slt{2}}^{(0:0)} } \cbr{\partial_{\st{0}}\nf_{\st{0}\vv\slt{2}}^{(1:1)}+\cbr{1+\dfrac{ v^{(2:0)}_{\st{0},\slt{1}}}{g}}\nf_{\st{0}\vv\slt{2}}^{(1:1)} }\,. \quad&\symb
        \end{cases} 
    \end{equation}
    
By differentiating in $\st{0}$ and eliminating $v^{(1:1)}_{\st{0},\slt{1}}$ and its time derivative through~\eqref{eq:v11} itself and~\eqref{eq:vf_appf}, one gets Eq.~\eqref{eq:diff2order}. The explicit expression of its right hand side reads
    \begin{equation}
    \label{eq:rhsF}
        F_{\slt{0}}=\begin{cases}
           \nf_{0\vv\slt{2}}^{(0:0)}\,\partial_{\nq} \cbr{2 \,v^{(0:0)}_{\ttf,\slt{1}}+U_{\star}+\ln\nf_{0\vv\slt{2}}^{(0:0)}   }+\dfrac{4}{1+g}\,\partial_{\nq} \cbr{v^{(2:0)}_{\st{0},\slt{1}}\,\nf_{0\vv\slt{2}}^{(0:0)}}\quad&\syma\\[0.3cm]
            \dfrac{\omega^2}{2}\,\nf_{0\vv\slt{2}}^{(0:0)}\,\partial_{\nq}\cbr{v^{(0:0)}_{\ttf,\slt{1}}+\ln\nf_{0\vv\slt{2}}^{(0:0)}}+\dfrac{1}{g}\,\partial_{\nq}\cbr{v^{(2:0)}_{\st{0},\slt{1}}\,\nf_{0\vv\slt{2}}^{(0:0)}}-\dfrac{\partial_{\nq}\nf_{0\vv\slt{2}}^{(0:0)} }{2\,g}\,.\quad&\symb\,
        \end{cases}
    \end{equation}
The dependence of the value function on $\st{1}$ can be actually dropped, since no resonant equation holds for $v^{(0:0)}_{\slt{0}}$ and $v^{(2:0)}_{\slt{0}}$ on that time scale.
This result allows us to find~\eqref{eq:sys1:f11} through the use of the Green function~\eqref{eq:green}:
$$
\nf_{\st{0}\vv\slt{2}}^{(1:1)}=\int_{\mathbb{R}}\mathrm{d}s \, G_{\st{0},s}F_{s\vv\slt{2}}\,.
$$

Once an explicit expression for $\nf_{\st{0}\vv\slt{2}}^{(1:1)}$ is known (i.e., Eq.~\eqref{eq:sys1:f11}), it can be substituted into Eq.~\eqref{eq:f11bc} to get the relation
$$
\frac{G^{(0:2)}\,\omega^2}{G^{(2:2)}}\,\nf^{(0:0)}_{0\vv\slt{2}}\,\partial_q \zeta_{\slt{2}}=\begin{cases}
           \frac{4}{(1+g)}\,\partial_{\nq}\cbr{v^{(2:0)}_{\ttf\vv\slt{2}}\,\nf^{(0:0)}_{0\vv\slt{2}}}+ C_1\,\nf^{(0:0)}_{0\vv\slt{2}}\quad&\syma\\[0.3cm]
            \frac{1}{g}\,\partial_{\nq}\cbr{v^{(2:0)}_{\ttf\vv\slt{2}}\,\nf^{(0:0)}_{0\vv\slt{2}}}-\frac{1}{2g}\,\partial_{\nq}\nf^{(0:0)}_{0\vv\slt{2}}+ C_2\,\nf^{(0:0)}_{0\vv\slt{2}}\,,\quad&\symb\,
        \end{cases}
$$
where $C_1$ and $C_2$ are constants that can be evaluated considering the limits for $\nq \to \pm\infty$. One then has
\begin{equation}
    \label{eq:v20f00}
    \partial_{\nq}\cbr{v^{(2:0)}_{\ttf\vv\slt{2}}\nf^{(0:0)}_{0\vv\slt{2}}}=\begin{cases}
           -\frac{(1+g)}{4}\,\frac{G^{(0:2)}\omega^2}{G^{(2:2)}}\,\nf^{(0:0)}_{0\vv\slt{2}} \cbr{\partial_{\nq}\zeta_{\slt{2}}-\kappa_{\slt{2}}}\quad&\syma\\[0.3cm]
            -g \,\frac{G^{(0:2)}\omega^2}{G^{(2:2)}}\,\nf^{(0:0)}_{0\vv\slt{2}} \cbr{\partial_{\nq}\zeta_{\slt{2}}-\kappa_{\slt{2}}}+\frac{\nf^{(0:0)}_{0\vv\slt{2}}}{2}\,.\quad&\symb\,
        \end{cases}
\end{equation}

In Section~\ref{sec:sys2} we derive the differential equation~\eqref{eq:sol_o2_v}, which allows closing the system, providing the $\st{2}$-dependence of $\nf_{0\vv\slt{2}}^{(0:0)}$. To this aim, one first needs to substitute the expression for $v^{(1:1)}_{\st{0}\vv\slt{2}}$  given by Eq.~\eqref{eq:v02}, obtaining
\begin{equation}
\label{eq:step1}
         \partial_{\st{0}}v_{\st{0}\vv\slt{2}}^{(0:2)}+\partial_{\st{2}}v_{\ttf\vv\slt{2}}^{(0:0)}- 2 \cbr{\partial_{\nq}-\bcbr{\partial_{\nq}\nU^{(0)}}} \dfrac{v_{\st{0}\vv\slt{2}}^{(2:0)} \,\nf_{\st{0}\vv\slt{2}}^{(1:1)}}{\nf_{0\vv\slt{2}}^{(0:0)}} =\begin{cases}
            \dfrac{1+g}{2}\,\cbr{W_{\star}-W^{(0)}  }\quad&\syma\\[0.3cm]
            g\,\cbr{-\partial_{\nq}\ln \nf_{0\vv\slt{2}}^{(0:0)}- W^{(0)}}\,.\quad&\symb\,
        \end{cases}
\end{equation}
where we took into account Eq.~\eqref{eq:fp_appf2} and we introduced
$$
W^{(0)}=\partial_{\nq}^2\nU^{(0)}-\dfrac{1}{2}\bcbr{\partial_{\nq}\nU^{(0)}}^2
$$
and
$$
W_{\star}=\partial_{\nq}^2\nU_{\star}-\dfrac{1}{2}\bcbr{\partial_{\nq}\nU_{\star}}^2\,.
$$
Because of the boundary conditions~\eqref{eq:bcv} one has
\begin{equation}
\label{eq:trick1}
 \int_{0}^{\tf}\, \mathrm{d}\st{0} \,v^{(0:2)}_{\st{0}\vv\slt{2}}=0\,. 
\end{equation}
Besides, using~\eqref{eq:eq2hermite} at order 0,
$$
 2\,v^{(2:0)}_{\st{0}\vv\slt{2}}=\begin{cases}
           \partial_{\st{0}}v^{(2:0)}_{\st{0}\vv\slt{2}}\quad&\syma\\[0.3cm]
            \partial_{\st{0}}v^{(2:0)}_{\st{0}\vv\slt{2}}+1\quad&\symb\,,
        \end{cases}
$$
hence
\begin{equation}
\label{eq:trick2}
    \int_0^{\tf}\mathrm{d} \st{0} \,\cbr{\partial_{\st{0}}  \nf_{\st{0}\vv\slt{2}}^{(1:1)} + \nf_{\st{0}\vv\slt{2}}^{(1:1)} } \cbr{2\,v^{(2:0)}_{\st{0}\vv\slt{2}} \,\nf_{\st{0}\vv\slt{2}}^{(1:1)} }=\begin{cases}
           0\quad&\syma\\[0.3cm]
            \int_0^{\tf}\mathrm{d} \st{0} \,\cbr{\nf_{\st{0}\vv\slt{2}}^{(1:1)} }^2\quad&\symb\,.
        \end{cases}
\end{equation}
We also notice, recalling Eq.~\eqref{eq:diff2order}, that
\begin{equation}
\label{eq:trick3}
\int_0^{\tf}\mathrm{d} \st{0} \,\cbr{\partial_{\st{0}}\nf_{\st{0}\vv\slt{2}}^{(1:1)} }^2=-\int_0^{\tf}\mathrm{d} \st{0} \,\nf_{\st{0}\vv\slt{2}}^{(1:1)} \,\partial_{\st{0}}^2\nf_{\st{0}\vv\slt{2}}^{(1:1)} = \int_0^{\tf}\mathrm{d} \st{0} \,\nf_{\st{0}\vv\slt{2}}^{(1:1)} \cbr{\omega^2\,\nf_{\st{0}\vv\slt{2}}^{(1:1)}-F_{\slt{0}}}\,
\end{equation}
where $F_{\slt{0}}$ obeys Eq.~\eqref{eq:rhsF}. Taking into account Eqs.~\eqref{eq:fp_appf2},~\eqref{eq:f11bc},~\eqref{eq:trick1},~\eqref{eq:trick2} and~\eqref{eq:trick3}, one obtains from Eq.~\eqref{eq:step1} 
\begin{subequations}
    \begin{eqnarray}
         \syma\quad\partial_{\st{2}}v_{\ttf\vv\slt{2}}^{(0:0)}&=&\frac{1+g}{2}\cbr{W_{\star}+ \frac{\partial_{\nq}^2\nf_{0\vv\slt{2}}^{(0:0)}}{\bcbr{\nf_{0\vv\slt{2}}^{(0:0)}}^2} -\frac{1}{2}\cbr{\frac{\partial_{\nq}\nf_{0\vv\slt{2}}^{(0:0)}}{\nf_{0\vv\slt{2}}^{(0:0)}}}^2}\nonumber\quad\\
         &+&\frac{1+g}{4\,\tf}\int_0^{\tf}\mathrm{d}\st{0}\,\frac{2\,\partial_{\nq}\nf_{\st{0}\vv\slt{2}}^{(1:1)}-\nf_{\st{0}\vv\slt{2}}^{(1:1)}\,\partial_{\nq}\zeta_{\slt{2}}}{\nf_{0\vv\slt{2}}^{(0:0)}}+\int_0^{\tf}\mathrm{d}\st{0}\,\frac{\nf_{\st{0}\vv\slt{2}}^{(1:1)}\,e^{-2(\tf-\st{0})}}{\tf\bcbr{\nf_{0\vv\slt{2}}^{(0:0)}}^2}\partial_{\nq}\cbr{\nf_{0\vv\slt{2}}^{(0:0)}\,v_{\ttf\vv\slt{2}}^{(0:0)}}\quad\nonumber\\
        \symb\quad\partial_{\st{2}}v_{\ttf\vv\slt{2}}^{(0:0)}&=& \frac{1+g}{\tf}\int_0^{\tf}\mathrm{d}\st{0}\,\frac{\partial_{\nq}\nf_{\st{0}\vv\slt{2}}^{(1:1)}-\nf_{\st{0}\vv\slt{2}}^{(1:1)}\,\partial_{\nq}\zeta_{\slt{2}}}{\nf_{0\vv\slt{2}}^{(0:0)}}\nonumber\\
        &+&\int_0^{\tf}\mathrm{d}\st{0}\,\frac{\nf_{\st{0}\vv\slt{2}}^{(1:1)}e^{-2\,(\tf-\st{0})}}{\tf\bcbr{\nf_{0\vv\slt{2}}^{(0:0)}}^2}\cbr{\partial_{\nq}\cbr{\nf_{0\vv\slt{2}}^{(0:0)}\,v_{\ttf\vv\slt{2}}^{(0:0)}}-\frac{1}{2}}\,.\quad \nonumber
    \end{eqnarray}
\end{subequations}
Now we can substitute the explicit expressions for $\nf_{\st{0}\vv\slt{2}}^{(1:1)}$ and $\partial_{\nq}\cbr{\nf_{0\vv\slt{2}}^{(0:0)}v_{\ttf\vv\slt{2}}^{(0:0)}}$ provided by Eqs.~\eqref{eq:sys1:f11} and~\eqref{eq:v20f00}. The integrals are evaluated by making repeated use of Eq.~\eqref{eq:Gkl}. We arrive at
\begin{subequations}
    \begin{eqnarray}
         \syma\quad\partial_{\st{2}}v_{\ttf\vv\slt{2}}^{(0:0)}&-&\frac{1+g}{2}\cbr{W_{\star}+ \frac{\partial_{\nq}^2\nf_{0\vv\slt{2}}^{(0:0)}}{\bcbr{\nf_{0\vv\slt{2}}^{(0:0)}}^2} -\frac{1}{2}\cbr{\frac{\partial_{\nq}\nf_{0\vv\slt{2}}^{(0:0)}}{\nf_{0\vv\slt{2}}^{(0:0)}}}^2}\nonumber\quad\\
         &=&\frac{A}{2}\,\partial_{\nq}^2 \zeta_{\slt{2}}  +\frac{A}{2}\,\partial_{\nq} \zeta_{\slt{2}}\ln\nf_{0\vv\slt{2}}^{(0:0)} -\frac{B}{2}\,\kappa_{\slt{2}}\ln\nf_{0\vv\slt{2}}^{(0:0)} +\frac{A}{4}\cbr{\partial_{\nq}\zeta_{\slt{2}}}^2-\frac{B}{4}\,\kappa_{\slt{2}}\cbr{2\,\partial_{\nq}\zeta_{\slt{2}}-\kappa_{\slt{2}}}\quad\nonumber\\
        \symb\quad\partial_{\st{2}}v_{\ttf\vv\slt{2}}^{(0:0)}&=& A\,\partial_{\nq}^2 \zeta_{\slt{2}}  +A\,\partial_{\nq} \zeta_{\slt{2}}\,\ln\nf_{0\vv\slt{2}}^{(0:0)} -B\,\kappa_{\slt{2}}\ln\nf_{0\vv\slt{2}}^{(0:0)} -A\cbr{\partial_{\nq}\zeta_{\slt{2}}}^2-B\,\kappa_{\slt{2}}\cbr{2\,\partial_{\nq}\zeta_{\slt{2}}-\kappa_{\slt{2}}}\,.\quad\nonumber
    \end{eqnarray}
\end{subequations}

By recalling Eq.~\eqref{eq:solf-int}, we obtain Eq.~\eqref{eq:solv-int}.
It is now possible to compute the time derivative of $\kappa_{\slt{2}}$, by making use of~\eqref{eq:solf-int} and~\eqref{eq:solv-int}:
\begin{align}
\label{eq:derk}
 \partial_{\st{2}}  \kappa_{\slt{2}} &= \int_{\mathbb{R}}\mathrm{d}q \,\cbr{\cbr{\partial_{\st{2}}\nf_{0\vv\slt{2}}^{(0:0)}} \partial_q\zeta_{\slt{2}}+\nf_{0\vv\slt{2}}^{(0:0)} \partial_q\cbr{\partial_{\st{2}}\zeta_{\slt{2}}}}\nonumber\\
 &=\frac{\alpha^2}{A} \int_{\mathbb{R}}\mathrm{d}q \, \cbr{\partial_q^2 U_{\star}}\cbr{\nf_{0\vv\slt{2}}^{(0:0)}\, \partial_q U_{\star}- \partial_q \nf_{0\vv\slt{2}}^{(0:0)}}\,.
\end{align}
The right hand side vanishes for case \hyperref[itm:b]{EP}, and also for case \hyperref[itm:a]{KL} when either $U_{\star}$ is a linear function of $\nq$ (including the physically relevant case $U_{\star}=0$), or $U_{\star}$ is symmetric with symmetric boundary conditions.
By taking into account this result and the definition~\eqref{eq:sigma}, Eq.~\eqref{eq:solv-int} is straightforwardly recast into Eq.~\eqref{eq:sol_o2_v}.

We finally observe that, recalling~\eqref{eq:mus} and~\eqref{eq:ksigma}, one has
$$
\dot{\mu}_{\st{2}}^{({1})}=-(A-B)\kappa_{\st{2}}\,.
$$
Therefore,
$$
\dot{\mu}_{\st{2}}^{({1})}=0
$$
when the right hand side of~\eqref{eq:derk} vanishes.

\bibliography{bridge.bib}

\end{document}